\documentclass[11pt,a4paper]{article}
\usepackage{amsmath}
\usepackage{graphicx}
\usepackage{xcolor}
\usepackage[caption=false]{subfig}
\usepackage{mathrsfs,mathtools}
\usepackage{physics,amssymb}
\usepackage{siunitx}
\usepackage{bm}
\usepackage{braket}
\usepackage{listings}
\DeclareMathOperator{\sinc}{sinc}
\usepackage{cases}
\usepackage{comment}
\usepackage{soul}
\usepackage{cancel}
\usepackage{cases}
\usepackage[utf8]{inputenc}
\usepackage{url}
\usepackage{float}
\usepackage{longtable}
\usepackage[normalem]{ulem}
\usepackage{xspace}
\usepackage{aas_macros}
\setstcolor{red}
\usepackage{jcappub}
\hypersetup{colorlinks=true
,urlcolor=DARKBLUE
,anchorcolor=DARKBLUE
,citecolor=DARKBLUE
,filecolor=DARKBLUE
,linkcolor=DARKBLUE
,menucolor=DARKBLUE
,linktocpage=true
,pdfproducer=medialab
}

\newcommand{\calC}{\mathcal{C}}

\definecolor{MONZA}{HTML}{CF000F}
\definecolor{DARKBLUE}{HTML}{00008b}
\definecolor{DARKMAGENTA}{HTML}{8b008b}
\definecolor{DARKCYAN}{HTML}{008B8B}
\definecolor{DARKORANGE}{HTML}{FF8C00}

\begin{document}

\title{Formation of trapped vacuum bubbles during inflation, and consequences for PBH scenarios}

\author[a]{Albert Escriv\`a}

\author[b]{, Vicente Atal}

\author[c]{and Jaume Garriga}

\affiliation[a]{\mbox{Division of Particle and Astrophysical Science, Graduate School of Science,}  Nagoya University. Nagoya 464-8602, Japan}

\affiliation[b]{Estudio Ocho Casas, Calle Blanco 196, Papudo, Chile}

\affiliation[c]{Departament de F\'isica Qu\`antica i Astrofi\'sica, i  Institut  de  Ci\`encies  del  Cosmos, Universitat de Barcelona. Mart\'i i Franqu\'es 1, 08028 Barcelona, Spain.}

\emailAdd{escriva.manas.albert.y0@a.mail.nagoya-u.ac.jp}
\emailAdd{vicente.atal@gmail.com}
\emailAdd{jaume.garriga@ub.edu}

\date{\today}
\abstract{A class of inflationary scenarios for primordial black hole (PBH) formation include a small barrier in the slope of the potential. There, the inflaton slows down, generating an enhancement of primordial perturbations. Moreover, the background solution overcomes the barrier at a very low speed, and large backward quantum fluctuations can prevent certain regions from overshooting the barrier. This leads to localized bubbles where the field remains ``trapped'' behind the barrier. In such models, therefore, we have two distinct channels for PBH production: the standard adiabatic density perturbation channel and the bubble channel. Here, we perform numerical simulations of bubble formation, addressing the issues of initial conditions, critical amplitude and bubble expansion. Further, we explore the scaling behaviour of the co-moving size of bubbles with the initial amplitude of the field fluctuation. We find that for small to moderate non-Gaussianity $f_{\rm NL} \lesssim 2.6$, the threshold for the formation of vacuum bubbles agrees with previous analytical estimates \cite{2020JCAP...05..022A} to $5\%$ accuracy or so. We also show that the mass distribution for the two channels is different, leading to a slightly broader range of PBH masses when both contributions are comparable. The bubble channel is subdominant for small $f_{\rm NL}$, and becomes dominant for $f_{\rm NL}\gtrsim 2.6$. We find that the mass of PBHs in the bubble channel is determined by an adiabatic overdensity surrounding the bubble at the end of inflation. Remarkably, the profile of this overdensity turns out to be of type-II. This represents a first clear example showing that overdensities of type-II can be dominant in comparison with the standard type-I. We also comment on exponential tails and on the fact that in models with local type non-Gaussianity (such as the one considered here), the occurrence of alternative channels can easily be inferred from unitarity considerations.}
\maketitle
\flushbottom
%%%%%%%%%%%%%%%%%%%%%%%%%%%%%%%%%%%%%%%%%%%%%%%%%%%%%%%%%%%%%%%%%%%%%%%%%%%
\section{Introduction}
Primordial black holes (PBH) forming during the radiation dominated era \cite{1967SvA....10..602Z,Hawking:1971ei,1974MNRAS.168..399C, 1975ApJ...201....1C,1979A&A....80..104N} (see \cite{Escriva:2022duf} for a recent review), 
%which are black holes that could have been generated at the early times of the Universe, are one of the most promising candidates for constituting the missing dark matter
constitute a promising candidate for dark matter
\cite{Chapline:1975ojl,2016PhRvD..94h3504C, 2017JPhCS.840a2032G, 2020ARNPS..70..355C, Carr:2020gox, 2021JPhG...48d3001G, Carr:2021bzv}. Although their dynamics does not involve any exotic physics beyond General Relativity, their formation requires certain assumptions about the initial conditions. These may be prepared during inflation, or at phase transitions in the early universe. PBH's have not been observed so far, but future gravitational wave observations may establish their existence \cite{2018CQGra..35f3001S,2016PhRvX...6d1015A, 2021arXiv211103606T,Murgia:2019duy}. 

A scenario which has received a great deal of attention in the literature is the collapse of large adiabatic perturbations generated during inflation. This requires a specific feature in the inflationary potential, enhancing the power spectrum by many orders of magnitude on a relatively narrow range of short distance scales, well below the CMB scales. The mass of the resulting more statistically relevant PBH's would be roughly of the order $M\sim (GH)^{-1}$, where $H$ is the Hubble radius at the time when the short scale falls within the horizon during the radiation era \cite{PhysRevD.50.7173}.

A host of alternative mechanisms have been proposed, with very diverse statistical properties and PBH mass distributions 
%Some examples 
(see \cite{Escriva:2022duf} for a detailed list and a brief description). Of some relevance for the present work are spherical domain walls \cite{Garriga:1992nm,2017JCAP...04..050D}, and vacuum bubbles \cite{Garriga:2015fdk,Garriga:2012bc,Kusenko:2020pcg,He:2023yvl,Deng:2018cxb} which may be produced by quantum tunneling during inflation. In such scenarios, tunneling is assumed to be a Poissonian process which can happen with nearly constant probability per unit time and volume\footnote{Such scenarios involve a multifield inflationary potential. For the case of domain walls, the simplest example is two field model with a standard slow roll potential in the inflaton direction $\phi$, and with a discrete symmetry which acts in an orthogonal direction $\varphi$. In the $\varphi$ direction the potential has the shape of a double well. In this way, the overall potential contains two parallel valleys. While the inflaton rolls down one of the valleys it has a finite probability per unit time and volume for transiting the neighboring valley by quantum tunneling accross the barrier. This results in the formation of a localized region of the new phase, which is separated from the old one by a domain wall. Under the assumption of near scale invariance, the rate of transition is nearly independent of time. This is only valid approximately, since tunneling rates are exponentially sensitive to parameters, and can vary significantly throughout the inflationary phase. If the double well in the $\varphi$ direction is not degenerate, the discrete symmetry is broken, and we may expect the field in the new valley to get stuck in a different false vacuum. This leads to the formation of relic vacuum bubbles. Such scenarios are very different from the one we consider here, where we are considering single field dynamics, and barrier penetration is not involved.}. Once formed, walls and vacuum bubbles are stretched to large sizes by the inflationary expansion, leading to a nearly scale invariant distribution of relics.
After inflation, the interior of bubbles larger than a critical size will continue inflating in an ambient radiation-dominated Universe. This causes a particular form of gravitational collapse, where the bubble migrates from the parent universe and continues inflating in a baby universe\footnote{When the co-moving size of the vaccuum bubble falls within the horizon, its gravitational field carves a transient wormhole, similar to the Einstein-Rosen bridge of the extended Schwarzschild solution, and the bubble migrates to a baby universe. The bridge eventually pinches off at a black hole singularity. The black hole has two future event horizons, one facing the baby universe and one facing the ambient universe we inhabit~\cite{Garriga:2015fdk, 2017JCAP...04..050D, 2019JCAP...09..073A, 2020JCAP...05..022A, 2020JCAP...09..023D}.}. Parent and baby become causally disconnected soon after 
the co-moving scale of the bubble falls within the horizon and a trapped region forms. From then on, and from the point of view of the parent universe, one is left with a PBH whose mass is, again, of order $M\sim (GH)^{-1}$ a the time of formation. Domain walls above a certain critical size undergo a similar fate, due to their repulsive gravitational field.

Interestingly, vacuum bubbles may also arise naturally in models producing PBHs from large adiabatic fluctuations. In particular, this happens in single field models with a small barrier in the slope of the potential \cite{2019JCAP...09..073A,Mishra:2019pzq,ZhengRuiFeng:2021zoz,Wang:2021kbh,Rezazadeh:2021clf,Iacconi:2021ltm}. The primary role of the barrier is to slow down the inflaton, thus producing an enhancement of curvature perturbations. On the other hand, large backward fluctuations can prevent localized domains from overshooting the barrier, resulting in vacuum bubbles where the field remains trapped behind the barrier \cite{2019JCAP...09..073A,2020JCAP...05..022A}. As a consequence, two different channels for PBH production coexist \cite{2019JCAP...09..073A,2020JCAP...05..022A} : the conventional one from large adiabatic fluctuations, and the bubble channel. The relative importance of both is determined by the sharpness of the barrier. This is in correspondence with the degree of non-Gaussianity of curvature perturbations at those scales, which can be characterized by the parameter $f_{\rm NL}$ in an expansion at low amplitudes. It was estimated in \cite{2020JCAP...05..022A} that PBHs from vacuum bubbles would dominate for relatively sharp barriers, with $f_{\rm NL} \gtrsim 3.5$. 

In a nutshell, this estimate follows from a non-perturbative relation $\zeta=\zeta(\zeta_G)$ between the curvature perturbation $\zeta$ and a Gaussian random field $\zeta_G$, which was obtained analytically by using the $\delta N$ formalism. The relation breaks down for $\zeta_G \gtrsim 1/f_{\rm NL}$, where $\zeta_G$ does not map into a finite or real value of $\zeta$, Thus, while the distribution for $\zeta_G$ is normalized to one, the total probability for obtaining a finite value of $\zeta$ is less than one. Unitarity then suggests that values of $\zeta_G \gtrsim 1/f_{\rm NL}$ must correspond to an alternative channel. In the present case, it is natural to infer that this corresponds to large backward fluctuations producing vacuum bubbles. In order to confirm the accuracy of this argument, however, a numerical simulation is needed.

In what follows, we focus on models with a small barrier in the slope of the potential. A difference with the tunneling scenarios discussed in Refs.~\cite{Garriga:2015fdk, 2017JCAP...04..050D, 2020JCAP...09..023D} is that, there, it is assumed that bubble nucleation happens at a nearly constant rate during inflation. Consequently the PBH mass function is rather broad. By contrast, in the present scenario, bubbles can only form near the time when the background solution is near the top of the barrier. This results in a sharply peaked co-moving size distribution. However, the width of this distribution can only be determined by numerical simulations, which have not been done so far.

With these issues in mind, the purpose of the present paper is a numerical study of the formation and evolution of trapped vacuum bubbles. Initial conditions will be derived from the spectrum of quantum fluctuations near the time when the background solution overshoots the barrier. We will then investigate the critical amplitude of the field fluctuation leading to bubble formation. We will also determine the scaling behaviour of the co-moving size of bubbles, as a function of the amplitude above the critical value. This will allow us to compare the numerical results with those inferred from analytical estimates. Finally, using the numerical results, we will estimate the abundance and mass function of PBH's from the two competing channels.

The plan of the paper is as follows: In Section \ref{sec:intro}, we describe the set-up and conventions. In Section \ref{sec:bubble_fomration}, we describe the setup for bubble formation, including the use of consistent initial conditions. In Section \ref{sec:strategy} we discuss the numerical strategy and the details of the simulations we have implemented. In Section \ref{sec:num_results} we present our results. In particular, we discuss the numerical solution for the curvature fluctuations generated during inflation, the bubble dynamics, and the results for the PBH abundance and mass function. Section \ref{ref:conclusions} is devoted to summary and conclusions, including perspectives for future work.

\section{Set-up} \label{sec:intro}

In this section, we explain the basic setup and ingredients to study  PBH formation from the collapse of adiabatic fluctuations. In particular, we consider the generation of the power spectrum of the curvature fluctuations, %\ref{subsec:inflationary_model} 
its statistical connection with the mean profile of the cosmological fluctuation, the role of non-Gaussianity,
%\ref{subsec:non_gaussian} 
and finally the estimation of the threshold and PBH mass.%\ref{subsec:adiabatic_collapse}.

\subsection{Homogeneous inflationary dynamics}
\label{subsec:inflationary_model}
Let us consider an inflationary scenario driven by a single scalar field, whose action (in Planck units, where $M_{\rm pl}=1/\sqrt{8 \pi G}$ is the reduced Planck Mass) is given by
\begin{equation}
    S = \int d^4 x \sqrt{-g} \left[ \frac{R}{2}  -\frac{1}{2} \partial_{\mu} \phi \partial^{\mu} \phi -V(\phi) \right],
\end{equation}
where $R$ is the Ricci scalar, and $\phi$ is the inflaton. 
For definiteness, we consider a Starobinsky-type potential \cite{Starobinsky:1980te} with a small barrier (or ``bump") given by a Gaussian shape, which is modulated by parameters $A$ (amplitude of the barrier), $\phi_{0}$ (roughly location of the peak of the bump) and $\sigma$ (the with of the bump): 
\begin{equation}
\label{eq:pot_starobinsky}
V(\phi)=V_0^4\left(1-e^{-\sqrt{2/3}\phi}\right)^2\left[1+A e^{-\left(\phi-\phi_0 \right)^2/ \left(2\sigma^2 \right) }  \right] \ .
\end{equation}
We anticipate that the inflationary dynamics will be very sensitive to the specific values of the parameters, as was already noticed in \cite{2019JCAP...09..073A}. 
\begin{figure}[t]
\centering
\includegraphics[width=3.0 in]{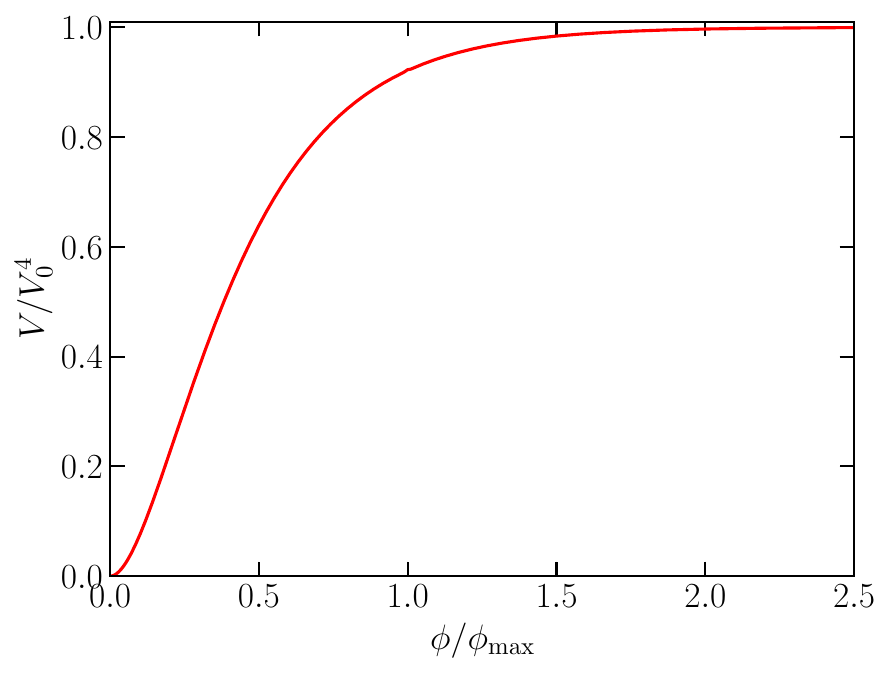}
\includegraphics[width=3.0 in]{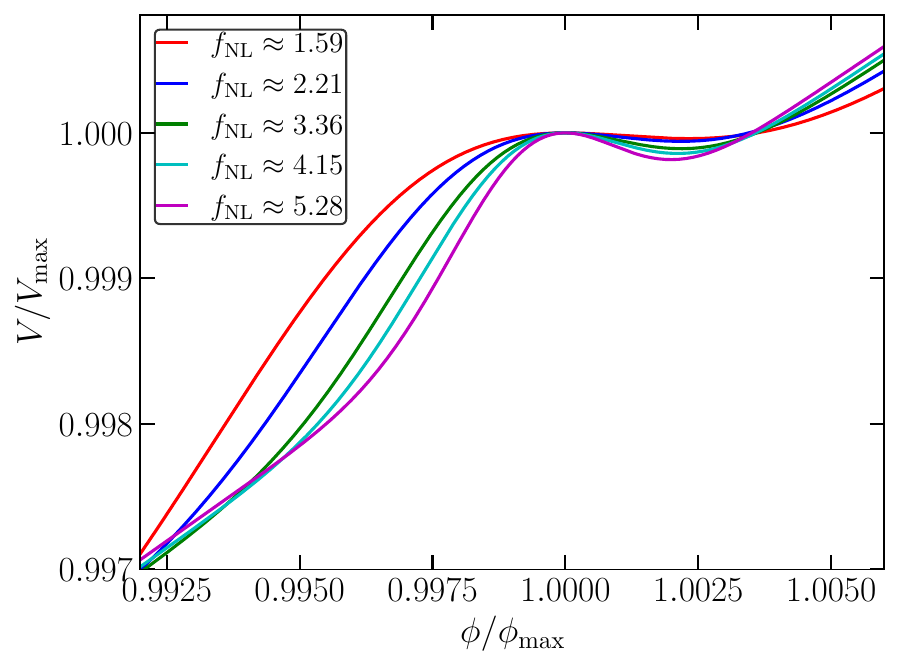}
\caption{Left-panel: Inflationary potential $V(\phi)$ of Eq.\eqref{eq:pot_starobinsky} in terms of $\phi/\phi_{\rm max}$. Right-panel: Shape of the inflationary potential around the bump for different values of $f_{\rm NL}$. The value $\phi_{\rm max}$ indicates the location of the local maxima of the bump and $V_{\rm max}= V(\phi_{\rm max})$. The parameters chosen can be found in Table \ref{table:nuc}.}
 \label{fig:potential}
\end{figure}
A schematic picture of the potential is shown in Fig.\ref{fig:potential}. The homogeneous dynamics of the inflaton field is obtained by solving the Klein-Gordon field equation on a FLRW (Friedmann-Lemaitre-Robertson-Walker) spacetime
\begin{equation}
    ds^{2} = -dt^2+a^2(t) (dr^2+r^2 d\Omega^2),
\end{equation}
where $t$ is the cosmic time, and $a(t)$ is the scale factor. The Hubble rate is defined as $H=d \ln a/dt$, and for our purposes, it will be convenient to use the number of e-foldings $N$ as our time variable. This is defined as $N = \int H dt$. The dynamics of the homogeneous background solution $\phi_{ \rm bkg}(N)$ is given by:
\begin{equation}
\label{eq:phi_homogenea}
    \ddot{\phi}_{ \rm bkg}+ 3 \dot{\phi}_{ \rm bkg}-\frac{1}{2}\dot{\phi}^{3}_{ \rm bkg} + \frac{V_{\phi}(\phi_{ \rm bkg})}{V(\phi_{\rm bkg})} \left( 3 - \frac{\dot{\phi}_{ \rm bkg}^2}{2} \right) = 0.
\end{equation}
Here, and in what follows, a dot will indicate derivative with respect to $N$, that is $\dot{\ } \equiv d/dN$, $V_{\phi}$ is the derivative of the potential with respect to the field, and
we have used $\dot{H}= -H \dot{\phi}_{\rm bkg}^2 /2$ together with the Friedman equation,
\begin{equation}
    H^2 = \frac{V(\phi_{\rm bkg})}{3-\frac{\dot{\phi}_{\rm bkg}^2}{2}}.
    \label{eq:H}
\end{equation}
%We denote $\dot{\phi}_{b}$ and $\ddot{\phi}_{b}$ as the first and second derivative of the field $\phi_b$ in terms of the number of e-folds $N$ respectively and 
 The initial conditions for solving Eq.\eqref{eq:phi_homogenea} are $\phi_{\rm bkg}(N_{\rm ini})$ where $N_{\rm ini}$ stands for the initial value of $N$,  and $\dot{\phi}_{\rm bkg}(N_{\rm ini})$, which is obtained using the slow-roll approximation at $N_{\rm ini}$, $\dot{\phi}_{\rm bkg}(N_{\rm ini})=-V_{\phi}(\phi_{\rm bkg}(N_{\rm ini}))/V(\phi_{\rm bkg}(N_{\rm ini}))$. It is useful to characterize the inflationary dynamics using the so-called Hubble slow-roll parameters defined as $\epsilon_{i+1}= d\ln \epsilon_{i}/ dN$. The first two are given by $\epsilon_{1} = \dot{\phi}^{2}_{\rm bkg}/2$ and $\epsilon_{2}=\dot{\phi}_{\rm bkg} \ddot{\phi}_{\rm bkg}/ \epsilon_{1}$. In the usual slow-roll (SR) phase, both $\epsilon_1$ and $|\epsilon_2|$ are small, but a drastic reduction of the velocity of the inflaton $\dot{\phi}_{\rm bkg}$ (leading to large negative values of $\epsilon_2$) can produce a significant enhancement of the primordial power spectrum responsible for the production of PBHs \cite{Germani:2017bcs,germani-vicente,Ballesteros:2017fsr,Ballesteros:2020qam}.

\subsection{Perturbations}

Scalar perturbations around the FLRW metric can be written in general as,
\begin{equation}
    ds^2 = a^{2}(\tau) \left[  -(1+2 A) d\tau^{2} + B_{,i} d\tau d x^i+\left[ (1-2 \Psi)\delta_{ij}+ E_{,ij} \right] dx^{i} dx^{j} \right]  
    \label{eq:pertub_metric}
\end{equation}
where $\tau = \int dt/a(t)$ is the conformal time, $A$ is the lapse function, $B_{,i}$ is the shift, $\Psi$ determines the curvature scalar of $\tau=const.$ surfaces, and $E$ can be gauged away by spatial reparametrization. We are not including the tensor part since this will not play a role in our discussion. We may now define the gauge invariant combination \cite{Mukhanov:1990me,Maldacena:2002vr},
\begin{equation}
\zeta_G = -\left(\Psi + \frac{\delta \phi}{ \dot{\phi}} \right), 
\end{equation}
which can be thought of as the spatial curvature of $\phi=const.$ hypersurfaces. Since this is defined in linear theory, we call it the Gaussian curvature perturbation. Throughout this work, we will consider the constant mean curvature gauge (or "flat slicing"), where $\Psi=0$, so that $\zeta_G \equiv  -\delta \phi /\dot{\phi}$.

In Fourier space, the amplitude of Gaussian curvature fluctuations $\zeta_{G}$ is characterized by the power spectrum $P_{\zeta_G}(N,k)$, defined by
\begin{equation}
    \langle \zeta_{G}(N,\mathbf{k}) \, \zeta_{G}(N,\mathbf{k'})   \rangle =   \frac{2 \pi^2}{k^3} \mathcal{P}_{\zeta_G} (N,k) (2 \pi)^3 \delta^{(3)}(\mathbf{k}+\mathbf{k'}),
\end{equation}
where $\mathbf{k}$ denotes the wavenumber vector and $k$ its modulus.
The evolution of the modes $\zeta_G(N,\mathbf{k})$ is given by the Mukhanov-Sasaki (MS) equation \cite{Sasaki:1983kd,1988ZhETF..94....1M}:
\begin{equation}
    \label{eq:MS_equation}
    \ddot{\zeta}_{G}+\left(1-\frac{1}{2} \dot{\phi}^{2}_{\rm bkg} +2 \frac{{\dot{z}}}{z}\right)\dot{\zeta}_{G} + \left( \frac{k}{a H} \right)^{2}\zeta_{G} = 0,
\end{equation}
where 
%$z$ is the so-called MS variable. No, no, this is not the MS variable.
$z = a \dot{\phi}_{\rm bkg}$ with  $\dot z= a (\dot{\phi}_{\rm bkg}+\ddot{\phi}_{\rm bkg})$, and $H$ is given directly by Eq.\eqref{eq:H} in terms of $\phi_{\rm bkg}$.

The positive frequency modes $\zeta_{G}$ can be obtained numerically by solving Eq.\eqref{eq:MS_equation} and using the inflationary homogeneous solution for $\phi_{\rm bkg}(N)$ obtained from Eq.\eqref{eq:phi_homogenea}. We use the Bunch-Davies initial conditions \cite{doi:10.1098/rspa.1978.0060} in Eq.\eqref{eq:MS_equation}, at a time characterized by $N_i$ when the modes are well inside the horizon ($k \gg aH$),
\begin{align}
    \label{eq:BD_vacuum}
    \textrm{Re}[\zeta_{G}] &= \frac{1}{\sqrt{2k}} \frac{1}{z(N_{i})}, \,\,\,\, \,\,\,\,\,\,\,\,\,\,\,\,\,\,\,\,\,\,\,\,\,\, \,\,\,\, \,\,\,\, \,\,\,\,\,\,\,\,  \textrm{Im}[\zeta_{G}] = 0, \,\,\,\,  \\  \nonumber
    \textrm{Re}[\dot{\zeta}_{G}] &= -\frac{1}{2k}\frac{1}{z(N_{i})} \left(  \frac{\ddot{\phi}_{\rm bkg}(N_{i})}{\dot{\phi}_{\rm bkg}(N_{i})}+1 \right), \,\,\,\,  \textrm{Im}[\dot{\zeta}_{G}] = -\sqrt{\frac{k}{2}} \frac{1}{a(N_{i}) H(N_{i}) z(N_{i})}.
    \label{eq:BD_vacuum}
\end{align}
From $\zeta_{G}$ we can compute the dimensionless power spectrum of the curvature fluctuation, which is given by
\begin{equation}
    \label{eq:PS_adiabatic}
    \mathcal{P}_{\zeta_G}(N,k) = \frac{k^3}{2 \pi^2}\left. \vert \zeta_{G}(N,k)\right\vert^{2}.
    %_{k \ll aH},
\end{equation}
The power spectrum becomes frozen at sufficiently large scales, much larger than the Hubble radius $k \ll aH$.

At the linear order in the Hubble parameters and assuming the slow-roll approximation, curvature fluctuations are frozen soon after horizon crossing. In this case, their power spectrum at super horizon scales is given by \cite{Mukhanov:2005sc},
\begin{equation}
\label{eq:slow_roll_equations}
\mathcal{P}_{\rm SR}(k)=\left.\frac{H^2}{8 \pi^2 \epsilon_{1}}\right\vert_{k=aH}.
\end{equation}
In our case, this expression is valid on CMB scales. At the much shorter PBH scales, departures from slow-roll will be important, and curvature perturbations may have a significant time dependence after horizon crossing. For that reason, their amplitude at the end of inflation will be computed numerically as outlined above\footnote{For an analytical approach, see \cite{Tasinato:2020vdk}.}. The parameters of the inflationary potential in Eq.\eqref{eq:pot_starobinsky} are chosen so that they fulfill the CMB constraints, and result in a sufficient duration of inflation. In particular: i) the power spectrum at the CMB scales at a reference scale $k_{\rm pivot}= a(N_{\rm pivot})H(N_{\rm pivot}) =0.05\textrm{Mpc}^{-1}$ should be $\mathcal{P}_{\zeta_G}(k_{\rm pivot}) = (2.2 \pm 0.1) \cdot 10^{-9}$ with a spectral index $0.9565 \leq n_{\rm s} \leq 0.9733$ and tensor ratio $r \lesssim 0.06$ at the $95\%$ confident level \cite{Planck:2018jri,Planck:2018vyg} and, ii) the number of e-folds between the end of inflation and the pivot scale must be $44.02 \leq N_{\rm end} - N_{\rm pivot} \leq 54.88$ \cite{Dodelson:2003vq,Liddle:2003as,Haro:2020qos}.

\subsection{Local type non-Gaussianity}
\label{subsec:non_gaussian}

The intrinsic non-Gaussianity $\delta \phi$ which is present before entering the attractor regime in the flat slicing can be estimated to be of order $\epsilon$ \cite{Domenech:2016zxn},which is less than $10^{-2}$ in our model. This means that $\zeta_G\equiv -\delta\phi/\dot\phi$ is also nearly Gaussian around the time of horizon crossing. On the other hand, the relation between $\zeta_G$ and the actual curvature perturbation $\zeta$, which determines the magnitude and fate of overdensities in the radiation dominated era, is highly non-linear in the range of interest. Therefore $\zeta$ is {\em not} a Gaussian random field.

Fortunately, a local relation $\zeta=\zeta(\zeta_G)$ can be found which applies to the present situation \cite{2019JCAP...09..073A}, by using the $\delta N$ formalism. For the sake of clarity, let us briefly review the derivation. Since the height of the barrier is tiny compared to the overall potential, we can safely assume that $H$ is approximately constant. In the vicinity of the local maximum $\phi=\phi_{\rm max} \approx \phi_0$, the potential can be approximated by a quadratic term, and the inflaton behaves as a free field with constant negative mass squared $m^2 \equiv V_{\phi\phi}(\phi_{\rm max})<0$ in de Sitter. The background solution can then be written as
\begin{equation}
    \phi_{\rm bkg}(N)-\phi_{\rm max} = \tilde A\ \left[e^{-\lambda_- \Delta N}-e^{-\lambda_+ \Delta N}\right],
\end{equation}
where we have introduced the notation 
\begin{equation}
\lambda_{\pm}=\frac{3}{2}\pm \alpha, \quad 
\alpha \equiv \sqrt{9/4 - m^2/H^2}>3/2,
\end{equation}
and
 $\Delta N= N-N_{\rm max}$. The integration constant $\tilde A$ is related to the speed at the top of the barrier $\tilde A = (2\alpha)^{-1}\dot\phi_{\rm max}$. Consider now some $N_{\star}$ such that 
 \begin{equation}
 \exp[2\alpha (N_{\star}-N_{\rm max})] \gg 1.\label{attr}
 \end{equation}
 Since $2\alpha>3$, it is clear that this will be satisfied very soon after the maximum, for $$N_{\star}\geq N_{\rm max} + {\cal O}(1).$$
 At later times, the hyperbolic function behaves like a growing exponential, and we have 
\begin{equation}
    \phi_{\rm bkg}(N)-\phi_{\rm max} \approx \tilde A\ e^{-\lambda_- \Delta N}, \quad (N\geq N_{\star}). \label{attractor1}
\end{equation}
In the long wavelength limit, when $k^2/a^2 \ll |m^2| + H^2$, a generic perturbation can be written as 
$ 
\delta\phi = \tilde B e^{-\lambda_+ \Delta N}+ \tilde C  e^{-\lambda_- \Delta N}.
$ 
Near the time of horizon crossing, one expects that growing and decaying components may have comparable magnitude. If this happens before reaching the top, then one expects that by $N\approx N_{\rm max}$ the growing component will be at least comparable to the decaying one. In this case, we have $\tilde B \lesssim \tilde C$, and therefore
\begin{equation}
\delta\phi \approx \tilde C  e^{-\lambda_- \Delta N} \propto \dot\phi_{\rm bkg} .\quad (N\geq N_{\star})\label{attractor2}
\end{equation}
Therefore, for $N>N_{\star}$ we have $\delta \dot\phi =-\lambda \delta\phi \propto \dot\phi$. Since the momentum perturbation is determined by the field perturbation, we call this the attractor behaviour. 

From then on, the Gaussian curvature perturbation stays constant, and is given by 
\begin{equation}
\zeta_G(N>N_{\star}) = \left.{ -\frac{\delta \phi}{ \dot\phi_{\rm bkg}}}\right|_{N_{\star}}.
\end{equation}
Here, $\delta\phi$ is evaluated on a flat slice with $N=N_{\star}$. Alternatively, we may write
\begin{equation}
\phi_{\rm bkg}(N)+\delta\phi(N,{\bf x})=\phi_{\rm bkg}(N+\delta N({\bf x})), \label{attractor3}
\end{equation}
so that the perturbed field $\phi_{\rm bkg}+\delta\phi$ is constant on the slice with $N+\delta N=const.$ As usual, $\delta N$ has the meaning of the decrease in the number of e-foldings from a flat slice until the end of inflation, due to the field perturbation.

It follows from (\ref{attractor1}) and (\ref{attractor3}) that for $N>N_{\star}$ we have $\lambda_-\delta\phi/\dot\phi_{\rm bkg} = 1-e^{-\lambda_-\delta N}$. The non-linear curvature perturbation $\zeta=-\delta N$ can then be expressed as \cite{2019JCAP...09..073A,2020JCAP...05..022A}
\footnote{Similar expressions were recently discussed in \cite{Pi:2022ysn}, where more general potentials are considered. Applied to the present case, the formalism of \cite{Pi:2022ysn} can be shown to reproduce Eq.\eqref{eq:zeta_NG}. Note that we are not neglecting $\delta\dot\phi$ on the initial flat slice. Also, we are not requiring that 
the solution be in the attractor regime already at the maximum of the potential. 
Rather, as pointed out after Eq.\eqref{attr}, we expect the attractor to hold shortly after.}
\begin{equation}
\label{eq:zeta_NG}
    \zeta = - \mu_{\star} \ln(1-\frac{\zeta_G}{\mu_{\star}}),\quad\quad (N>N_{\star})
\end{equation} 
where we have introduced 
\begin{equation}
\label{eq:mu_star_vv4}
\mu^{-1}_{\star}\equiv-\lambda_- 
=  \sqrt{\frac{9}{4}-3 \frac{V_{\phi \phi}(\phi_{\rm max})}{V(\phi_{\rm max})}} -\frac{3}{2}>0.
\end{equation}
In the last equality we have substituted $m^2$ and $H$ in terms of the inflationary potential and its derivatives at the local maximum\footnote{Eq.\eqref{eq:zeta_NG} will not necessarily hold for modes that cross the horizon much later than $N_{\rm max}$. Although an attractor will eventually be reached, and $\zeta_G$ will freeze, this may happen in a regime where the quadratic approximation to the potential is no longer valid. Then, due to the more complicated form of the background solution $\phi_{\rm bkg}(N)$, the relation between $\zeta_G$ and $\zeta$ may in general be different.}. 

It is customary to characterize local non-Gaussianity by the coefficient $f_{\rm NL}$ of the quadratic term in a perturbative expansion \cite{2009ApJ...706L..91V},
\begin{equation}\label{pertex}
 \zeta \approx \zeta_G+\frac{3}{5}f_{\rm NL}\ \zeta^2_G +...
 \end{equation}
  From (\ref{eq:zeta_NG}) with $\zeta_G\ll \mu_{\star}$ we can identify, 
\begin{equation}
\label{eq:mu_star}
   f_{\rm NL} = \frac{5}{6 \mu_{\star}}.
\end{equation}
This agrees with a perturbative determination of non-Gaussianity, which can be obtained diagramatically \cite{germani-vicente,2018JCAP...05..012C}. Here, however, we are interested in the case where both $f_{\rm NL}$ and $\zeta_G$ are ${\cal O}(1)$, and a truncation of the series is not useful. Hence, in what follows we will rely on the full non-perturbative expression (\ref{eq:zeta_NG}), using $f_{\rm NL}$ only as a proxy for different shapes of the barrier (see the right panel of Fig. \ref{fig:potential}). A sharper bump around $\phi_{\rm max}$ corresponds to higher values of $f_{\rm NL}$, and vice versa.

\subsection{Probability distribution and the profile of high peaks}
\label{subsec:high_peaks}
%We may now connect the power spectrum of the curvature fluctuations generated during the inflationary dynamics with the statistical description of the mean profile for the Gaussian curvature fluctuation $\zeta_G(r)$. 

The relation (\ref{eq:zeta_NG}) has a major impact in the probability distribution of adiabatic perturbations, affecting both the amplitudes and profiles of high peaks. These are key ingredients in assessing the resulting distribution of PBH, which require some discussion.

Here, and in what follows, we take the point of view that $\zeta_G$ is the fundamental random variable, with a normalized Gaussian distribution, $P_G[\zeta_G]$. One can of course derive from it a PDF for $\zeta$,
\begin{equation}
P[\zeta] = P_G[\zeta_G(\zeta)]\ {d\zeta_G\over d\zeta}. \label{pzeta}
\end{equation}
From (2.20), we have $\zeta_G=\mu (1-e^{-\zeta/\mu_{\star}})$. In the limit $\zeta\to \infty$, we have $\zeta_G\to \mu_{\star}$, and the first factor in the right hand side of (\ref{pzeta}) remains finite. The Jacobian, on the other hand, is given by
$(d\zeta_G/d\zeta)=e^{-\zeta/\mu_{\star}}$, leading to an "exponential tail" in the distribution 
\begin{equation}
P[\zeta] \propto e^{-\zeta/\mu_{\star}}. \quad(\zeta\to \infty)
\end{equation}
This is in contrast with the stronger Gaussian suppression of $P_G$ at large $\zeta_G$, a fact has been noticed e.g. in \cite{Pi:2022ysn} and references therein. This behaviour, however, stems from the vicinity of a point 
$\zeta_G=\mu_{\star}$ where the Jacobian is singular. 

In our view, therefore,
what is really interesting about the tail is not so much the limit $\zeta\to \infty$, 
but what lies beyond that limit, in the regime $\zeta_G>\mu_{\star}$. In other words, although $P_G[\zeta_G]$ is normalized, $P[\zeta]$ is not because the full range of $\zeta$ corresponds to a restricted range of $\zeta_G$:
\begin{equation}
    \int P[\zeta]\ D\zeta = \int_{\zeta_G<\mu_{\star}} P_G[\zeta_G]\ D\zeta_G < 1.
\end{equation}
The broader lesson is that a singularity in a local non-Gaussianity relation $\zeta(\zeta_G)$ indicates the presence of alternative channels, which restore unitarity\footnote{Related interesting examples of local non-gaussianity leading to two different channels for PBH formation are studied in \cite{Cai:2022erk,Kawaguchi:2023mgk}.}. This will be our focus here.

PBH originate from rare fluctuations of the Gaussian random field $\zeta_G$, with amplitude $\mu$ much larger than the standard deviation, $\mu\gg\sigma_0$. Here, we use the notation
\begin{equation}
\label{eq:sigma}
    \sigma_{n}^2(N) = \int k^{2n} \mathcal{P}_{\zeta_G}(N,k) d \ln k.
\end{equation}
Although perturbations are frozen at sufficiently large scales, for the sake of future reference here we keep the possibility that some of the modes may still be evolving. The $N$ dependence of $\mathcal{P}_{\zeta_G}$ and derived quantities will not be displayed explicitly in what follows, but it should be implicitly understood. Also, in practice, we cut off the momentum integrals at the UV. This regulator is equivalent to a top-hat window function, which eliminates short distance contributions not relevant to the problem.

Rare fluctuations with amplitude $\mu=\nu \sigma_0$, where $\nu\gg 1$ tend to be spherically symmetric. The mean profile for given $\nu$ takes the form \cite{1986ApJ...304...15B}
\begin{equation}
\label{eq:high_peaks}
<\zeta_G(r) \mid \nu
%, \textrm{peak}
> = \mu \Psi_{\zeta_G}(r) 
\end{equation}
where we have introduced the normalized two-point correlation function:
\begin{equation}
    \Psi_{\zeta_G}(r) = \frac{1}{\sigma^2_{0}} <\zeta_G (r)\zeta_G(r=0)>= \frac{1}{\sigma^2_0} \int \mathcal{P}_{\zeta_G}(k)\sinc(kr)d \ln k.
\end{equation}
 This function has a maximum at $r=0$, with $\Psi_{\zeta_G}(r=0)=1$, and typically decays to zero for large $r$. The deviations from the mean profile are characterized by the variance\footnote{Here we are not requiring that the value of $\nu$ corresponds to a peak. Expressions with this restriction are more cumbersome, and can also be found in \cite{1986ApJ...304...15B}. The difference can be appreciable for low to moderate $\nu$. However, the distinction seems hardly relevant for $\nu \sim 10$, which is the case of interest for PBH formation. Places where $\nu\sim 10$ are so rare, that for practical purposes they have to be either at a maximum, or very close to one.} 
 \begin{equation}
 <(\Delta \zeta_G)^2> = \sigma_0^2[1-\Psi_{{\zeta_G}}^2(r)].\label{errorbar}
 \end{equation}
 Hence, the profile can be approximated as 
\begin{equation}
    \zeta_G(r) = \mu \Psi_{\zeta_G}(r) \pm \Delta \zeta_G. \label{bundle}
\end{equation}
The correction $\Delta \zeta_G$ of order $\sigma_0$ need not be spherically symmetric, and has the variance given by (\ref{errorbar}). 
For $\mu\gg \sigma_0$ this term is small by comparison, and so we are led to the conclusion that high peaks are nearly spherically symmetric. Furthermore, such variance goes to zero as we approach $r=0$, where $\Psi_{\zeta_G}=1$. Hence, the non-sphericity is only appreciable far away from the center.  For given $\mu$, 68\% of all configurations of $\zeta_G$ are contained within the bundle (\ref{bundle}) of width $\Delta \zeta_G$ around the mean value. Here, we shall simply consider the mean configuration as a fair representative for each bundle\footnote{In Refs.\cite{2019JCAP...09..073A,2020JCAP...05..022A} the images $\{\zeta[\zeta_G]\}$ of such bundles under the mapping (\ref{eq:zeta_NG}) were considered. 
For the case $ \nu=5 $, it was found that such images have some dispersion $\Delta \mu$ in the critical value of the amplitude $\mu=\mu_c$ 
leading to PBH formation. The differences with the mean configuration were of order $\Delta\mu/\mu \sim$ 10\% at low $f_{\rm NL}\lesssim 1$, 
but quickly dropped to less than 2.5\% or so for $f_{\rm NL}\gtrsim 2$. Here we are considering $\nu\sim 10$ instead, and so we expect the relative dispersion to be smaller. A fair sampling of the bundle is computationally expensive and oustide the scope of the present work. Hence, we leave this as a subject for further research.}.

The amplitude $\mu$ follows the Gaussian distribution,
\begin{equation}
    P(\mu)=\frac{1 }{\sqrt{2 \pi \sigma^2_0}}e^{-\frac{\mu^2}{2 \sigma^2_0}},
\end{equation}
and in the limit  $\nu \gg 1$, the number density of peaks $\mathcal{N}_{\rm pk}(\nu)$ is given by \cite{1986ApJ...304...15B}
\begin{equation}
\label{eq:peak_number}
\mathcal{N}_{\rm pk}(\nu) d \nu =  \left( \frac{\sigma_1}{\sqrt{3} \sigma_0}\right)^{3} (\nu^3-3 \nu) e^{-\frac{1}{2}\nu^2} d\nu ,
\end{equation}
where the $\sigma_{n}$ are computed following Eq.\eqref{eq:sigma}. We will estimate PBH abundances and the mass function following this statistical description accounting for peaks in the curvature fluctuation $\zeta_{G}$. 
%\cite{Garriga}.

As argued in \cite{2019JCAP...09..073A,2020JCAP...05..022A}, perturbations for which $\mu \ge \mu_{\star}$ correspond to portions of the universe that fall back in the false minimum of the potential behind the barrier, leading to trapped vacuum bubbles. These do not admit a description in terms of $\zeta$, simply because inflation does not end inside the bubbles and $\delta N$ is formally divergent. On the other hand, perturbations with $\mu < \mu_{\star}$ might form PBHs from the collapse of the adiabatic fluctuations. Let us now consider the two channels for PBH production in more detail.

\subsection{Collapse of adiabatic fluctuations: threshold and PBH mass}
\label{subsec:adiabatic_collapse}
From now on, $\mu_{a}$ denotes the amplitude of the adiabatic perturbation. The associated profile $\Psi_a\equiv \Psi_{\zeta_G}$ and variance $\sigma_a\equiv \sigma_0(N\gg N_{\star})$ are calculated from the power spectrum once all relevant wavelengths are frozen in, near the end of inflation.
Sufficiently large super-horizon fluctuations $\zeta$ above the critical threshold value $\mu_{a} > \mu_{a,c}$ will collapse forming black holes after horizon reentry in the radiation dominated era. 
These fluctuations will initially be at super-horizon scales, fulfilling that $\epsilon \equiv R_H/R_m \ll 1$, where $\epsilon$ is a dimensionless parameter relating two scales: the Hubble horizon $R_H$ and the length-scale of the fluctuations $R_m$. In the limit $\epsilon \rightarrow 0$ and under the assumption of spherical symmetry\footnote{
A discussed above, in the limit $\nu \gg 1$, fluctuations can be considered approximately spherically symmetric. This consideration extends to the presence of NGs if the peaks are sufficiently large. See previous footnote.}, it was shown in \cite{Shibata:1999zs} that the spacetime metric locally corresponds to
\begin{equation}
    ds^2 = -dt^2+a^{2}(t)e^{2 \zeta(r)} \left(  dr^2+r^2 d \Omega^2 \right).
\end{equation}
Notice that $\zeta(r)$ is time independent at super-horizon scales since $ \partial \zeta/ \partial t \sim \mathcal{O}(\epsilon^2)$ \cite{Shibata:1999zs,Tanaka:2007gh,Lyth:2004gb,Sugiyama:2012tj}.

A useful quantity to measure the strength of a spherically symmetric fluctuation is the compaction function (introduced in \cite{Shibata:1999zs}), that quantifies the mass excess enclosed in the area radius $R(r,t)$ relative to the FLRW background $M_{\rm bkg}=4 \pi \rho_{\rm bkg}R^3/3$,
\begin{equation}
\label{eq:compact}
    \mathcal{C}(r,t) = 2 \frac{M(r,t)-M_{\rm bkg}(r,t)}{R} .
\end{equation}
At super-horizon scales, the compaction function turns out to be time-independent and is related to the curvature fluctuation $\zeta$ \footnote{Adiabatic fluctuations are characterized by a single variable. Historically, the density contrast $\delta \equiv (\rho-\rho_b)/\rho_b$ has often been used in discussions of critical collapse. However it is more convenient to use the curvature perturbation $\zeta$ instead. In the limit of slow-roll parameters the latter variable is Gaussian, while $\delta$ is not. Here $\zeta$ will not be Gaussian, but it is a local function of a Gaussian variable $\zeta_G$. From $\zeta$ one can always obtain $\delta$, but this operation is unnecessary. The numerical code is written directly in terms of $\zeta$.} as \cite{refrencia-extra-jaume}
\begin{equation}
\label{eq:compact_superhorizon}
    \mathcal{C}(r)= \frac{2}{3} \left[1-(1+r\zeta'(r))^{2}\right]+O(\epsilon^2),
    \end{equation}
where the factor $2/3$ accounts for a radiation-dominated Universe. In \cite{universal1}, it was shown that the averaged critical compaction function $\bar{\mathcal{C}}_{c}$ integrated up to its peak value $r_m$ is a good estimator to characterize the threshold for PBH formation giving a universal critical value 
\begin{equation}
\bar{\mathcal{C}}_{c}=2/5.
\end{equation}
This estimator was found to be independent of the shape of the curvature fluctuation within a $\sim 2\%$ deviation, for a wide class of cosmological relevant curvature profiles (see \cite{2021JCAP...01..030E} for a generalization to other perfect fluid equations of state, different from radiation)\footnote{Such a result allows to build an analytical formula for the threshold values in terms of a dimensionless parameter that characterizes the shape around the compaction function peak \cite{universal1}. See \cite{universal1} for details.}. More explicitly, $\bar{\mathcal{C}_{c}}$ is defined as
\begin{equation}
\label{eq:barCm}
\bar{\calC}_{c}=\frac{3}{r^3_m(\mu_{a,c}) e^{3 \zeta(r_m(\mu_{a,c}))}} \int_{0}^{r_m(\mu_{a,c})} \mathcal{C}_{c}(r)(1+r \zeta') e^{3 \zeta(r)} r^2 dr .
\end{equation}
For a Gaussian random field, the position of the peak of the compaction function $r_m$ (obtained solving $\zeta'(r_m)+r_m \zeta''(r_m)=0$ from Eq.\eqref{eq:compact_superhorizon}), does not depend on the amplitude of the fluctuation. However, in the presence of NGs, $r_m$ does depend on the amplitude $\mu_a$, since the shape of the non-gaussian peak depends non-linearly on that parameter. Therefore we have included a dependence $r_m(\mu_{a,c})$ in Eq.\eqref{eq:barCm}.
Notice that two types of fluctuations can be differentiated. Fluctuations of type-I (the "standard" ones) satisfy that the areal radius $R=a \,r \, e^{\zeta}$ is monotonically increasing in $r$. Instead, fluctuations of type-II \cite{Kopp:2010sh} fulfil that there is a point $r_{II}$ such that $R'(r_{II})=0$, which makes $R$ a non monotonic increasing function $R'<0$. This condition translates to $1+r_{II}\zeta'(r_{II})=0$ which implies that the peak value of the compaction function is $2/3$, its maximum possible value in a radiation-dominated Universe. Fluctuations of type-II are usually neglected in the literature and considered unimportant for the estimation of PBH abundances since they are highly over-threshold and, therefore, statistically suppressed. The behaviour and dynamics of fluctuations of type-I have been studied in detail thanks to numerical simulations (see \cite{2022Univ....8...66E} for a review), but this is not the case for type-II (except the analytical exploration done in \cite{Kopp:2010sh}), for which numerical studies are needed. 
%$where in the presence of non-gaussianities, the compaction function peak $r_m(\mu_a)$ will also be dependent on the amplitude of the fluctuation $\mu_{a}$, due to the non-linear relation between $\zeta_G$ and $\zeta$ and therefore we have included a dependence $r_m(\mu_{a,c})$ in Eq.\eqref{eq:barCm}.
In this work we use the remarkable result of $\bar{\mathcal{C}}_{c}=2/5$ to obtain the threshold $\mu_{a,c}$ without the need to perform systematic numerical simulations\footnote{We have indeed verified using simulations of the gravitational collapse of the adiabatic fluctuations using the publicly available code of \cite{escriva_solo} that for the profiles $\zeta$ under consideration the thresholds are within $\sim 1\%$ deviation compared with the ones obtained following the analytical approach. For the smaller case of $f_{\rm NL}$ that we consider, we have $\sim 1\%$ deviation, and it becomes smaller when increasing the value of $f_{\rm NL}$ since $\mu_{a} \rightarrow \mu_{\star}$. Indeed, numerical simulations for $f_{NL} >3.5$ has been shown to be numerically challenging. Therefore in this work we use the analytic estimate based on the averaged critical compaction function to obtain $\mu_{a,c}$.} of the gravitational collapse of the adiabatic fluctuations from the adiabatic channel. This analytical approach has been used successfully already in the context of non-gaussianities in \cite{2020JCAP...05..022A,Kitajima:2021fpq,2022JCAP...05..012E}. The details about the numerical approach are given in section \ref{sec:strategy}.

The PBH mass from the collapse of adiabatic fluctuations follows a critical collapse regime for fluctuations amplitudes $\mu_a$ close to the critical value $\mu_a \rightarrow \mu_{a,c}$ \cite{Niemeyer1,2002CQGra..19.3687H,musco2013}. In particular, taking into account the relation of horizon crossing $r_m e^{\zeta(r_m)}=k^{-1}$ (the fluctuation reenters the cosmological horizon at the time $t_H$, for which $a(t_H)H(t_H)r_m e^{\zeta(r_m)}=1$), the PBH mass is given by \cite{Kitajima:2021fpq}
\begin{equation}
\label{eq:mass_function}
M_{\rm PBH}(\mu_{a}) = \mathcal{K}_{a}(\mu_{a,c}) M_{k}(k) x^2_m(\mu_a) e^{2 \zeta \left(r_m(\mu_{a}) \right)} (\mu_{a}-\mu_{a,c})^{\gamma_{a}} ,
\end{equation}
where $x_m(\mu_a) = r_m(\mu_a) k $. The term $M_k(k)$ is the mass of the cosmological horizon when the single mode $k$ (given in units of $\textrm{Mpc}^{-1}$) reenters it \cite{2019PhRvD.100b3537T,Kitajima:2021fpq}
\begin{equation}
M_{k}(k)  \approx  10^{20} \left( \frac{g_{\star}}{106.75} \right)^{-1/6}\left( \frac{k}{1.56 \cdot  10^{13} \textrm{Mpc}} \right)^{-2} 5.027 \cdot 10^{-34} M_{\odot},
\end{equation}
where $\gamma_{a}$ is a critical universal exponent (independent of the shape of the curvature fluctuation) found to be $\gamma_{a} \approx 0.356$ for radiation \cite{1994PhRvL..72.1782E,Niemeyer1,musco2013}. We also take $g_{\star} \sim 106.75$ (the effective number of degrees of freedom for the energy density of the cosmic fluid at the at the time the scale $k$ reenters the horizon) uniformly in our PBH mass range as in \cite{Kitajima:2021fpq}. For simplicity, we are going to consider a constant value for $\mathcal{K}_{a}(\mu_{a,c})=\mathcal{K}_{a}$=6\footnote{This simplification is commonly done in the literature, since the constant $\mathcal{K}_{a}$ appears just as a linear factor in the PBH mass, and it's precise value $\mathcal{O}(1)$ is does not significantly change the PBH abundance and the mass function.}, which corresponds to the value we have found using numerical simulations for $f_{\rm NL} \approx 1$ (the value of $\mathcal{K}$ depends on the shape of the fluctuation and $f_{\rm NL}$ modulates it). For other cases, we expect a small deviation within a factor $[0.5,3]$ (see Fig.9 in \cite{Escriva:2021pmf}). In \ref{section:mass_function}, we  give details about the procedure followed to estimate the corresponding mass function to PBH abundances.

\section{Bubble-formation}
\label{sec:bubble_fomration}
\begin{figure}[h]
\centering
\includegraphics[width=3.5 in]{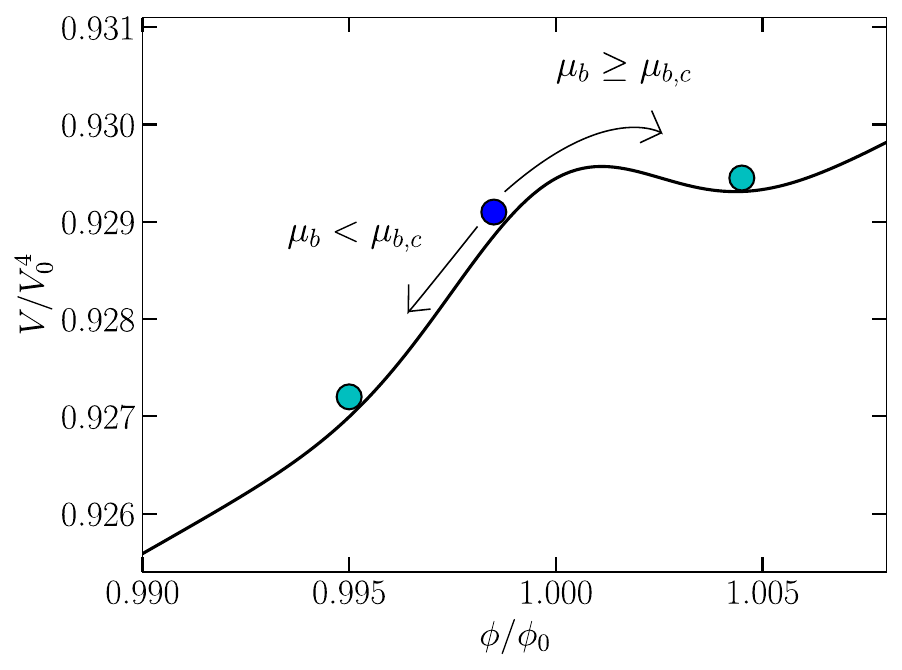}
\caption{Schematic picture of a bubble formation. For fluctuations above the threshold $\mu_{b} > \mu_{b,c}$ a localized region will become trapped producing a vacuum bubble. In the opposite case, for sub-critical fluctuations $\mu_{b}< \mu_{b,c}$ the inflaton overshoots the barrier.}
\label{fig:fluctuation_potential}
\end{figure}

Let's move now to study the channel of PBH production from vacuum bubbles, which needs a detailed discussion. For sufficiently large backward quantum fluctuations, the inflaton field cannot overshoot the barrier and exit inflation, forming trapped vacuum bubbles. This is  schematically represented in Fig.\ref{fig:fluctuation_potential}. In Ref. \cite{2020JCAP...05..022A} $\mu_{a}=\mu_{\star}$ was considered to be the threshold for vacuum bubble formation, so that for $\mu_{a}> \mu_{\star}$ a localized region becomes trapped in the false vacuum. Here, we intend to simulate this process in order to check the validity of this estimation. Therefore, we should define a new amplitude variable $\mu_{b}$ to characterize the fluctuations leading to bubble formation, different from $\mu_{\star}$. The dynamics of the inflaton field under spherical symmetry with a radial dependence $\phi(N,\tilde{r})$ is approximately\footnote{The approximation is that the Hubble rate is determined by the background solution, but the gravitational field of the bubble is ignored. This is a very good approximation during bubble formation, since the height of the barrier is insignificant compared with the overall size of the potential. Nonetheless, it should be kept in mind that once a bubble forms, the field in its interior is stuck in the false vacuum. Hence, near the end of inflation and during the radiation era, the inside will eventually expand at a faster rate than the environment.} given by

\begin{equation}
    \label{eq:non-homogeneus}
    \ddot{\phi}+\dot{\phi} \left( 3- \frac{1}{2} \dot{\phi}^2_{\rm bkg} \right) -\left(\frac{a_I H_I}{a(N) H(N)}\right)^{2} \Delta \phi +\frac{1}{H^{2}}\frac{V_{\phi}(\phi)}{V(\phi)} = 0,
\end{equation}
where $\phi_{\rm bkg}$ corresponds to the background dynamics solution  obtained solving Eq.\eqref{eq:phi_homogenea} and $\Delta$ is the Laplacian operator in spherical coordinates. 
We have included a rescaling on the radial coordinate $r$ in terms of the comoving Hubble scale $a_{I} H_{I}$ as $\tilde{r}=r \cdot (a_{I} H_{I})$, where the sub-index $I$ refers to the time when the initial conditions are set. The consideration of what initial conditions should be introduced is essential, since the dynamics will be strongly dependent on that. 

As shown in Fig.\ref{fig:fluctuation_potential}, we consider that at a specific time $N_{\star}$ after the background field has overshoot the potential barrier, the accumulated inflaton fluctuation at a specific point corresponds to a backward jump of magnitude $\delta\phi$. At this time we have
\begin{align}
\label{eq:initial_conditions}
    \phi(N_{\star},\tilde{r}) &= \phi_{\rm bkg}(N_{\star})+\delta \phi(N_{\star},\tilde{r}), \\
   \dot{\phi}(N_{\star},\tilde{r}) &= \dot{\phi}_{\rm bkg}(N_{\star})+\delta \dot{\phi}(N_{\star},\tilde{r}).
\end{align}
The field perturbation $\delta \phi$ is calculated from the power spectrum of the field $\phi$ at the time $N_{\star}$, which we denote as $\mathcal{P}_{\delta \phi}(N_{\star},k)$. This is computed numerically by solving the MS equation, and evaluating the curvature modes right at $N_{\star}$ 
\begin{equation}
    \label{eq:PS_bubble}
    \mathcal{P}_{\delta \phi}(N_{\star},k) = \frac{k^3}{2 \pi^2} \dot{\phi}^2_{\rm bkg} (N_{\star})  \mid \zeta_{G}(N_{\star}, k) \mid^{2}.
    %_{k = a(N_{\star})H(N_{\star})}^2.
\end{equation}
By analogy with the case of adiabatic curvature perturbations, we can compute the two-point correlation function $\Psi_{b}(\tilde{r})$ associated with $\delta \phi$ as
\begin{equation}
\label{eq:correaltion_bubble}
    \Psi_{b}(N_{\star},\tilde{r}) = \frac{1}{\sigma^2_{b}}\int_{k_i}^{k_f} \mathcal{P}_{\delta \phi}(N_{\star},k) \sinc(k \tilde{r}) d \ln k,
\end{equation}
where $\sigma_{b}$ is the variance of the power spectrum $\mathcal{P}_{\delta \phi}$, and we can define $\nu_b=\mu_b/\sigma_{b}$. Then the initial value of the field fluctuation is given by 
\begin{equation}
\label{eq:initial_condition_space}
\delta \phi(N_{\star},\tilde{r}) = \mu_{b}  \Psi_{b}(N_{\star},\tilde{r}).
\end{equation}
The limits of integration $k_f$ and $k_i$ deserve some discussion. 
These are chosen so that we include all modes that have been enhanced by the dynamics while the scalar field  overshoots the barrier, regardless of whether they are frozen in or not at the time $N_{\star}$. On the other hand, we exclude UV contributions whose amplitude is the same as in flat space. More explicitly, we will consider that a mode has been enhanced at a given scale when the power spectrum is much larger than the one for a scalar field in a Minkowski vacuum, at the same physical scale
\begin{equation}
\mathcal{P}_{\delta \phi}(N_{\star},k)\gg {k^2\over a^2 (2\pi)^2}. \label{exc}
\end{equation}
For a massless field, this holds on superhorizon scales, $k\ll aH$. In our case, $\mathcal{P}_{\delta \phi}(N,k)$ may start growing before horizon crossing, since the effective mass squared is negative, $m^2 =V_{\phi\phi}<0$, near the top of the barrier. This triggers exponential growth of modes with $k^2 \ll a^2 |m^2|$. By the time $N_{\star}$ some modes can be excited with a large variance even if they are still sub-horizon. In practice, we will use the condition 
(\ref{exc}) in order to determine the limits of integration in Eq.\eqref{eq:correaltion_bubble}.

%\end{equation}

In quantum theory, it is not possible to simultaneously specify the field 
$\delta\phi$ and its conjugate momentum $\delta\pi = a^3H\delta \dot \phi$, due to the uncertainty principle. However, here we are dealing with modes whose variance is strongly enhanced relative to the zero point fluctuation, as in Eq.\eqref{exc}. Also, the anticommutator of field and momentum conjugate is much larger than their commutator\footnote{For the enhanced modes, it can be checked that,
$\langle\{\delta\pi({\rm\bf k}), \delta\phi(\rm\bf k)\}\rangle \gg [\delta\pi({\rm\bf k}), \delta\phi(\rm\bf k)]$, so the correlation between field and momentum conjugate far exceeds the level of their intrinsic quantum uncertainty.}. Hence, it seems appropriate to treat such excitations as a classical ensemble. As discussed in Subsection \ref{subsec:non_gaussian}, there is a specific attractor regime for $N \approx N_{\star}$ where $\dot{\phi}_{\rm bkg} \approx - \lambda_{-} (\phi_{\rm bkg}-\phi_{\rm max})$, which also holds for the perturbations \cite{2019JCAP...09..073A}
\begin{equation}
\label{eq:initial_condition_velocity}
\delta \dot{\phi} \approx -\lambda_{-} \delta \phi,
\end{equation}
as discussed after Eq.\eqref{attractor2}. The attractor behavior will not be exact, and we have to numerically monitor to what extent it is satisfied. To optimize our strategy, we define $N_{\star}$ as the time that minimes the departure from the attractor regime, i.e.
\begin{equation}
\label{eq:Delta_N}
    {\Delta_{\rm att}(N_{\star})}= \frac{\dot{\phi}_{\rm bkg}(N_{\star})+\lambda_{-}(\phi_{\rm bkg}(N_{\star})-\phi_{\rm max})}{\dot{\phi}_{\rm bkg}(N_{\star})}.
\end{equation}
As we shall see, for small $f_{\rm NL}$ the attractor holds to good  approximation for a brief interval of e-folds. However, it becomes less well defined for large non-Gaussianity, $f_{\rm NL}\gtrsim 3.5$ (see caption in Fig.\ref{fig:modes_evolution}). This will make our choice of initial conditions less reliable at high values of $f_{\rm NL}$. 

\section{Numerical strategy and details}\label{sec:strategy}
To perform the numerical simulations we have developed a numerical code based on \textit{Python} \cite{van2009python} that allows us to solve in independent blocks the different numerical procedures and simulations that are needed. We take the profit of some implemented procedures to solve differential equations using \textit{Numpy} and \textit{Scipy} libraries \cite{harris2020array,2020SciPy-NMeth}. 
By order of application, we give general details of the implementation as follows:

\begin{itemize}
\item \textbf{Block I (homogeneous solution)}: For each set of parameters of Eq.\eqref{eq:pot_starobinsky} (see Tables \ref{table:nuc} and \ref{table:fixedfraction}), we first solve the background dynamics of the field Eq.\eqref{eq:phi_homogenea} numerically to obtain $\phi_{\rm bkg}(N)$ and $\dot{\phi}_{\rm bkg}(N)$. The background solution is then used to solve the MS equation numerically using Eq.\eqref{eq:BD_vacuum} as the initial conditions for modes well inside the cosmological horizon $k \sim 10^{3} a(N_{\rm crossing})H(N_{\rm crosing})$. From there we obtain the power spectrum $\mathcal{P}_{\zeta_G}$ (the modes $\zeta_{G}$ are evaluated at super-horizon scales, when the modes are frozen with $k \sim 10^{-4} a(N_{\rm crossing})H(N_{\rm crosing})$) and $\mathcal{P}_{\delta \phi}$ (we evaluate $\zeta_{G}$ at $N_{\star}$). Using them, we compute the corresponding two point-correlation functions $\Psi_{a}(\tilde{r})$ and $\Psi_{b}(N_{\star} , \tilde{r})$ making numerically an anti-Fourier transformation. We rescale the radial coordinate as the comoving Hubble horizon at $N_{\star}$ given by $1/(a(N_{\star})H(N_{\star}))$, where $N_{\star}$ is found numerically as the value that minimizes Eq.\eqref{eq:Delta_N}.
\item \textbf{Block II (adiabatic channel)}: Using $\Psi_{a}(\tilde{r})$ we build the mean Gaussian curvature shape $\zeta_{G}$ and the corresponding non-Gaussian curvature $\zeta$ with amplitude $\mu_{a}$ and corresponding parameter $f_{\rm NL}$. We perform a bisection on $\mu_{a}$ to find the critical value $\mu_{a,c}$ such that the averaged critical compaction function is equal to $\bar{\mathcal{C}}_{c}=2/5$ Eq.\eqref{eq:barCm}. Using this value, we can compute the mass spectrum following Eq.\eqref{eq:mass_function} and the abundance of the number of peaks Eq.\eqref{eq:peak_number}. In particular, we compute $x_{m}(\mu_a)=r_{m}(\mu_a) k_{\rm max}$ finding for each $\mu_{a}$ the corresponding location of the maximum of the compaction function $r_{m}(\mu_a)$ and taking into account the position of the maximum of the power spectrum $\mathcal{P}_{\zeta_G}(k_{\rm max})$ for each realization of $f_{\rm NL}$.
\item \textbf{Block III (bubble channel)}: Using $\Psi_{b}(N_{\star},\tilde{r})$ computed from $\mathcal{P}_{\delta \phi}$, we set up the initial condition for bubble formation following the prescription of section \ref{sec:bubble_fomration}. To solve numerically Eq.\eqref{eq:non-homogeneus} we discretize it in the radial coordinate $\tilde{r}$ using finite difference central scheme (second-order accuracy), and we use Runge-Kutta $4$ method to evolve the equation in time. We also implement the following boundary conditions $\phi'(N,\tilde{r}=0)=\phi'(N,\tilde{r} =\tilde{r}_{\rm end})=0$ (where $' \equiv \partial /\partial \tilde{r}$ and being $\tilde{r}_{\rm end}$ the last point of the grid)\footnote{In discretizing the grid we have a limitation in implementing the boundary condition at infinity, which would be the ideal situation since $\phi(N,\tilde{r} \rightarrow \infty) = \phi_{\rm bkg}(N)$. We implement this condition at the last point of our grid at sufficiently large radii where the fluctuation $\Psi_{b}(N_{\star},\tilde{r}_{\rm end})$ is very small compared with the background solution $\phi_{\rm bkg}(N)$.}. In general, we have used $dN\sim 10^{-3.3}$ (time-step) with $d\tilde{r} \sim 10^{-3.6}$ (spatial-resolution) to evolve the equations. We solve the dynamics until the very end of inflation $N_{\rm end}$. At that time, we obtain the profile in $\phi(N,\tilde{r})$ from which we can infer the bubble size and study the tail profile of the scalar field, which will have important consequences on the mass function estimation as we will see in section \ref{section:mass_function}. The threshold $\mu_{b,c}$ can be obtained through a bisection method, and repeating simulations with different values for $(\mu_{b}-\mu_{b,c})$, we can determine the size of the bubble in terms of the amplitude of the fluctuation.
\end{itemize}

Following the numerical procedure detailed above, we obtain parameters for the potential Eq.\eqref{eq:pot_starobinsky} fulfilling two criteria. First, in Table \ref{table:nuc}, we show parameters for which the threshold $\nu_{b,c}=\mu_{b,c}/\sigma_{b}$ takes three different values $\nu_{b,c}  \sim 8,10,12$. This allows us to study the formation of the bubbles as a function of the threshold $\nu_{b,c}$. Second, in Table \ref{table:fixedfraction} we show parameters such that the total fraction of PBH in the form of DM is roughly one. For these parameters we can compute the mass function from the two channels of PBH production in a cosmologically relevant case.

In both cases we find iteratively the value of $A$ fixing the other parameters values $\sigma,\phi_{0},V_{0}$ to fulfill the CMB constraints and have the peak $k_{\rm max}$ of the power spectrum $\mathcal{P}_{\zeta_G}$ at the asteroid mass range $M_{k}(k_{\rm max}) \in [10^{-13}, 10^{-12}] M_{\odot}$ (see Table \ref{table:nuc}, \ref{table:fixedfraction} for details).

\section{Numerical results}\label{sec:num_results}
In this section we present the numerical results we have obtained following the procedures and techniques pointed out in the previous sections and using the parameters of Tables \ref{table:nuc}, \ref{table:fixedfraction}.

\subsection{Enhancement of cosmological fluctuations}
\label{sec:num_results_homogeneus}

\begin{figure}[h]
\centering
\includegraphics[width=2.5 in]{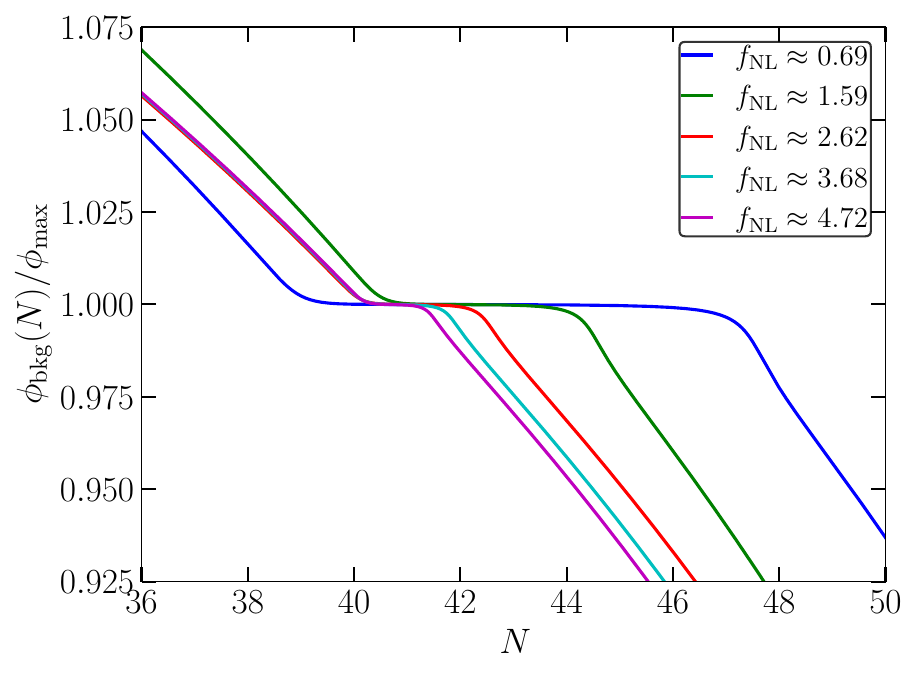}
\includegraphics[width=2.5 in]{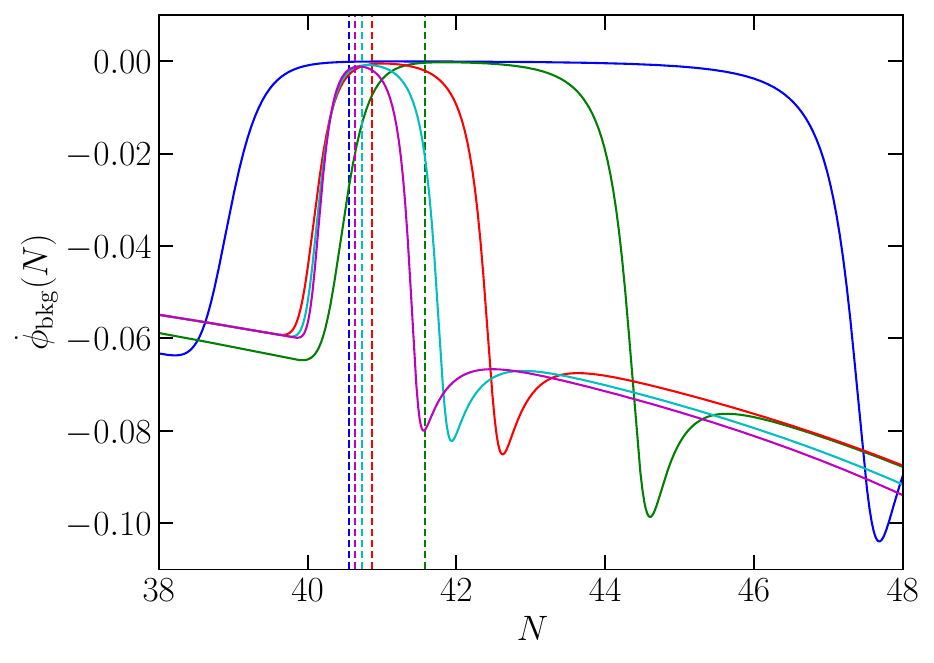}
\includegraphics[width=2.5 in]{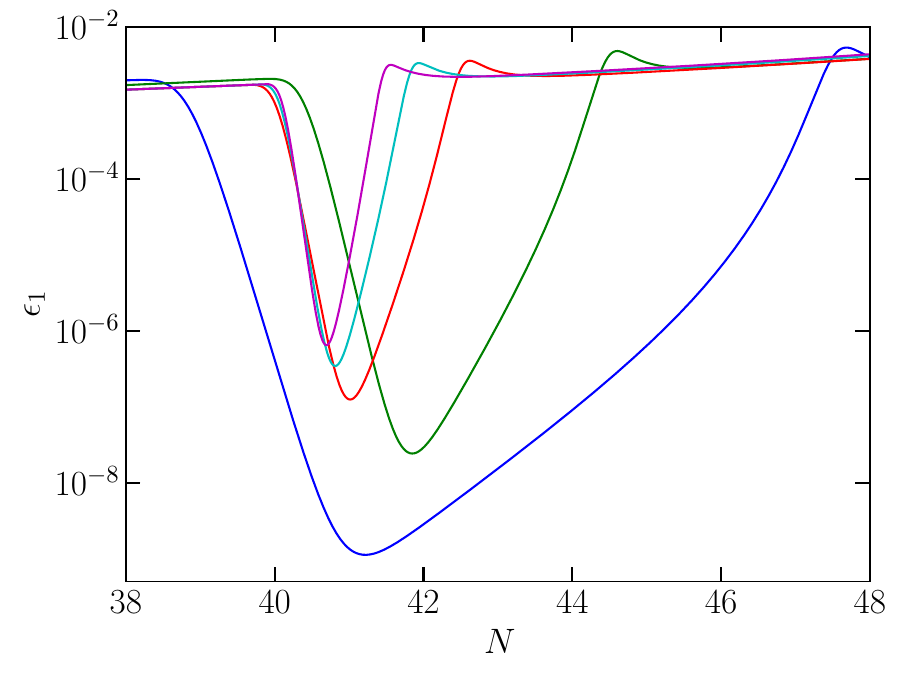}
\includegraphics[width=2.5 in]{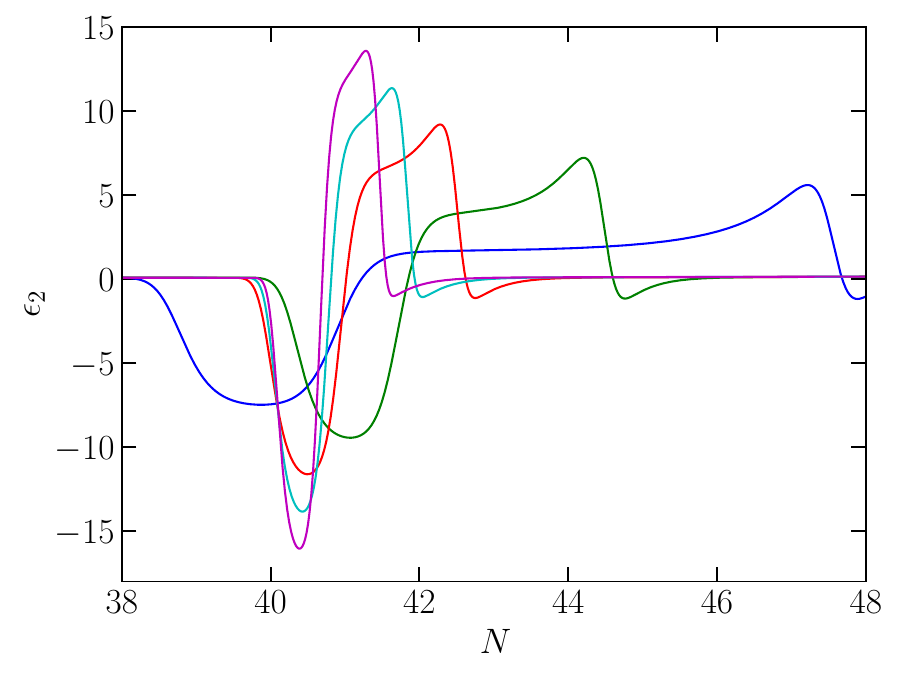}
\caption{Top panels: Dynamics of the inflaton field $\phi(N)$ (left) and $\dot{\phi}_{\rm bkg}(N)$ (right) for different values of $f_{\rm NL}$. The dashed vertical lines indicates the location of $N_{\rm max}$, $\phi_{\rm bkg}(N_{\rm max}) \equiv \phi_{\rm max}$. Bottom panels: Evolution of the Hubble slow-roll parameters $\epsilon_1$ (left), $\epsilon_{2}$ (right) respectively for different $f_{\rm NL}$ values. In all cases, the parameters taken correspond to $\nu_{b,c} \sim 8$ from Table \ref{table:nuc}.}
\label{fig:homogeneus_background}
\end{figure}

The homogeneous solution of the inflationary dynamics of Eq.\eqref{eq:phi_homogenea} is shown in Fig.\ref{fig:homogeneus_background} for some specific values of $f_{\rm NL}$ following the Table \ref{table:nuc}, in particular for the case $\nu_{b,c} \sim 8$. The inflaton dynamics develops a plateau phase when the inflaton overshoots the barrier during a relatively small number of e-folds. The deceleration of the inflation is evident in the panel for $\dot{\phi}_{\rm bkg}(N)$, where the minimum velocity happens slightly before the inflaton approaches the maximum $\phi_{\rm max}$. The duration of this phase is smaller when increasing $f_{\rm NL}$. This is due to the sharper profile of $V$ around the local maximum of the barrier (see right panel of Fig.\ref{fig:potential}). In the bottom panels of Fig.\ref{fig:homogeneus_background} is shown the evolution of the $\epsilon_{1}$, $\epsilon_{2}$ parameters. The corresponding deceleration of the inflation induces negative values in $\epsilon_{2}$.

\begin{figure}[h]
\centering
\includegraphics[width=3.0 in]{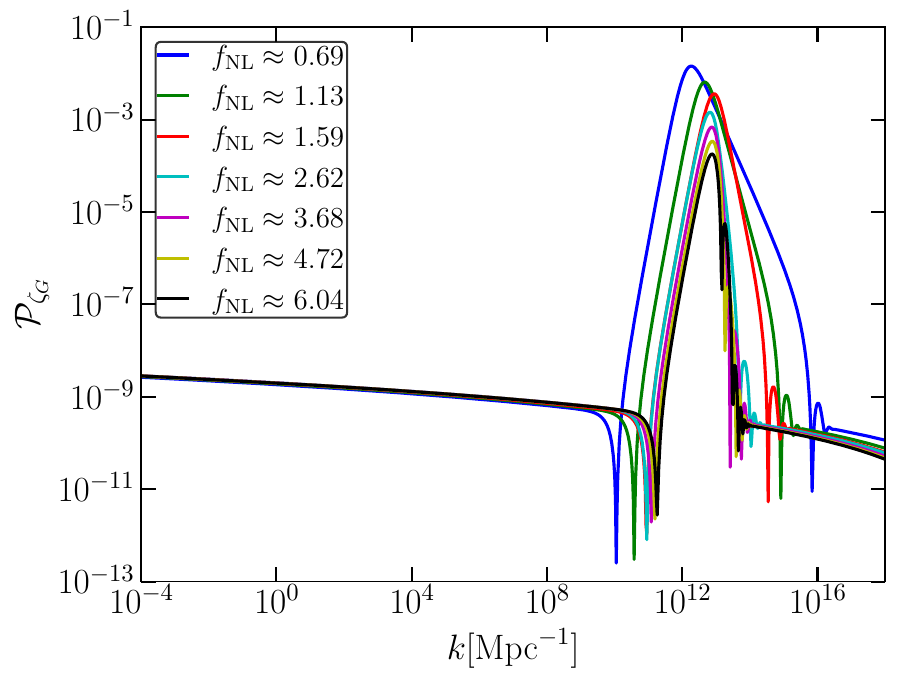}
\includegraphics[width=3.0 in]{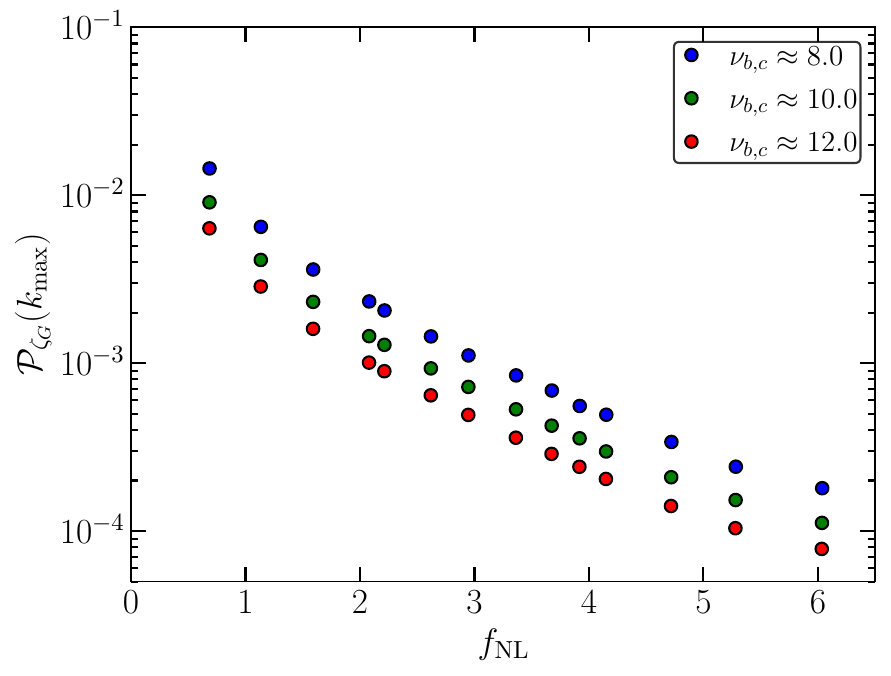}
\caption{Left panel: Power spectrum of the Gaussian curvature fluctuation $\zeta_G$ for different values of $f_{\rm NL}$ for the case $\nu_{b,c} \sim 8$ (see Table \ref{table:nuc}). Right-panel: Reduction of the peak value of $\mathcal{P}_{\zeta_G}(k_{\rm max})$ in terms of $f_{\rm NL}$ for different values of $\nu_{b,c}$.}
\label{fig:PS}
\end{figure}

The power spectrum of the Gaussian curvature fluctuation $\mathcal{P}_{\zeta_G}$ obtained solving the MS equating is shown in Fig.\ref{fig:PS} for different $f_{\rm NL}$ values for the case $\nu_{b,c} \sim 8$. The large enhancement of the PS from the pivot scale $k=0.05 \textrm{Mpc}^{-1}$ to the PBH scales ($ k_{\rm max} \in [10^{12}, 10^{13}]\textrm{Mpc}^{-1}$) due to the significant reduction of $\dot{\phi}_{\rm bkg}$ is clear. The power spectrum $\mathcal{P}_{\zeta_G}$ becomes more monochromatic by increasing $f_{\rm NL}$, since the duration of the deceleration of the inflaton is smaller.

Notice that the peak amplitude $\mathcal{P}_{\zeta_G}(k_{\rm max})$ decreases when we increase $f_{\rm NL}$ for the same fixed value $\nu_{b,c} \sim 8$, which is shown in the right panel of Fig.\ref{fig:PS}. The same qualitative behaviour is found when fixing the fraction of PBHs to account for all the dark matter following the parameters of Table \ref{table:fixedfraction}. For instance the peak $\mathcal{P}_{\zeta_G}(k_{\rm max})$ is reduced by roughly a factor $5$ when increasing $f_{\rm NL}$ from $1.59$ to $3.92$.

This is an interesting feature of non-gaussian models that can be exploited to avoid, for instance, spectral distortion constrains \cite{Atal:2020yic}. It is also relevant for the suppression of various non-gaussian contributions to some observables, as in the case of the induced gravitational waves \cite{Atal:2021jyo}, and could play a role in recent discussions of loop corrections to the power spectrum (see \cite{Kristiano:2022maq,Motohashi:2023syh,Fumagalli:2023hpa,Tada:2023rgp} and references therein). The reason behind the protection against non-Gaussian correction is simple. These contributions are typically proportional to $f^{2}_{\rm NL} \mathcal{P}_{\zeta_G}(k_{\rm max})$. In the case of large $f_{\rm NL}$, when vacuum bubbles dominate, the critical amplitude for PBH production goes as $\mu_{a,c} \sim \nu_{a,c} \,\sigma_{a}\sim 1/f_{\rm NL}$ (since $\mu_{\star} \sim 1/f_{\rm NL}$ as shown in Eq.\eqref{eq:mu_star}). This implies that $f^{2}_{\rm NL} \mathcal{P}_{\zeta_G}(k_{\rm max})\sim f^{2}_{\rm NL}\sigma_a^2 \sim \nu_{a,c}^{-2}$ is roughly independent on the non-Gaussianity parameter, and smaller than one.

\begin{figure}[h]
\centering
\includegraphics[width=3.0 in]{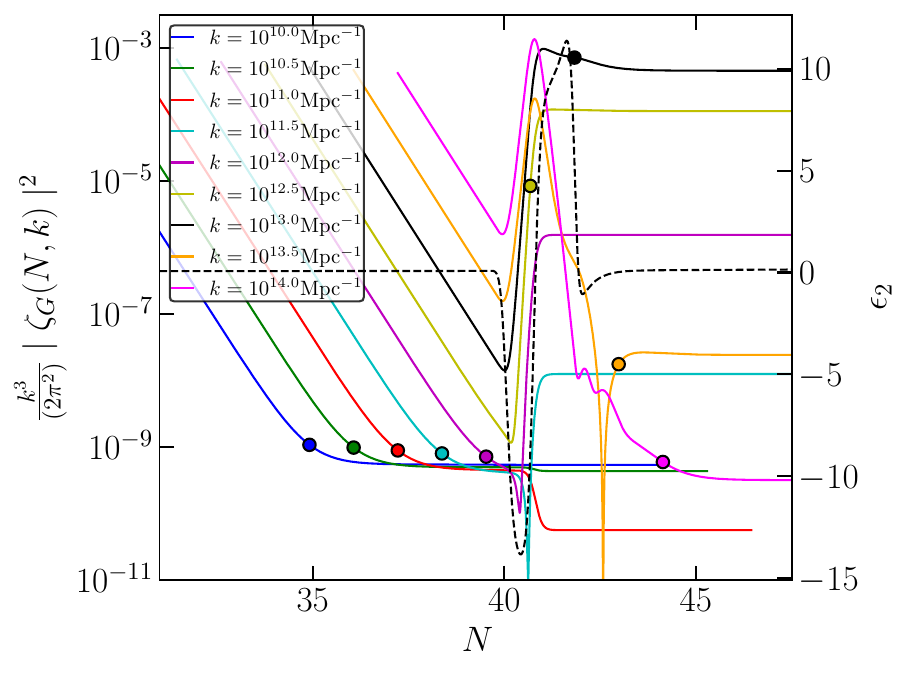}
\includegraphics[width=3.0 in]{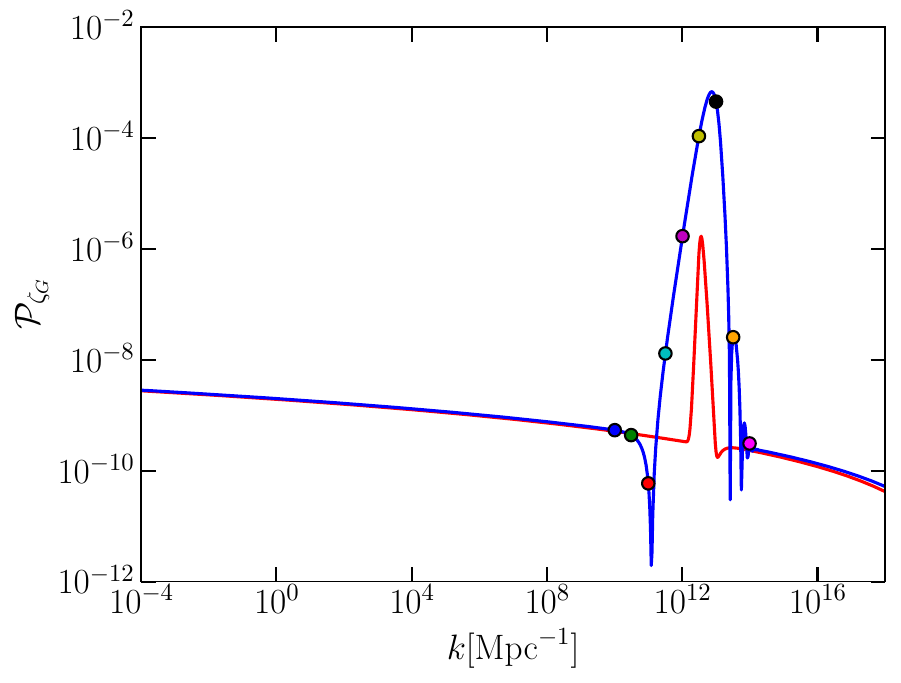}
\caption{Left panel: Evolution of the modes $\zeta_{G}$ in terms of the number of e-folds $N$ for some specific $k$ modes. The solid dots indicate the time when the modes cross the cosmological horizon at $k=a(N_{\rm crossing})H(N_{\rm crossing})$. We show $\epsilon_2$ as a dotted line. Notice that the modes that most contribute to the enhancement of $\mathcal{P}_{\zeta_G}$ become approximately frozen already at $N_{\star} \approx 41.18$, as given by the definition of Eq.\eqref{eq:Delta_N}. This is in agreement with the expected behaviour once the attractor regime sets in. This is not exact as for instance the black dot is not completely frozen at the asymptotic value. The effect is more dramatic for the orange and pink dots but these are very subdominant in the power spectrum. Right-panel: Power spectrum of $\mathcal{P}_{\zeta_G}$ obtained solving the MS equation (blue line) together with the one obtained using the analytical approximation of Eq.\eqref{eq:slow_roll_equations} (red line). The dots correspond to the values of $k$ shown in the left panel. In both cases, we have considered the case $f_{\rm NL} \approx 3.68$ with $\nu_{b,c} \sim 8$.}
\label{fig:modes_evolution}
\end{figure}

On the other hand, in Fig.\ref{fig:modes_evolution}, we show the evolution of the curvatures modes $\zeta_{G}$ from the numerical solution of Eq.\eqref{eq:MS_equation} in terms of the number of e-folds, for different representative modes $k$. The modes start with Bunch-Davies initial conditions well inside the cosmological horizon and freeze once they cross it at sufficiently large super-horizon scales. Qualitatively, modes that encounter a sufficiently large deceleration/acceleration of $\dot{\phi}_{\rm bkg}$ when they are roughly at horizon crossing can be enhanced/suppressed depending on the specific ratio $aH/k$. See \cite{Ballesteros:2020qam} for a detailed description in terms of the MS equation with another type of inflationary model.

Finally in the right panel of Fig.\ref{fig:modes_evolution} we show the comparison of the numerical results with the analytical estimation of Eq.\eqref{eq:slow_roll_equations} (red line), where can be observed the underestimation of the enhancement from the assumption of the SR phase when computing the PS.
%%%%%%%%%%%%%EXPLICAR

\begin{figure}[h]
\centering
\includegraphics[width=3.5 in]{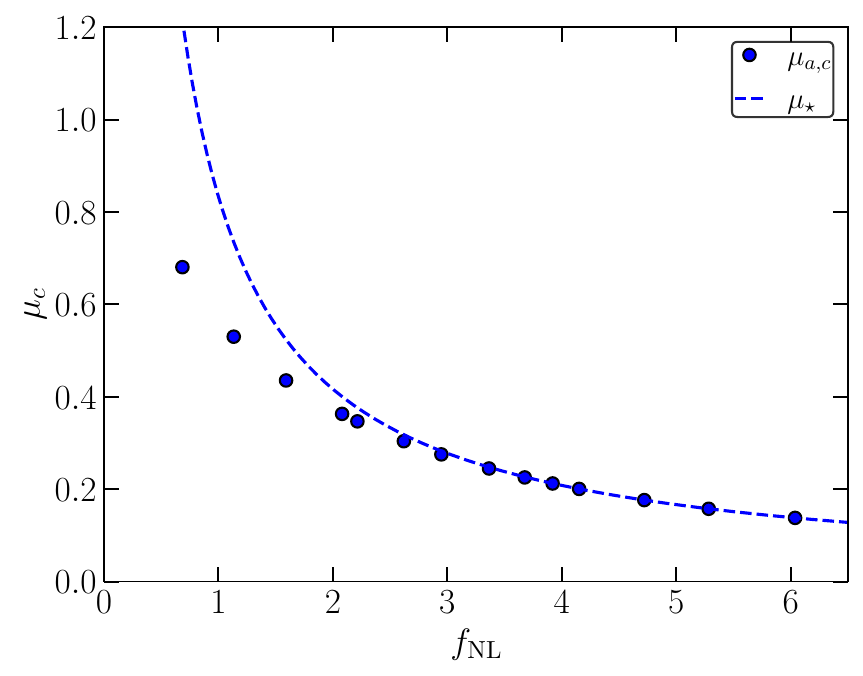}
\caption{Threshold values $\mu_{a,c}$ obtained using the approach of the average of the compaction function (solid dots) in terms of the values $f_{\rm NL}$ and compared with the analytical values $\mu_{\star}=5/(6 f_{\rm NL})$ (dashed line).}
\label{fig:adiabatic_thresholds}
\end{figure}

Making the anti-Fourier transformation of the power spectrum, we can obtain the two-point correlation function $\Psi_{a}(\tilde{r})$ to build the full non-linear gaussian curvature fluctuation $\zeta$. Then the compaction function $\mathcal{C}(r)$ can be obtained following Eq.\eqref{eq:compact_superhorizon} and using the numerical procedure of the averaged compaction function pointed out in section \ref{sec:strategy}, we can finally obtain the threshold values $\mu_{a,c}$ for the collapse of the adiabatic fluctuation $\zeta$. The result is shown in Fig.\ref{fig:adiabatic_thresholds}.

This is similar to what was obtained in \cite{2020JCAP...05..022A}, with the difference that there were used numerical simulations with the code of \cite{escriva_solo} and taking an analytical approximation for the shape of the power spectrum. As already pointed out in that work, we observe a reduction of the threshold $\mu_{a,c}$ as the level of NGs increases. In the large $f_{\rm NL}$ limit, we observe that $\mu_{a,c} \rightarrow \mu_{\star}$. In this regime, it is expected that the production of PBHs from the adiabatic channel will decrease due to the proximity of the critical threshold $\mu_{a,c}$ to the limiting value $\mu_{\star}$, which means that only a small range of allowed curvature fluctuations will be able to collapse forming PBHs. We will use $\mu_{a,c}$ to estimate the mass function and PBH mass in \ref{section:mass_function}. Notice that the values are independent on $\nu_{b,c}$ since the threshold for PBH formation from the adiabatic channel ($\mu_{a,c}$) mainly depends on the shape around the critical compaction function's peak, as shown and discovered in \cite{universal1}.

\begin{figure}[h]
\centering
\includegraphics[width=3.4 in]{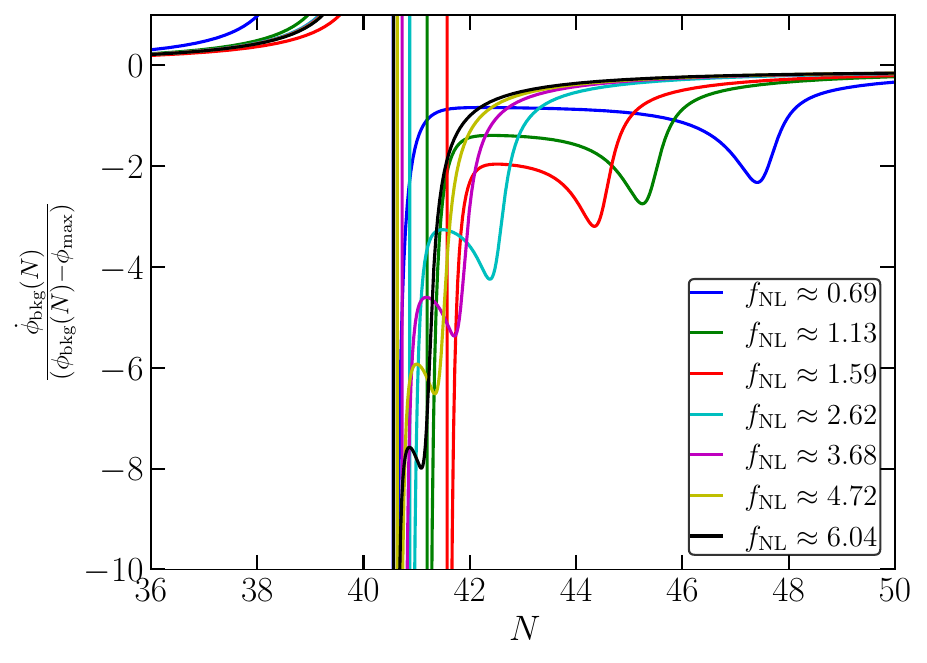}
\includegraphics[width=2.7 in]{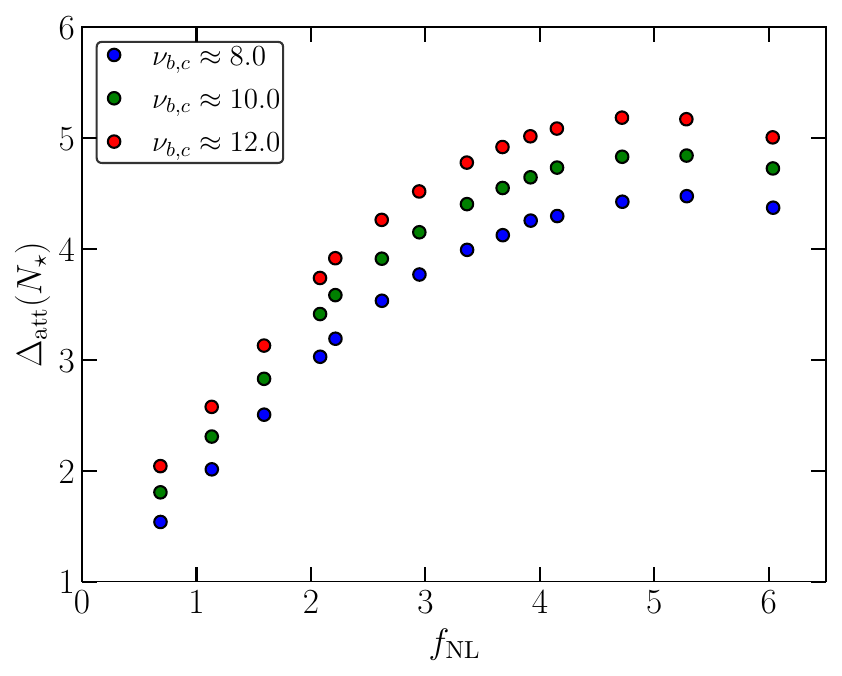}
\includegraphics[width=2.9 in]{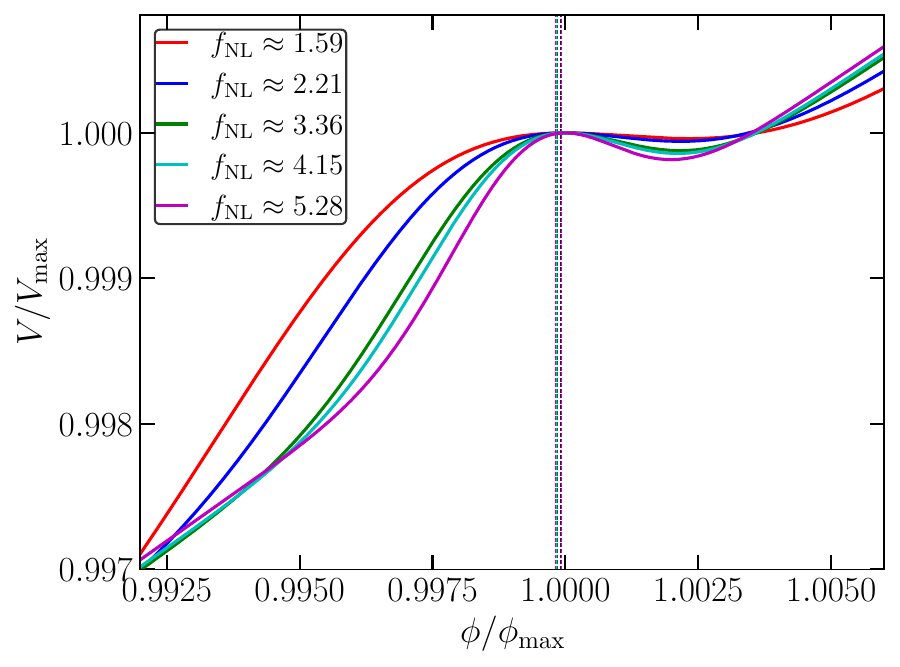}
\caption{Top panel: Values of $\dot{\phi}_{\rm bkg}/(\phi_{\rm bkg}-\phi_{\rm max})$ in terms of the number of e-folds and for different $f_{\rm NL}$ for the case $\nu_{b,c} \sim 8$ (Table \ref{table:nuc}). Notice that in the attractor regime we expect a constant flat plateau giving the value of $-\lambda_{-}$. This plateau is well defined for small $f_{\rm NL}$ but gradually disappears for large $f_{\rm NL}$. Left-bottom panel:  Deviation between the analytical attractor regime solution and the numerical one in percentage. Right-bottom panel: Shape of the potential for different $f_{\rm NL}$ and where the vertical dashed lines indicate the location of $\phi_{\rm bkg}(N_{\star})$ in $V$.}
\label{fig:atractor_regime}
\end{figure}

Lets consider now the case of bubble formation. We first find the value of $N_{\star}$ minimizing the deviation $\Delta_{\rm att}(N_{\star})$ introduced in Eq.\eqref{eq:Delta_N}. We found that the attractor regime is fulfilled only approximately for a specific small range of e-folds. This approximation worsens when $f_{\rm NL}$ increases. This is shown in Fig.\ref{fig:atractor_regime}, where the ratio $\dot{\phi}_{\rm bkg}/(\phi_{\rm bkg}-\phi_{\rm max})$ is plotted in terms of $N$. In the attractor regime we expect a constant flat plateau giving the value of $-\lambda_{-}$. This is satisfied approximately for small $f_{\rm NL}$, but for large $f_{\rm NL} $ there is no clear plateau. The deviation in percentage between the numerical solution and the attractor regime, which is a few per cent, is also displayed. For small $f_{\rm NL}$ the numerical results using the initial conditions Eqs.\eqref{eq:initial_condition_space},\eqref{eq:initial_condition_velocity} match the analytical estimates very well. For large $f_{\rm NL}$ Eq.\eqref{eq:initial_condition_velocity} is less well justified but we still expect it to be a good approximation for modes which are exponentially growing near the top of the potential.

\begin{figure}[h]
\centering
\includegraphics[width=3.2 in]{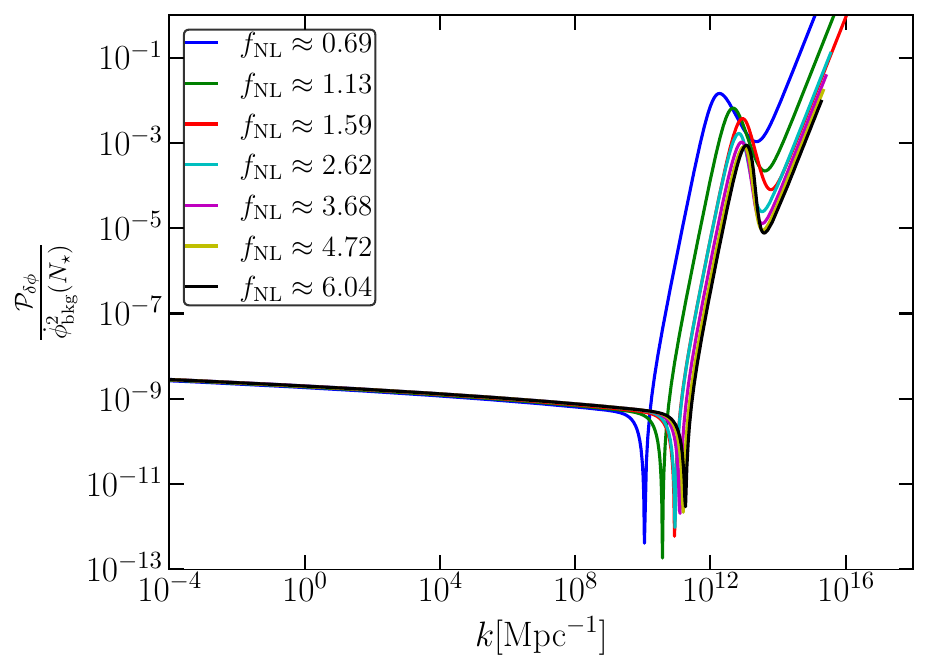}
\caption{Power spectrum $\mathcal{P}_{\delta \phi}(N_{\star},k)$ in units of $\dot{\phi}^2_{\rm bkg}(N_{\star})$ for different values of $f_{\rm NL}$ and the case $\nu_{b,c} \sim 8$ from the parameters in Table \ref{table:nuc}.}
\label{fig:PS_deltaphi}
\end{figure}

On the other hand, in Fig.\ref{fig:PS_deltaphi} it is shown the power spectrum $\mathcal{P}_{\delta \phi}$ evaluated at $N_{\star}$ and reescaled with $\dot{\phi}^2_{\rm bkg}(N_{\star})$. At large scales $k$, the power spectrum matches with the power spectrum associated with the curvature fluctuation $\mathcal{P}_{\zeta_G}$ when we rescale it with the factor $\dot{\phi}_{\rm bkg}(N_{\star})$, since that modes $k$ are already frozen at super-horizon scales when they are evaluated at $N_{\star}$. The local maxima we observe at short scales are attributed to the enhancement of the perturbation field $\delta \phi$ due to the reduction of the inflaton velocity. In this case, not all modes have exited the cosmological horizon. At shorter scales than the local maxima, we find the expected growth of the curvature modes of the scalar field in a Minkowski vacuum $\sim k^2$, which corresponds to modes that are still inside the cosmological horizon. To take realistically into account the growth of the power spectrum, we take the limit of the integrals $k_f$ and $k_i$ from Eq.\eqref{eq:correaltion_bubble} such that the $k_f$ corresponds to the local minima of $\mathcal{P}_{\delta \phi}$ after the local maxima, to avoid the unrealistic contribution from modes that are still growing within the cosmological horizon. The $k_{i}$ is taken as the corresponding scale such that  $\mathcal{P}_{\delta \phi}(k_f)=\mathcal{P}_{\delta \phi}(k_i)$ to account only for the enhancement due to the dynamics of the inflaton around the bump of $V$ and avoid infrared contributions, which in any case wouldn't change the result taking into account the large enhancement of the  $\mathcal{P}_{\delta \phi}$ at the peak compared with large scales.

%%%%%bubble case

\subsection{Bubble dynamics formation and size}

Using the $\Psi_{b}(N_{\star},\tilde{r})$ obtained numerically in the previous section allows us to build the initial conditions for bubble formation according to section \ref{sec:bubble_fomration} and then proceed to numerically solve the dynamical equation for the field $\phi(N,\tilde{r})$ Eq.\eqref{eq:non-homogeneus}.
\begin{figure}[h]
\centering
\includegraphics[width=1.9 in]{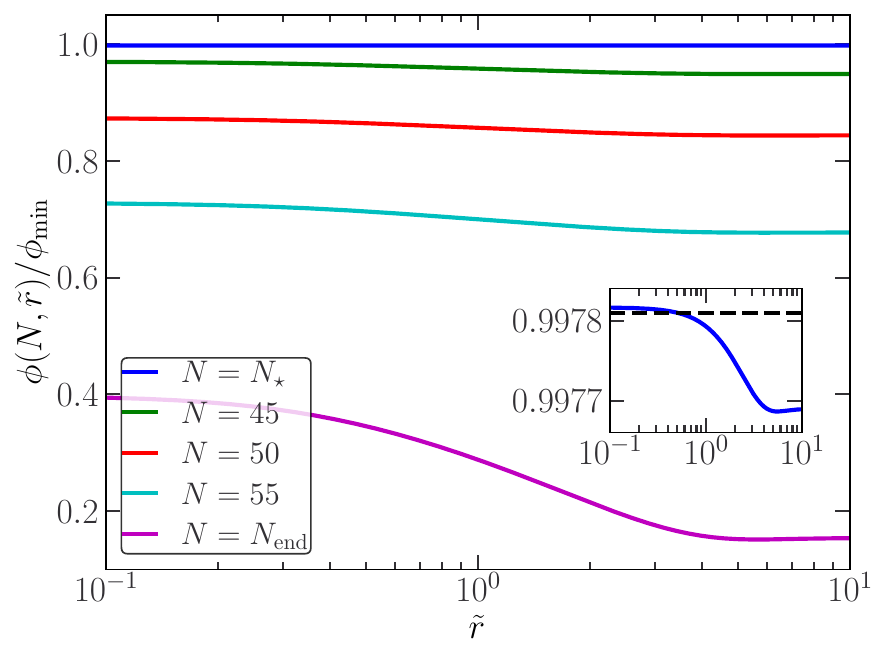}
\includegraphics[width=1.9 in]{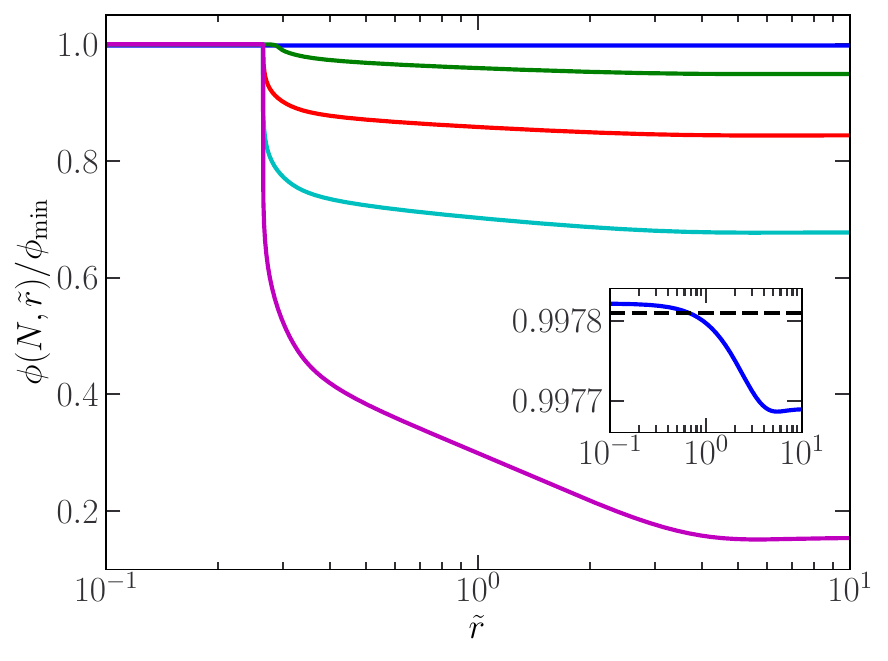}
\includegraphics[width=1.9 in]{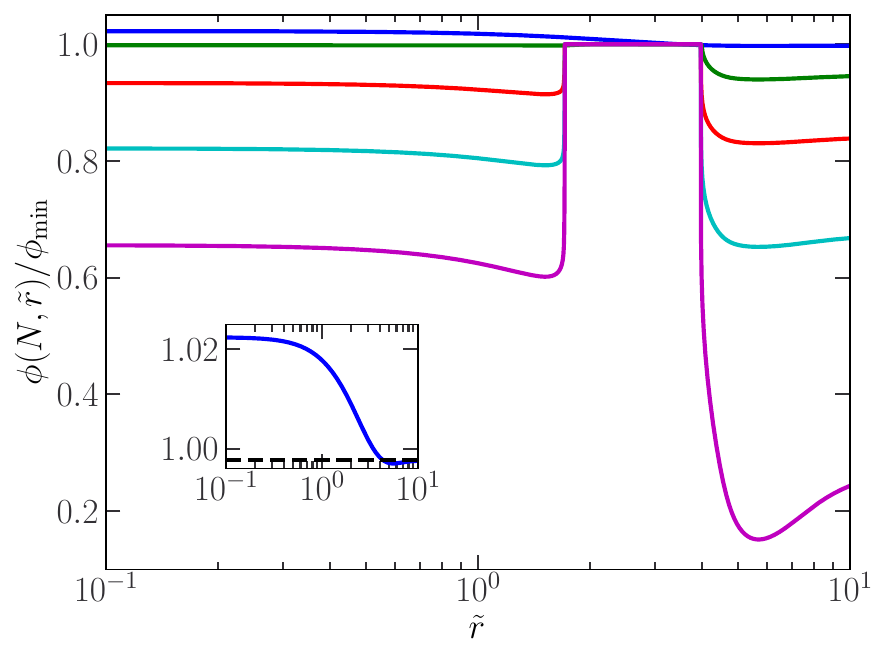}
\includegraphics[width=1.9 in]{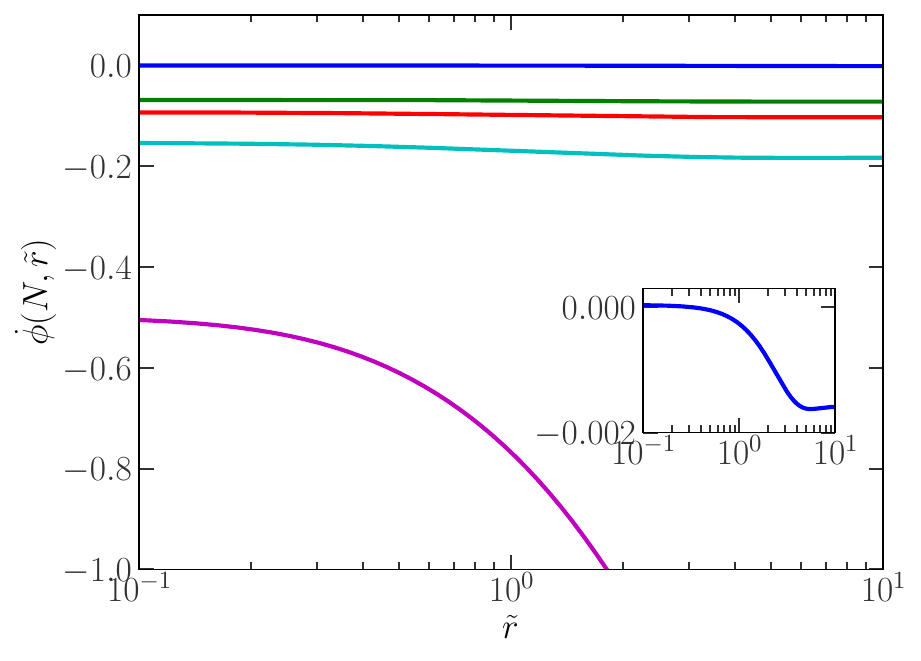}
\includegraphics[width=1.9 in]{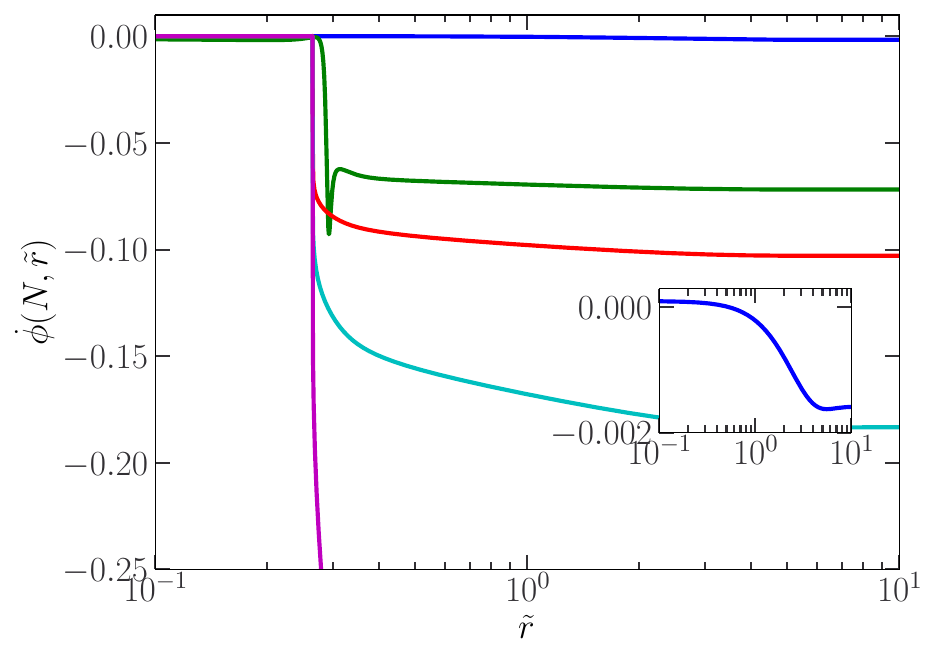}
\includegraphics[width=1.9 in]{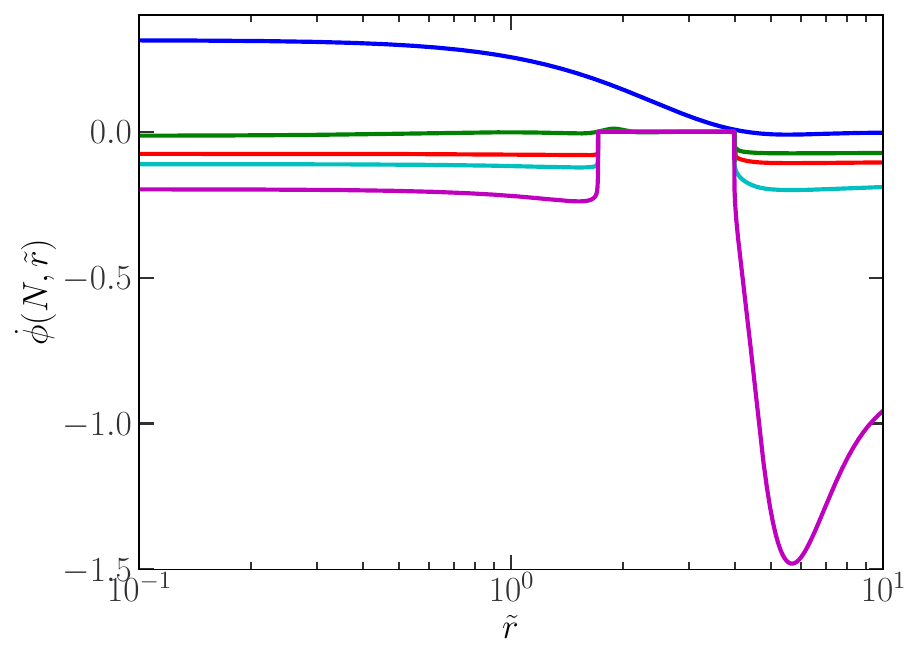}
\caption{Profile of the inflaton at different e-folds $N$. Panels in the first row correspond to the evolution of $\phi(N,\tilde{r})$ in units of $\phi_{\rm min}$, whereas in the second is shown the evolution of the velocity $\dot{\phi}(N,\tilde{r})$. Left panels corresponds to a subcritical case $\mu_{b}< \mu_{b,c}\, (\mu_{b} = \mu_{b,c}-10^{-5})$, middle to supercritical $\mu_{b}> \mu_{b,c}\, (\mu_{b} = \mu_{b,c}+10^{-5})$ and right to super-supercritical $\mu_b \gg \mu_{b,c} \, (\mu_{b} = \mu_{b,c}+10^{-1})$. The three cases correspond to the case $f_{\rm NL} \approx 2.62$ from Table \ref{table:fixedfraction} with $\mu_{b,c} \approx 5.243 \cdot 10^{-4}$. The subplots show the initial configuration at $N_{\star}$ for better visualization. In particular, the dashed line corresponds to $\phi_{\rm max}$ in units of $\phi_{\rm min}$.}
\label{fig:bubble_snapshots}
\end{figure}

An example of the dynamical formation of a bubble is shown in Fig.\ref{fig:bubble_snapshots} (panels in the top are for $\phi(N,\tilde{r})$ and in the bottom for $\dot{\phi}(N,\tilde{r})$), for the case $f_{\rm NL} \approx 2.62$ from Table \ref{table:fixedfraction} (other cases leads to a similar qualitative behaviour). We can make the following observations:
\begin{itemize}

\item In the left panel is shown a subcritical case ($\mu_{b}<\mu_{b,c}$), where the inflaton overshoots the barrier and exits inflation without the formation of bubbles. In this case, initially, the domain $\tilde{r} \gtrapprox 0.5$ of the inflaton overshoots the $\phi_{\rm max}$, and therefore will exit inflation (notice also a negative velocity). On the other hand, for $\tilde{r} \lessapprox 0.5$, the inflaton lies between the $\phi_{\rm max}$ and $\phi_{\rm min}$ of the potential with an initial small positive velocity, which makes the inflaton roll down to $\phi_{\rm min}$. After that, the contribution of the gradient of the scalar field and a negative velocity obtained when falling again into the minimum make the inflaton overshoot the barrier and avoid becoming trapped.

\item In the middle panel, instead, is shown the evolution of the bubble formation for a supercritical case ($\mu_{b} > \mu_{b,c}$). In this case, the large backward quantum fluctuation is sufficiently large to trap a localized region of the inflaton field, producing the false vacuum bubble. The size of the bubble is defined as the last point of the comoving size $\tilde{r}$ such that fulfil $\phi(N_{\textrm{end}},\tilde{r}\leq R_{\rm b}) = \phi_{\rm min}$ (being $R_{b}$ the size of the bubble in units of $1/(a(N_{\star})H(N_{\star}))$), since the region where the inflaton is trapped will correspond to the minima of $V$ and the field domain that exits the barrier will fulfill $\phi(N_{\textrm{end}},\tilde{r} > R_{b})<\phi_{\rm max}$. Notice that for $\dot{\phi}(N_{\textrm{end}},\tilde{r} \le R_{b})$, the velocity is zero since the inflaton is trapped. Interestingly, at the end of inflation, there is a tail for the scalar field, which doesn't precisely correspond to a domain wall shape. We will comment on that in more detail later on.

\item In the right panels, we have a super-supercritical fluctuation with $\mu_{b} \gg \mu_{b,c}$. We find the formation of a concentric shell instead of a central bubble. In this case, at $N_{\star}$, the large backward quantum fluctuation will displace the central region of the inflation far away from the minima $\phi_{\rm min}$ allowing this region to get enough velocity when falling again into the minima to, after that, overshoot the barrier. Instead, the tail that falls between $\phi_{\rm min}$ and $\phi_{\rm max}$ will become trapped similarly as in the previous case. Although this last scenario has an interesting dynamical behaviour, as we will see later on, such a case is largely statistically suppressed since $\mu_{b} \gg \mu_{b,c}$. Therefore, in this work, we will only consider standard spherically-centred bubbles.

\end{itemize}

\begin{figure}[h]
\centering
\includegraphics[width=3.0 in]{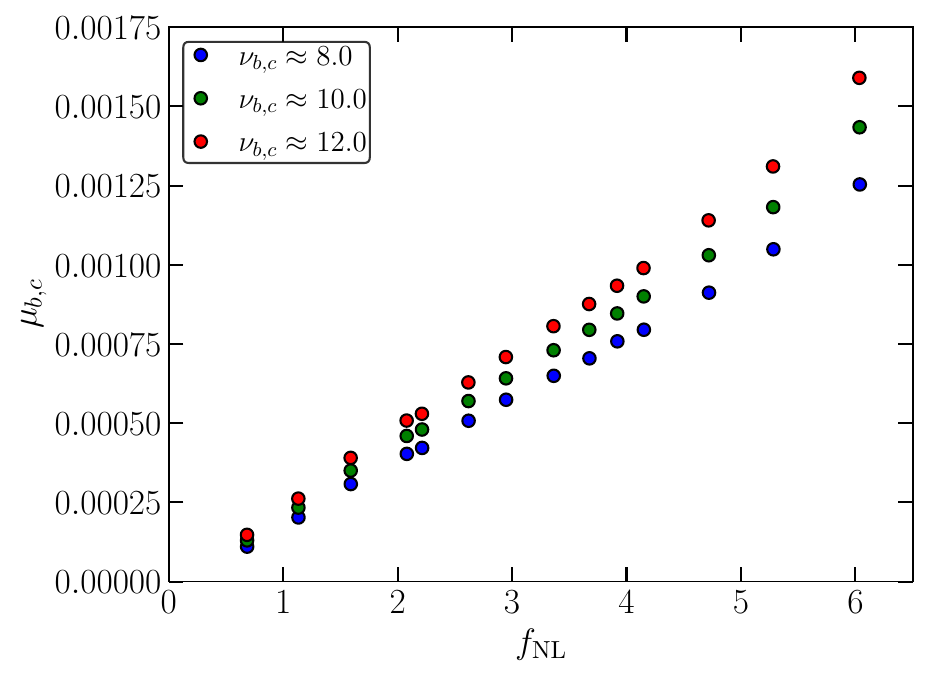}
\includegraphics[width=3.0 in]{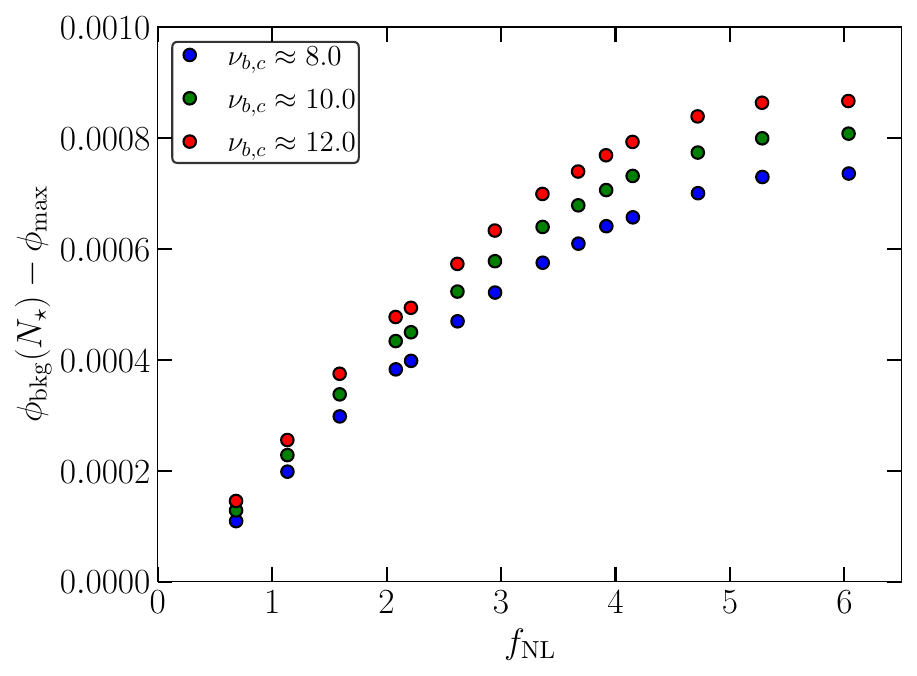}
\caption{Left panel: Threshold $\mu_{b,c}$ in terms of $f_{\rm NL}$ and for different $\nu_{b,c}$. Right panel: Difference in the field $\phi$ between the location of $N_{\star}$ and $\phi_{\rm max}$ in terms of $f_{\rm NL}$ and for different configurations $\nu_{b,c}$. The parameters chosen correspond to Table \ref{table:nuc}.}
\label{fig:mu_c_bubble}
\end{figure}

Using a bisection method, we obtain the thresholds $\mu_{b,c}$, which can be seen in Fig.\ref{fig:mu_c_bubble} for different $\nu_{b,c}$. In the left panel of Fig.\ref{fig:mu_c_bubble} $\mu_{b,c}$ increase in terms of $f_{\rm NL}$, which is in agreement with the fact that the distance between the point where we implement the initial condition $\phi_{\rm bkg}(N_{\star})$ and the position of the maximum of the potential $\phi_{\rm max}$ (see right panel) increases with $f_{\rm NL}$ up to remain roughly constant, and therefore the amplitude of the backward fluctuation should be larger to make the inflaton trapped. Another effect in combination with the previous one is that the shape of $V$ around $\phi_{\rm max}$ is sharper for larger $f_{\rm NL}$, then the gradients of $V$ increase the minimum $\mu_{b}$ needed to make the inflaton trapped and produce the bubble, due to the forces of the potential in the inflationary dynamics.

%%%%%bubble case
\begin{figure}[h]
\centering
\includegraphics[width=4.5 in]{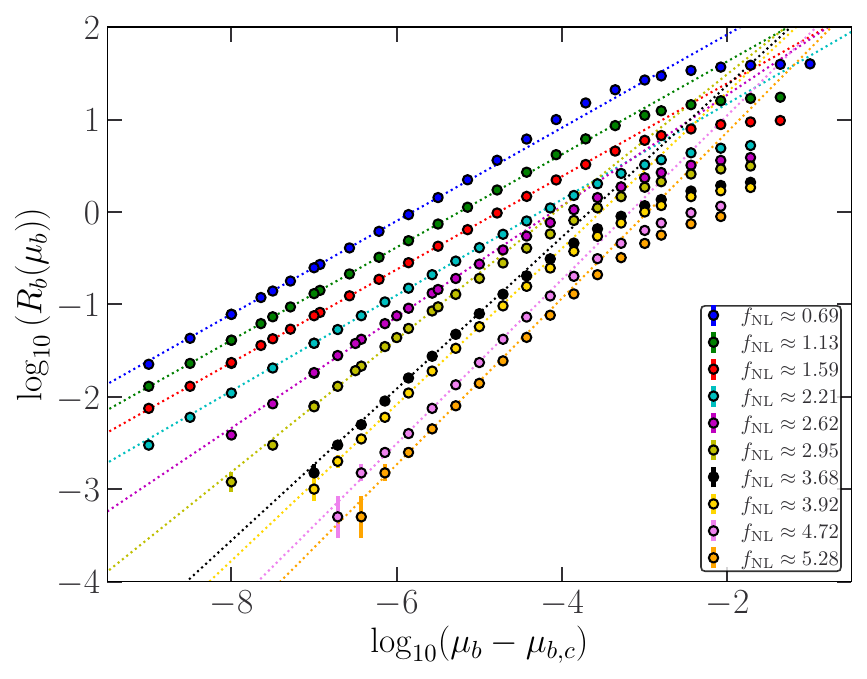}
\caption{Size of the bubbles in logarithmic scale for different amplitudes $\mu_{b}-\mu_{b,c}$ for the case $\nu_{b,c}=8$. The dashed lines indicate  Eq.\eqref{eq:bubble_scaling} with the fitting done in the range where the scaling law is more likely with $ \mu_{b} -\mu_{b,c} \lessapprox 10^{-1} \mu_{b,c}$. There are also shown the error bars due to the resolution of the grid.}
\label{fig:bubble_size}
\end{figure}

\begin{figure}[h]
\centering
\includegraphics[width=3.5 in]{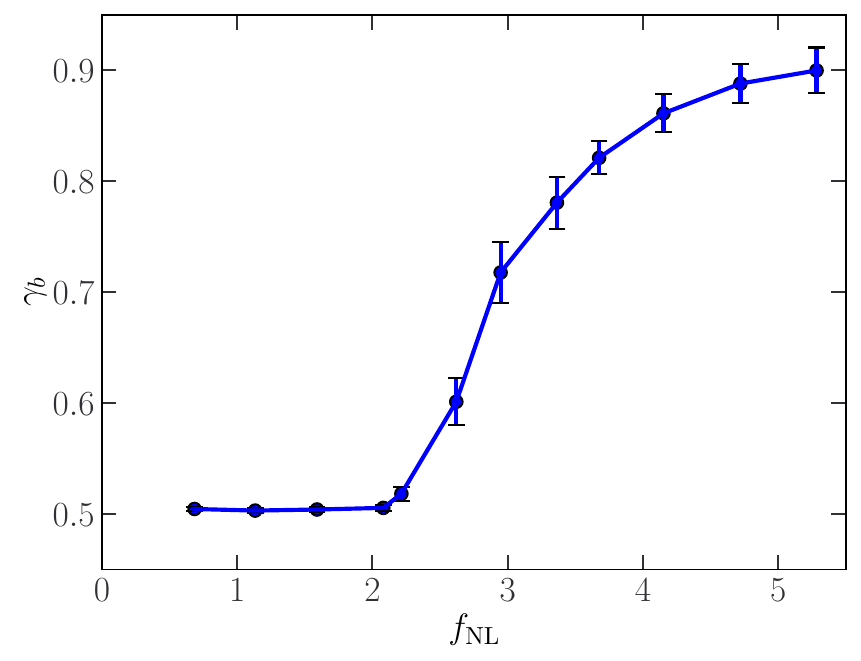}
\caption{Critical exponent $\gamma_{b}$ of the size of the bubbles $R_{b}(\mu_b)$ in terms of the non-Gaussian parameter $f_{\rm NL}$. There are also indicated error bars from the numerical fitting of the data to Eq.\eqref{eq:bubble_scaling}, mainly due to grid resolution limitation and the specific range of $\mu_b-\mu_{b,c}$ chosen where the fit is performed.}
\label{fig:critical_exponent}
\end{figure}

By making simulations for small values of $\mu_{b}-\mu_{b,c}$, we can explore the critical regime of the bubble sizes. This is shown in Fig.\ref{fig:bubble_size} for a specific case of $\nu_{b,c}=8$. We have found for this case that for fluctuations very close to the critical one with $ \mu_{b} -\mu_{b,c} \lesssim 10^{-1} \mu_{b,c}$ the bubble size follows a critical regime\footnote{This is reminiscent of the critical behaviour in gravitational collapse \cite{Gundlach:2007gc}. However, the present case is a different phenomenon where gravitational forces do not play a role.}
\begin{equation}
\label{eq:bubble_scaling}
    R_b(\mu_b) = \mathcal{K}_{b}(\mu_{b,c}) (\mu_{b}-\mu_{b,c})^{\gamma_{b}(f_{\rm NL})},
\end{equation}
where $\mathcal{K}_{b}$ is a constant that depends on the threshold $\mu_{b,c}$ for each configuration $f_{\rm NL}$ and $\gamma_{b}$ (shown in Fig.\ref{fig:critical_exponent}) seems to be in general a coefficient independent\footnote{The differences on $\gamma_b(f_{\rm NL})$ between the three cases tested $\nu_{b,c} \approx 8,10,12$ considered in Table \ref{table:nuc} lies within the error bars. For $f_{\rm NL} \lesssim 2$, where the precision for the exponent is quite high, we find indeed that the exponent is independent on $\nu_{b}$. We expect that this characteristic also hold for larger $f_{\rm NL}$, and it only should depend on the shape of $V$ around $\phi_{\rm max}$, which characterizes $f_{\rm NL}$ through Eqs.\eqref{eq:mu_star_vv4},\eqref{eq:mu_star}.} on the threshold $\mu_{b,c}$, and only dependent on $f_{\rm NL}$. For small $f_{\rm NL}$, the critical exponent saturates to a factor $\sim 0.50$, but then starts to increase for $f_{\rm NL} \gtrsim 2$. Due to a limitation of the resolution $d\tilde{r} \sim  10^{-3.6}$ (this mainly induces the error bars in the coefficient $\gamma_{b}$ shown in the figure), we cannot make simulations for very small $\mu_{c}-\mu_{b,c}$ to explore the bubble size for large $f_{\rm NL}$ since $R_b$ becomes smaller. In any case, we expect that the critical regime Eq.\eqref{eq:bubble_scaling} should hold for smaller $\mu_{b}-\mu_{b,c}$ than the ones tested, as it is more clearly shown for small $f_{\rm NL}$ where the bubble size is larger, and the critical regime is clearly identified. Still, a dedicated study with some specific technique like adaptive mesh refinement \cite{Wainwright:2013lea,1984JCoPh..53..484B} would probably be needed to confirm our intuition, specially for the cases of large $f_{\rm NL}$ where the simulations are more challenging.

Physically the spread in bubble sizes can be understood as follows. A small fluctuation with $\mu_b<\mu_{b,c}$ may form a very small bubble, but that one immediately recollapses under its tension and leaves nothing behind. The bubble with $\mu_b=\mu_{b,c}$, on the other hand, will be in unstable equilibrium between expansion and recollapse. Classically, it could stay in this "loitering" equilibrium point, with a fixed physical size, for an indefinite number of e-foldings before it decides to expand or recollapse. The longer the bubble is loitering, the smaller will be the co-moving size of the bubble. That is the reason why we have a tail of smaller co-moving size bubbles as the critical amplitude is approached.

In practice, however, we have to account for the fact that, while the bubble is loitering, quantum fluctuations of the field will keep accumulating, changing in practice the overall amplitude $\mu$ of the original field perturbation. The field fluctuates by the amount $(\delta\phi)^2 \sim (H/2\pi)^2$ per e-folding, and over a period $\Delta N$ such fluctuations add quadratically as a random walk, so we have 
\begin{equation}
\Delta\mu \sim {H\over 2\pi} (\Delta N)^{1/2} \gtrsim (\mu_b-\mu_{b,c}). \label{cutoff}
\end{equation}
The last inequality is meant to represent the boundary beyond which the fluctuation in the amplitude is large enough to make the bubble recollapse. Since the co-moving size of the bubble follows the scaling $R_b\sim e^{-\Delta N}$, for any given $R_b$ the relation (\ref{cutoff}) leads to a minimum amplitude $\mu_{b,\rm (cut-off)}$ above $\mu_{b,c}$, such that the required amount of loitering is still allowed by the quantum drift:
\begin{equation}
\label{eq:cut_off}
\mu_{b,\rm (cut-off)}-\mu_{b,c}  \approx \sqrt{-\ln(R_b(\mu_b))} \frac{H(N_{\star})}{2 \pi}.
\end{equation}
This represents a quantum boundary of the classical scaling regime.

\begin{figure}[h]
\centering
\includegraphics[width=2.9 in]{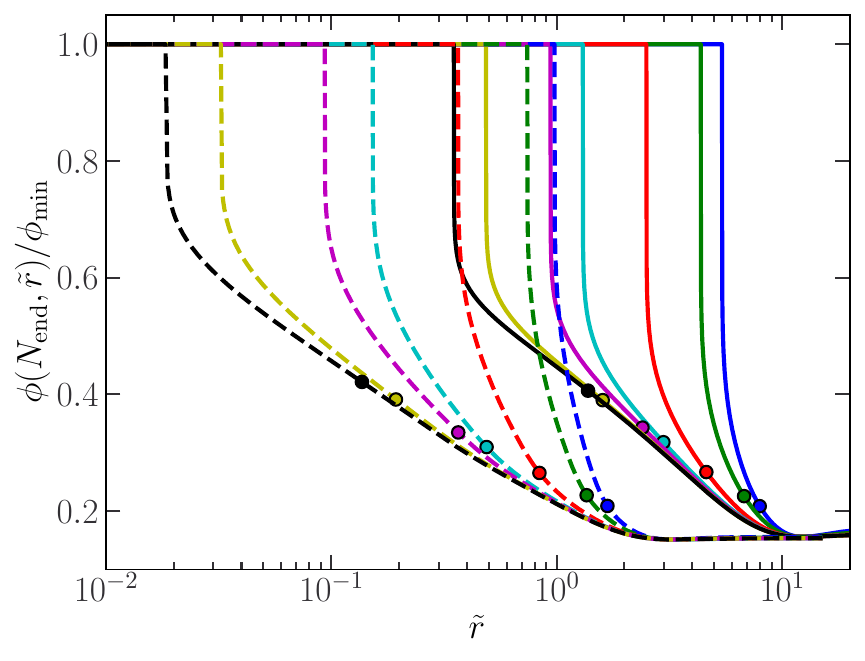}
\includegraphics[width=3.0 in]{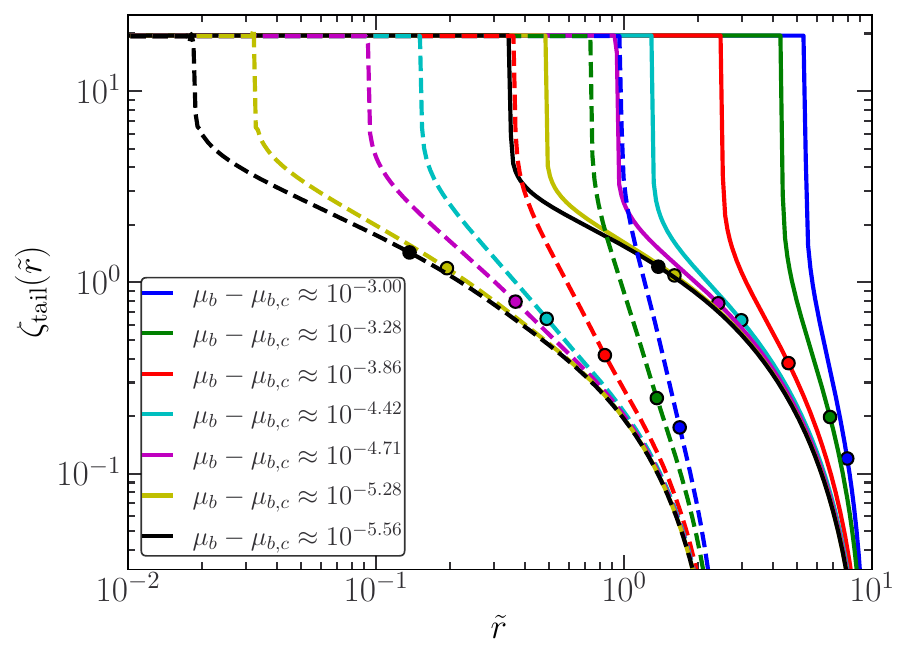}
\caption{Left-panel: Profile of the inflaton $\phi(N,\tilde{r})$ at the end of inflation $N_{\rm end}$. Right panel: adiabatic curvature fluctuation $\zeta_{\rm tail}$ that surrounds the scalar field obtained using the $\delta N$ formalism. In both cases, the solid line is the case for $f_{\rm NL} \approx 1.59$ and dashed for $f_{\rm NL} \approx 2.95$ from Table \ref{table:fixedfraction}. The dots show the location $\tilde{r}_{\rm tail}$ for what $1+\tilde{r}_{\rm tail} \zeta'_{\rm tail}(\tilde{r}_{\rm tail})=0$ (notice that we reescaled $r$ in terms of $\tilde{r}$ in the numerical simulations). The different colours represents different cases of $\mu_b-\mu_{b,c}$.}
\label{fig:profile_bubble_end}
\end{figure}

Finally, we have found that at the end of inflation\footnote{Notice that we are considering a homogeneous Hubble factor in Eq.\eqref{eq:non-homogeneus} since we are not including the backreaction to the spacetime metric. This would make a minor correction on the final profile of the scalar field at the end of inflation, where the friction due to the backreaction is not considered, and therefore the $\zeta_{\rm tail}$ could be slightly underestimated. We leave for future research a refinement on that to take into account this possible small correction.}, the bubble is surrounded by a tail of the scalar field $\phi$, as is shown in Fig.\ref{fig:profile_bubble_end} for different values $\mu_{b}-\mu_{b,c}$. This tail is more important for values $\mu_{b}$ closer to the critical $\mu_{b,c}$. Making use of the $\delta N$ formalism, we can compute the contribution of the curvature fluctuation from the tail $\zeta_{\rm tail}$ as $\zeta_{\rm tail} \equiv \delta N$ with $\phi(N_{\rm end},\tilde{r}) = \phi_{\rm bkg}(N_{\rm end}-\delta N)$. The profiles $\zeta_{\rm tail}$ are shown in the right panel of Fig.\ref{fig:profile_bubble_end}. 

Remarkably, we find that $\zeta_{\rm tail}$ is indeed a fluctuation of type-II, which fulfills that $1+\tilde{r}_{\rm tail} \zeta'(\tilde{r}_{\rm tail})=0$ and the scale $\tilde{r}_{\rm tail}$ is indicated as a dot in the Fig.\ref{fig:profile_bubble_end}. What we will see in section \ref{section:mass_function} is that, indeed, the $\zeta_{\rm tail}$ will dominate the final PBH mass with respect to the contribution of the bubble itself. We have observed that the $\tilde{r}_{\rm tail}$ seems to follow a scaling law in terms of the amplitude $\mu_{b}-\mu_{b,c}$ as $\tilde{r}_{\rm tail} \sim (\mu_{b}-\mu_{b,c})^{\gamma_{\rm tail}}$ (similar to Eq.\eqref{eq:bubble_scaling}) for the cases tested in Table \ref{table:fixedfraction}. We leave for future research a more detailed exploration of that.

\subsection{PBH production from the adiabatic and bubble channel}
\begin{figure}[h]
\centering
\includegraphics[width=3.0 in]{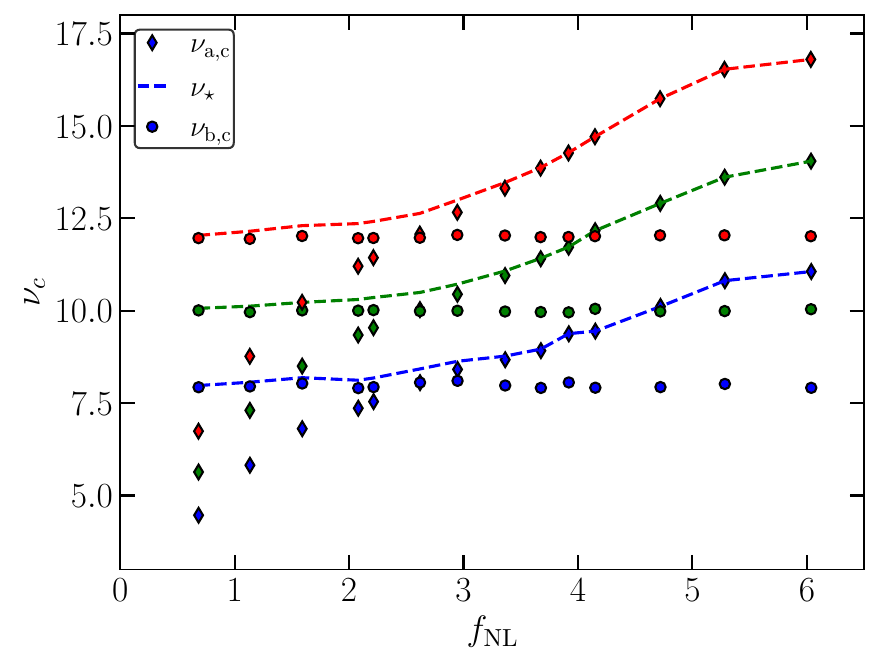}
\includegraphics[width=3.0 in]{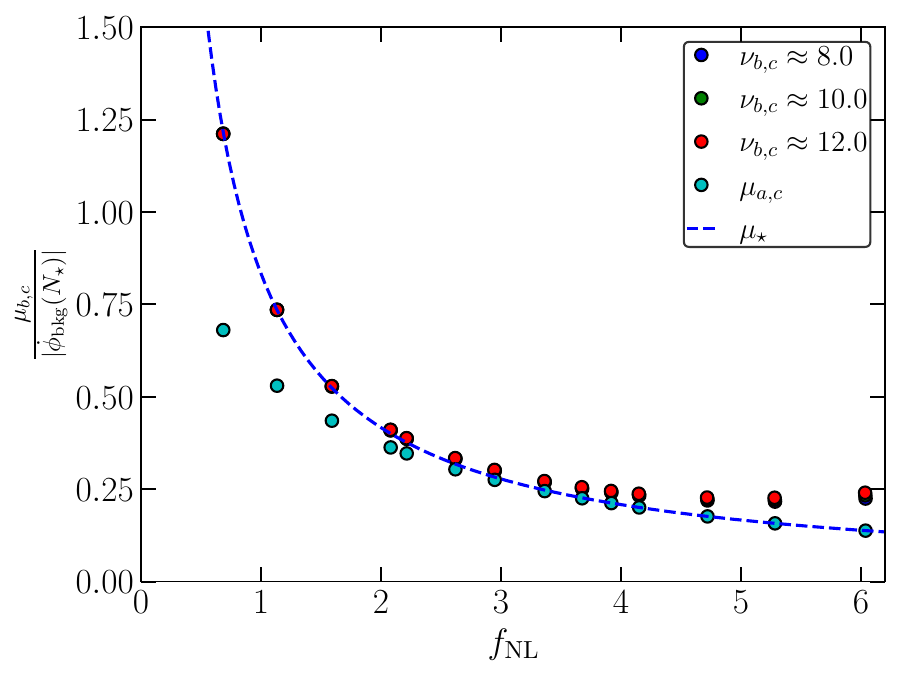}
\caption{Left-panel: Critical values $\nu_{c}$ for the two channels of PBH formation. Circles represents the thresholds for the bubble channel $\nu_{b,c}$, thin diamond the adiabatic channel $\nu_{a,c}$ and the dashed line the analytical estimate $\nu_{\star}=5/(6 f_{\rm NL})$. Right-panel: Values $\mu_{b,c}/\lvert\dot{\phi}_{\rm bkg}(N_{\star})\rvert$ compared with the analytical $\mu_{\star}$ and the numerical values $\mu_{a,c}$. In both panels, blue, green and red symbols correspond to the cases $\nu_{b,c} \sim 8,10,12$ respectively.}
\label{fig:nu_critical_values}
\end{figure}

\begin{figure}[h]
\centering
\includegraphics[width=3.1 in]{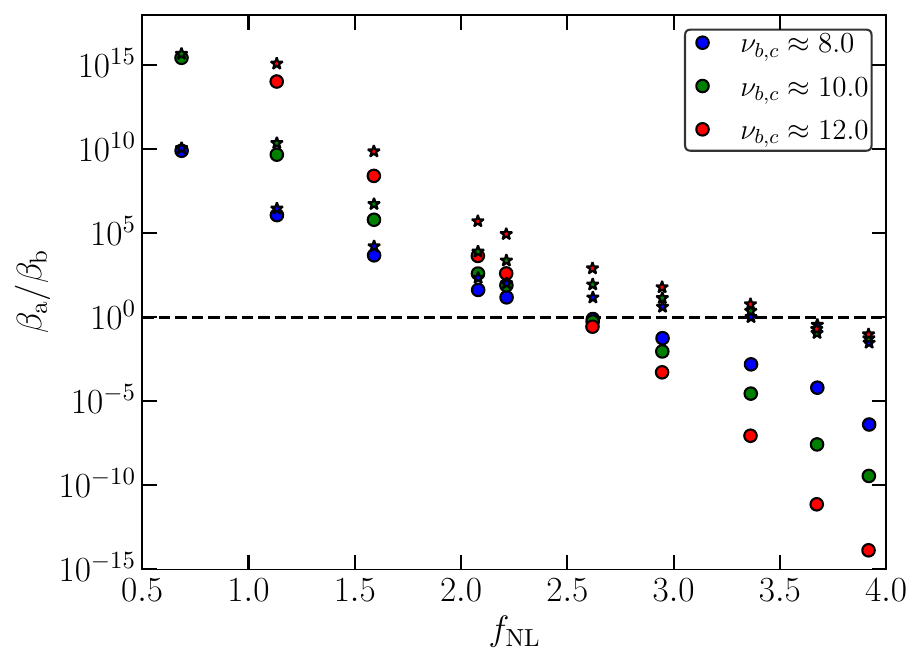}
\includegraphics[width=2.9 in]{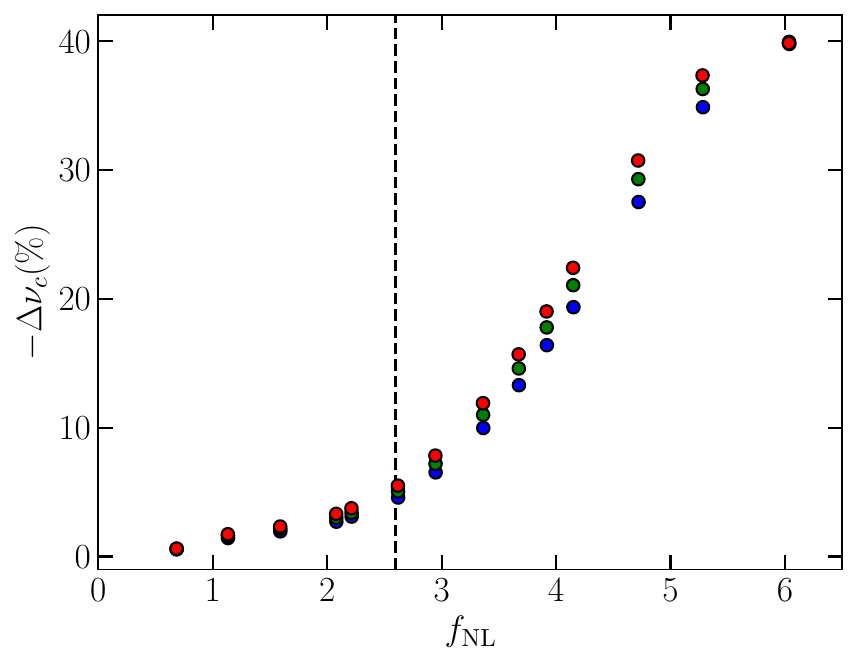}
\caption{Left panel: Relative abundance between the number of peaks generated from the adiabatic and the bubble channel. Dots represent the results of numerical simulations using $\nu_{b,c}$, whereas the stars are computed using the analytical value $\nu_{\star}$. Right-panel: Relative deviation $\Delta \nu$ between the numerical $\nu_{b,c}$ and analytical estimate $\nu_{\star}$ in percentage defined in Eq.\eqref{eq:deviation_nus}, in terms of $f_{\rm NL}$ and for different fixed values of $\nu_{b,c}$.}
\label{fig:relative_abundance_peaks}
\end{figure}

In the left panel of Fig.\ref{fig:nu_critical_values}, we show the numerical results for the thresholds $\nu_{c}$ for both channels of production. We choose three values for $\nu_{b,c} \approx 8,10,12,$ and with these parameters we compute the corresponding normalised threshold $\nu_{a,c}$ for the adiabatic channel. We also show the analytical estimate $\nu_{\star} = 5/(6 f_{\rm NL})$.
%The values are fixed for the same $\nu_{b,c}$ (bubble channel) and are compared with the other values from the adiabatic channel $\nu_{a,c}$ and the analytical estimate $\nu_{\star} = 5/(6 f_{\rm NL})$. 
In the case of small NGs, the values of the analytical estimate for the bubble channel $\nu_{\star}$ match with very good agreement with the numerical one $\nu_{b,c}$, being $\nu_{a,c}$ smaller. Instead $\nu_{a,c}$ increases with $f_{\rm NL}$ for given value of $\nu_{b,c}$. For $f_{\rm NL} \approx 2.6$, $\nu_{a}>\nu_{b}$, meaning that bubble formation becomes more likely than PBH formation through the adiabatic channel. Let us note that $\nu_{b,c}$ starts to differ from the analytical estimate $\nu_{\star}$, meaning that bubble formation is more likely than previously estimated. Notice that this qualitative behaviour is the same for the different values of $\nu_{b,c}$ chosen.
In the right panel instead, we show the thresholds $\mu_c$ for the adiabatic channel compared with the normalized threshold of the bubble channel in units of $\lvert\dot{\phi}_{\rm bkg}(N_{\star})\rvert$. The values for the bubble channel match with very good agreement with the ones inferred from the analytical estimate for small $f_{\rm NL}$, but we start to see a deviation for larger $f_{\rm NL}\gtrsim 3$. Notice that although we find for large $f_{\rm NL}$ that $\mu_{b,c}/\lvert \dot{\phi}_{\rm bkg}(N_{\star})\rvert>\mu_{a,c}$, in the left panel we have $\nu_{b,c}<\nu_{a,c}$. We should take into account that the integrated power spectrum is larger for $\mathcal{P}_{\delta \phi}/\lvert\dot{\phi}_{\rm bkg}(N_{\star})\rvert$ than for $\mathcal{P}_{\zeta_G}$. Therefore $\sigma_{b}/ \lvert \dot{\phi}_{ \rm bkg}(N_{\star}) \rvert> \sigma_{a}$, making the bubble-channel for PBH formation more likely. We can then estimate the relative abundance of the peaks generated by the adiabatic fluctuations and the bubbles using the values of Fig.\ref{fig:nu_critical_values} following the procedure of section \ref{subsec:high_peaks}. The abundance of the number of peaks from both channels is estimated as

\begin{align} 
\label{eq:abundance_peaks_ab}
    \beta_{\rm a} &= \int_{\nu_{a,c}}^{\nu_{\star}} \mathcal{N}_{\rm a}(\nu_{a}) d \nu_{a} = \int_{\nu_{a,c}}^{\nu_{\star}} \left( \frac{\sigma_{a,1}}{\sqrt{3} \sigma_{a,0}}\right)^{3} (\nu^3_a-3 \nu_a) e^{-\frac{1}{2}\nu^2_a} d\nu_a , \\
    \beta_{\rm b} &= \int_{\nu_{b,c}}^{\infty} \mathcal{N}_{\rm b}(\nu_b) d \nu_b=\int_{\nu_{b,c}}^{\infty}  \left( \frac{\sigma_{b,1}}{\sqrt{3} \sigma_{b,0}}\right)^{3} (\nu^3_{b}-3 \nu_{b}) e^{-\frac{1}{2}\nu^2_{b}} d\nu_{b} .
\end{align}
Notice that we are considering that the limits of integration for both cases are different. In the adiabatic channel, we integrate from $\nu_{a,c}$ (obtained using the analytical approach of the averaged compaction function) to the limiting value $\nu_{\star}$ for which the log-template relation is defined. On the other hand, in the bubble channel, we integrate from $\nu_{b,c}$ up to infinity. In Fig.\ref{fig:relative_abundance_peaks} we show the ratio $\beta_{\rm a}/\beta_{\rm b}$ in terms of the value of $f_{\rm NL}$.

We find that the number of peaks generated by the bubble channel dominates over those coming the adiabatic one for $f_{\rm NL} \gtrapprox 2.6$, being $f_{\rm NL} \approx 2.6$ (dots in the figure) the commensurable case with $\beta_{a} \approx \beta_{b}$. We also show the values computed considering the analytical estimate of $\nu_{\star}$ (star symbol) in the computation of Eq.\eqref{eq:abundance_peaks_ab}. The latter underestimates the production of bubbles at higher $f_{\rm NL}$. In particular for the case $f_{\rm NL} \approx 3.5$ bubbles are already $\sim 10^{3}$ more abundant by comparison.

On the other hand, in the right panel of Fig.\ref{fig:relative_abundance_peaks}, we show the relative deviation in percentage between $\nu_{b,c}$ and the analytical estimate $\nu_{\star}$ that we expect from the log relation between $\zeta_{G}$ and $\zeta$ (Eq.\eqref{eq:deviation_nus}). We define the relative deviation $\Delta \nu_{c}$ as
\begin{equation}
\label{eq:deviation_nus}
\Delta \nu_{c}(\%)=10^{2} \frac{\nu_{b,c}-\nu_{\star}}{\nu_{b,c}}.
\end{equation}
We find good agreement between the analytical estimate $\nu_{\star}$ and the numerical result $\nu_{b,c}$ for small $f_{\rm NL}$. This is the regime where we precisely expect that the analytical estimate should be more accurate, as mentioned in section \ref{subsec:non_gaussian}. Increasing $f_{\rm NL}$ the deviation $\Delta \nu$ is larger, which could be due to several reasons. First of all, for large $f_{\rm NL}$ the attractor solution in the quadratic part of the potential (see Fig.\ref{fig:atractor_regime}) is lost, and therefore we are not in the regime of applicability of Eq.\eqref{eq:zeta_NG}. Another reason may be the use of initial conditions $\delta \dot{\phi}_{\rm bkg} = -\lambda_{-} \delta (\phi_{\rm bkg}-\phi_{\rm max})$ at $N_{\star}$, again corresponding to the attractor regime. Even so, the deviation $\Delta \nu_{c}$ is within a factor $\sim 12\%$ for $f_{\rm NL}<3.5$. A more realistic comparison between $\nu_{\star}$ and $\nu_{b,c}$ for very large values of $f_{\rm NL}$ would require new methods and is left for future research.

\subsection{Mass function and spectrum of PBHs}
\label{section:mass_function}

In this last section, we are going to use the numerical results to estimate the mass function and spectrum from both channels of PBH production. The PBH mass spectrum of the adiabatic channel follows the critical regime when $\nu_a \rightarrow \nu_{a,c}$ as already shown in Eq.\eqref{eq:mass_function}

\begin{equation}
\label{eq:mass_scaling2}
M_{a}(\nu_a) = \mathcal{K}_{a}\sigma^{\gamma_a}_{a} x^2_m(\nu_a) e^{2 \zeta(r_m(\nu_a))}(\nu_a-\nu_{a,c})^{\gamma_a} M_k(k_{\rm max}),
\end{equation}
where $x_m(\nu_a) = k_{\rm max} r_m(\nu_a)$ being $k_{\rm max}$ the location of the peak of the power spectrum $\mathcal{P}_{\zeta_G}$. Then, the mass function contribution from the adiabatic channel is estimated using peak theory following \cite{Garriga,Kitajima:2021fpq}

\begin{equation}
\label{eq:mass_function_adiabatic}
    f_{\rm a}(M_a) = \frac{M_{\rm a}(\nu_a)\mathcal{N}_{\rm pk}(\nu_a(M_a))}{\rho_{\rm critical} \Omega_{\rm DM}} \Bigg | \frac{d \ln M_a(\nu_a)}{d \nu_a} \Bigg |^{-1},
\end{equation}
where $\rho_{\rm critical}=3 M^2_{\rm pl}H^2_{0}$ corresponds to the current critical energy density of the Universe and $\Omega_{\rm DM}$ the current fraction of dark matter. The last term in Eq.\eqref{eq:mass_function_adiabatic} is the inverse of the Jacobian, which in the critical regime is given by,

\begin{equation}
\label{eq:jacobian_adiabatic}
 \frac{d \ln M_a(\nu_a)}{d \nu_a}  =  \frac{2}{x_m(\nu_a)} \frac{d x_m(\nu_a)}{d \nu_a} +2 \frac{d \zeta(r_m(\nu_a))}{ d\nu_a}+\frac{\gamma_{a}}{\nu_{a}-\nu_{a,c}} .
\end{equation}

When $\nu_a \rightarrow \nu_{a,c}$ the Jacobian is dominated by the last term and therefore $f_{a}(M_a) \sim M_{a}^{1+1/\gamma_{a}}$. The contribution to the mass function for the bubble channel needs to take into account several considerations.

As we have mentioned, there is a surrounding adiabatic curvature fluctuation of type-II at the end of inflation $\zeta_{\rm tail}$. This fluctuation will source the main contribution to the PBH mass in comparison with the bubble itself, since the length-scale $\tilde{r}_{\rm tail}$ can be substantially larger than the comoving scale of the bubble size $R_b$. Once the baby Universe becomes causally disconnected from the parent Universe (see Fig.4 of \cite{Garriga:2015fdk} and Fig.3 of \cite{vacum_bubles} for a schematic picture of the inflating bubble and a causal diagram respectively) the fluctuation of type-II will be swallowed by the remaining event horizon. Notice that although the PBH mass contribution from the bubble channel is dominated by fluctuations of type-II, its dynamical formation is completely different from the standard adiabatic channel, due to the presence of the bubble.

In order to obtain a very accurate estimation of the resulting PBH mass, it would be necessary to perform a full numerical simulation to take into account the full shape of $\phi(N_{\rm end},\tilde{r})$ including the $\zeta_{\rm tail}$, which is left for future research. But we can already make a realistic estimation taking into account only the contribution to the PBH mass from the $\zeta_{\rm tail}$, which will give the dominant contribution. Fluctuations of type-II have not been explored numerically yet, and its precise mass spectrum is unknown (although one expects larger masses than type-I since they are fluctuations highly over-threshold). But as a first estimate following the common consideration done for fluctuations of type-I\footnote{In the appendix \ref{sec:uper_bound}, we make another estimate to take into account the possible effect of accretion using the average of the compaction function. The result does not differ substantially from the one we consider in this section.}, we will mainly consider that the PBH mass formed is basically the horizon mass when the comoving scale $r_{\rm tail}$ reenters the cosmological horizon. Specifically
\begin{equation}
 M_{b}(\nu_b) = F M_{k}(k_{\rm tail}(\nu_b)) ,  
\end{equation}
where the $k_{\rm tail}(\nu_b)$ wave-mode is  given by
\begin{equation}
k_{\rm tail}(\nu_b) =  \frac{1}{r_{\rm tail}(\nu_b) e^{\zeta_{\rm tail}(r_{\rm tail}(\nu_b))}}.
\end{equation}
Notice its dependence with the amplitude $\nu_b$ according to Fig.\ref{fig:profile_bubble_end}. In other words, the location of the peak of the compaction function will depend on the amplitude of the fluctuation $\nu_b$ that generates the bubble. The constant factor $F$ relates the comoving size at horizon crossing with the mass of the resulting black hole. In most PBH formation scenarios this factor is of the order of a few. Numerical simulations would be required to determine it with precision, which is outside the scope of the present work. The numerical result of \cite{vacum_bubles,2017JCAP...04..050D} found that the PBH mass formed from large vacuum bubbles is basically the horizon mass at the time when the comoving scale of the bubble reenters the cosmological horizon times a factor $F = 5.6$. That case is different from ours since we find the surrounding adiabatic curvature fluctuation $\zeta_{\rm tail}$ at the tail of the bubble profile instead of a perfect domain wall profile. Although that, we expect that the value $F$ will be similar in a range $(1-10)$, and we propose a factor $F \sim 3$ instead of $5.6$ to account for this possible variation.

\begin{figure}[h]
\centering
\includegraphics[width=3.0 in]{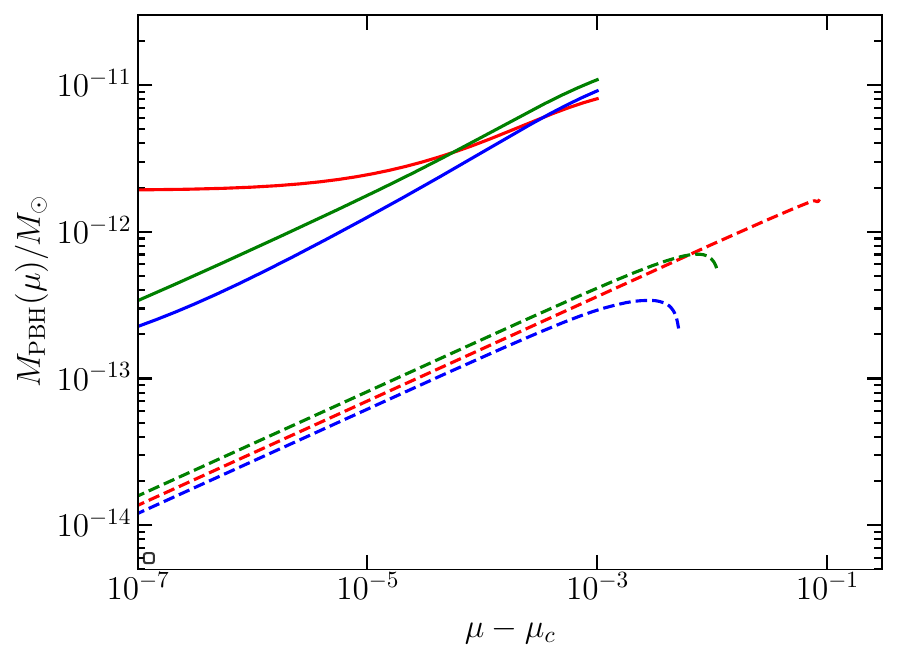}
\includegraphics[width=3.0 in]{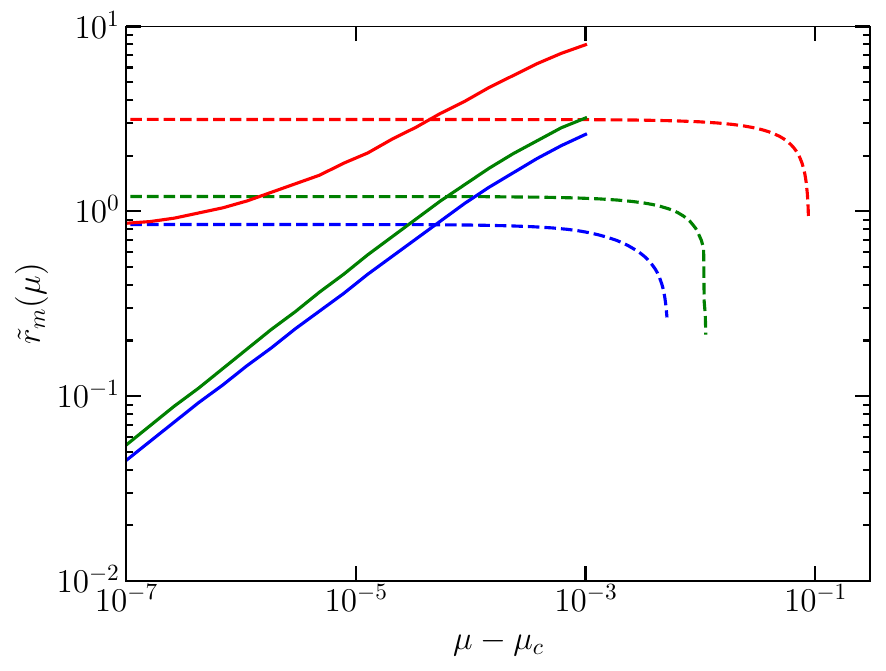}
\includegraphics[width=3.0 in]{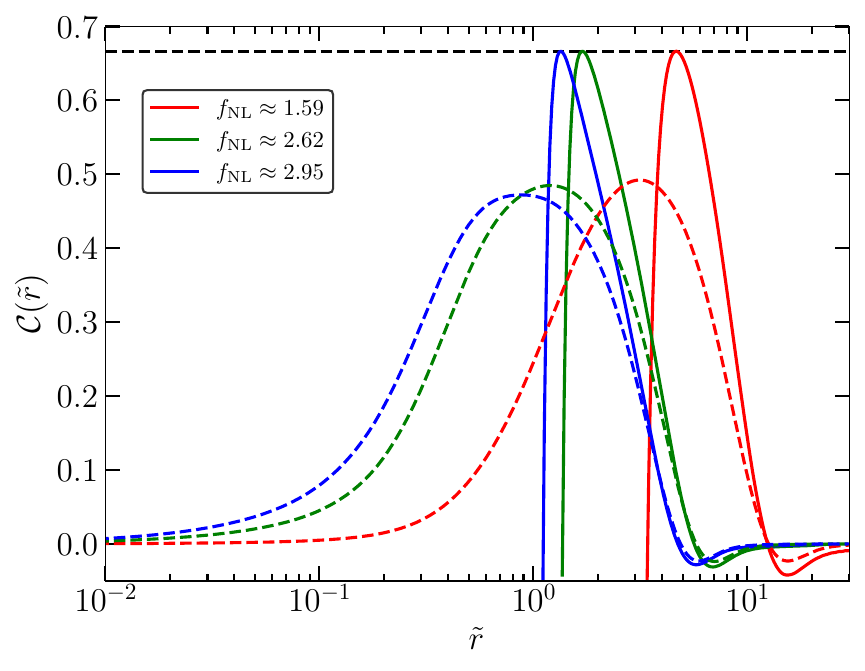}
\caption{Top-Left panel: PBH mass dependence in terms of the amplitude of the fluctuation $\mu$. In both cases, the amplitude $\mu$ corresponds to the adiabatic (dashed-line) and bubble channel (solid-line) corresponds to $\mu_{a}$ and $\mu_{b}$, respectively. This also holds for the other plots. Top-Right panel: location $\tilde{r}_m$ of the peak of the compaction function from the $\zeta_{\rm tail}$ and the one from $\zeta$ in terms of the amplitude $\mu$. Bottom panel: compaction functions from the bubble channel $\zeta_{\rm tail}$ and from the standard adiabatic fluctuation $\zeta$ for different cases of $f_{\rm NL}$ and with a fixed $\mu-\mu_{c} \approx 10^{-4}$.}
\label{fig:mass_depencen}
\end{figure}

In the top-left panel of Fig.\ref{fig:mass_depencen}, we show the mass dependence of both channels of PBH production in terms of the amplitude $\mu$. The mass increases in both cases, although in the adiabatic channel there is a reduction for very large $\mu_a$, consistent with the sharp reduction of the $\tilde{r}_m$ (since the compaction function becomes compressed towards the origin of coordinates) shown in the right-top panel. Notice that the PBH mass in the case of the adiabatic channel is substantially smaller than the bubble channel for very small $\mu_{a}-\mu_{a,c}$ due to the effect of the critical regime. Moreover $\tilde{r}_{\rm tail}$ increases when stronger backward quantum fluctuations are considered, which can also be visualized in Fig.\ref{fig:profile_bubble_end}. We also show the compaction functions in the bottom panel for the case $\mu-\mu_c \approx 10^{-4}$. As expected, the peak value of  $\mathcal{C}_{\zeta_{\rm tail}}$ is located at $2/3$ (the maximum value) since it corresponds to type-II fluctuation. In this case for fluctuations of type-II, should exist another local maximum with peak value $2/3$, which is hidden inside the bubble and matches with the location of the bubble size $R_{b}$.  

Then, the mass function for PBHs from the bubble channel is given by
\begin{equation}
\label{eq:mass_function_bubble}
    f_{\rm b}(M_b) = \frac{M_{ b}\mathcal{N}_{\rm pk}(\nu_b(M_b))}{\rho_{\rm critical} \Omega_{\rm DM}} \Bigg | \frac{d \ln M_b(\nu_b)}{d \nu_b} \Bigg |^{-1},
\end{equation}
where the Jacobian term in this case is computed as
\begin{equation}
 \frac{d \ln M_{b}(\nu_b)}{d \nu_b}  =   \frac{2}{r_{\rm tail}(\nu_b)} \frac{d r_{\rm tail}(\nu_b)}{d \nu_b} +2 \frac{d \zeta_{\rm tail}(r_{\rm tail}(\nu_b))}{d \nu_b}.
\end{equation}

\begin{figure}[h]
\centering
\includegraphics[width=4.5 in]{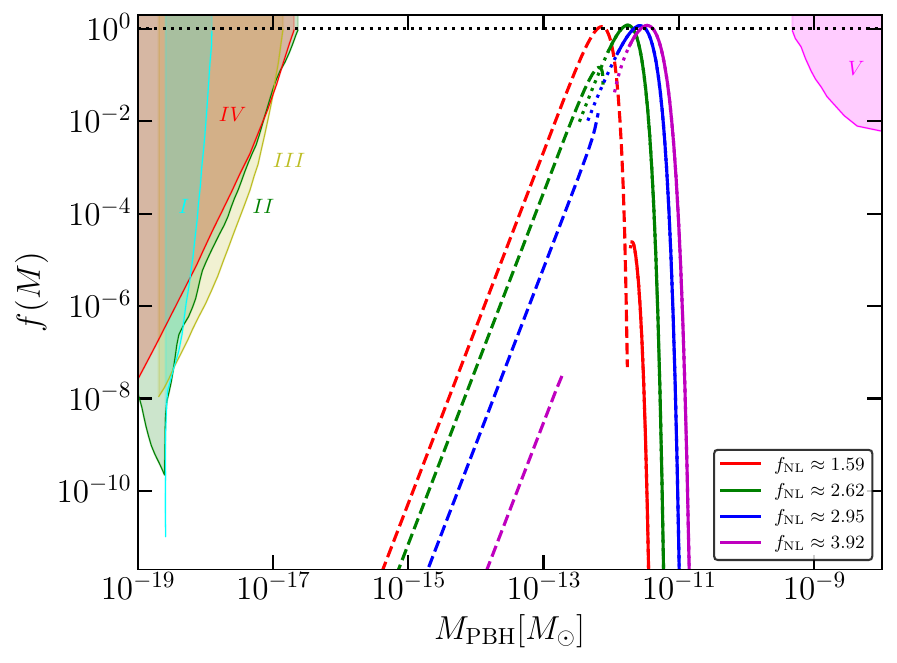}
\caption{Mass function $f(M)$ for the two channels of PBH production: false vacuum bubbles (solid line) and adiabatic channel (dashed line) in terms of different values of $f_{\rm NL}$. The mass range of $f(M)$ for the case $f_{\rm NL} \approx 2.95$ (blue line) and $f_{\rm NL} \approx 3.92$ (magenta line) has been shifted by a factor $\sim 2-4$ respectively for visualization purposes. The constraints of the fraction of PBHs in the form of dark matter are also shown, which have been obtained by digitalizing Fig.11 of \cite{Carr:2020gox}. Constraint $\boldsymbol{I}$ comes from Galactic $\gamma$-rays \cite{Carr:2016hva}, $\boldsymbol{II}$ from extragalactic $\gamma$-rays \cite{Carr:2009jm}, $\boldsymbol{III}$ from Voyager-1 $e^{\pm}$ \cite{Boudaud:2018hqb}, $\boldsymbol{IV}$ from CMB spectral distortions and anisotropies \cite{Acharya:2020jbv,Chluba:2020oip} and $\boldsymbol{V}$ from microlensing of stars in M31 by Subaru (HSC) \cite{Niikura:2017zjd}. See \cite{Carr:2020gox} for more details. The dotted lines correspond to values $\mu_{b}<\mu_{b,(\rm cut-off)}$, which are included for completeness, but are not taken into account when normalizing the mass function. In particular, the corresponding cut-off for $\mu_{b}$ according to Eq.\eqref{eq:cut_off} is given by $\mu_{b,\rm (cutoff)}-\mu_{b,c}  \approx  1.32 \cdot 10^{-6}$ (for $f_{\rm NL} \approx 1.59$), $\mu_{b,\rm (cutoff)}-\mu_{b,c}  \approx  1.66 \cdot 10^{-6}$ (for $f_{\rm NL} \approx 2.62$), $\mu_{b,\rm (cutoff)}-\mu_{b,c}  \approx  1.81 \cdot 10^{-6}$ (for $f_{\rm NL} \approx 2.95$) and $\mu_{b,\rm (cutoff)}-\mu_{b,c}  \approx  2.22 \cdot 10^{-6}$ (for $f_{\rm NL} \approx 3.92$).}
\label{fig:mass_function}
\end{figure}

The total fraction of PBHs in the form of dark matter taking into account both channels of PBH production is simply given by
\begin{equation}
    f_{PBH}^{\rm tot} = \int_{-\infty}^{\infty} f_{a}(M_a)d \log M_a + \int_{-\infty}^{\infty} f_{b}(M_b) d \log M_b.
\end{equation}
The result for the mass function\footnote{Notice that for visualization purposes, we have only considered the mass spectrum before the decrease shown in Fig.\ref{fig:mass_depencen} to avoid the apparent divergence from the Jacobian in Eq.\eqref{eq:mass_function_adiabatic} (similar consideration was done in \cite{Kitajima:2021fpq}), which is integrable and hardly contributes to $f_{\rm PBH}^{\rm tot}$. In Table \ref{table:fixedfraction}, we give the values of $f_{\rm PBH}^{\rm tot}$ taking into account the full range $\mu_{a,c}<\mu_{a}<\mu_{\star}$. Notice as well that we are assuming that the scaling law Eq.\eqref{eq:mass_scaling2} is fulfilled even for very large amplitudes up to $\mu_{a}<\mu_{\star}$, for what can be deviated few percentages as shown in \cite{escriva_solo}, and therefore this apparent divergent feature may not be found if numerical simulations are used to study the mass spectrum for very large $\mu_{a}$.} is shown in Fig.\ref{fig:mass_function}, which is compared with the current observational constraints\footnote{The constraints assume a monochromatic mass function, which is not strictly our case. Nevertheless we follow the standard practice in the literature to get an orientative plot.} of PBHs in the form of dark matter in this mass range (see the caption of Fig.\ref{fig:mass_function} for details).
As can be appreciated, for large $f_{\rm NL}$, the bubble channel (solid line) dominates over the adiabatic (dashed line). Since the contribution from the bubble channel is dominated by a fluctuation of type-II whereas in the adiabatic channel the main contribution comes from fluctuations of type-I, we show that fluctuations of type-II can contribute substantially more to the mass function than the standard fluctuations of type-I. To our knowledge, this is the first example where this behaviour has been noted.

Notice that in the adiabatic channel we can also have a spectrum of fluctuations of type-II for relatively small $f_{\rm NL}$ (for sufficiently large $f_{\rm NL}$ it is not possible to have fluctuations of type-II within a range of amplitudes $\mu_{a,c}<\mu_{a} < \mu_{\star}$). However, within our considerations and for the cases tested these fluctuations are statistically highly suppressed compared with fluctuations of type-I. The main point is that we should consider the bubble channel of PBH production together with the adiabatic one. Finally, in the case when both channels give a commensurable production of PBHs, accounting for the contribution of the bubble channel makes the mass function significantly broader.

\section{Conclusion and discussion}\label{ref:conclusions}

In this work, we have numerically studied the vacuum bubble formation process produced by trapping a localized region of the inflaton field, due to large backward fluctuations after it overshoots a bump in a single field Starobinsky potential. These bubbles represent a second channel for the production of PBHs and might coexist with PBHs created from the collapse of large adiabatic perturbations.

Using numerical simulations with consistent initial conditions, we have found the threshold for the creation of these bubbles, and thus, we have been able to accurately predict the relative abundance of adiabatic PBHs versus bubble PBHs. This can be expressed in terms of the strength of non-Gaussianity $f_{\rm NL}$, which parametrizes a local relation between the curvature perturbations $\zeta$ and the Gaussian variable $\zeta_G$ Eq.\eqref{eq:zeta_NG}.

Our results in the limit of small $f_{\rm NL}$ confirm that the log relation between the Gaussian curvature fluctuation and its non-Gaussian counterpart in the presence of a small bump is successful in predicting the channel of bubble formation, which is remarkably accurate within a deviation of $5\%$ for $f_{\rm NL} \lesssim 2.6$. For larger $f_{\rm NL} \gtrsim 3.5 $, the numerical results deviate more from the analytical estimate. This is to be expected, since in this regime the assumption of reaching the attractor in the quadratic part of the potential becomes less accurate (see caption of Fig.\ref{fig:modes_evolution}). It would be interesting to see whether this feature is model-independent or not.

We have confirmed the result found in \cite{2020JCAP...05..022A} that the number of peaks generating the bubbles is higher than those producing adiabatic PBHs in the case of large non-Gaussianity. Specifically we find that this happens for $f_{\rm NL} \gtrsim 2.6$ as shown in Fig.\ref{fig:nu_critical_values}. This is slightly smaller than the value $f_{\rm NL} \approx 3.5$ found in \cite{2020JCAP...05..022A}, indicating that vacuum bubbles are more easily formed than expected. The reason is a slight evolution of the power spectrum from the time $N_{\star}$ when bubbles form to the end of inflation, where adiabatic perturbations will set in.

On the other hand, we have found for the cases tested that the comoving size of the bubble at the end of inflation follows a critical regime for sufficiently small $\mu_{b}-\mu_{b,c}$ with a critical exponent $\gamma_b$ that increases for larger values of $f_{\rm NL}$. This makes the size of the bubbles significantly smaller for very small $\mu_{b}-\mu_{b,c}$ when considering models with large $f_{\rm NL}$. Even so, a numerical refinement method would be needed to reduce the error bars in Fig.\ref{fig:critical_exponent} and explore the bubble size behaviour for smaller values $\mu_{b}-\mu_{b,c}$, especially for large $f_{\rm NL}$.

Interestingly, at the end of inflation, the bubble is surrounded by an adiabatic curvature fluctuation $\zeta_{\rm tail}$, which surprisingly has been found to be of type-II. To the best of our knowledge, this is the first example where it is explicitly shown that fluctuations of type-II contribute substantially more to the PBH abundance than standard fluctuations of type-I. This is explicitly shown in the mass function $f(M)$ of Fig.\ref{fig:mass_function}, which for small $f_{\rm NL}$ is dominated by fluctuations of type-I (adiabatic channel) and for large $f_{\rm NL}$ it is dominated by fluctuations of type-II (bubble channel). A refinement of the PBH mass function with full simulations that take into account the gravitational collapse of fluctuations of type-II is left for future research.  

Finally, we have shown explicitly that, for a given abundance of PBHs, the amplitude of the peak of the power spectrum needs to be smaller as $f_{\rm NL}$ increases. In this work, we have chosen the parameters of the inflationary potential to realize a peak in the power spectrum in the asteroid mass range, for which there are currently no strong constraints, and there is compatibility for PBHs to account for all the dark matter. Still, it would be interesting to repeat the same analysis in the mass range of stupendous massive black holes \cite{Carr:2020erq,Atal:2020yic,Deng:2021edw}, where there are relatively strong constraints (see Figs.37,19 of \cite{Carr:2020gox,Escriva:2022duf} respectively) due to $\mu$ distortions from the CMB \cite{Kohri:2014lza}. Therefore, large non-gaussianities can help avoiding the restriction on the power spectrum peak amplitude \cite{Nakama:2017xvq,Unal:2020mts}, even when there is a sizable fraction of dark matter in the form of PBH. Such fraction will in fact come from the bubble channel of PBH production.

\acknowledgments
We thank Ken-ichi Nakao, Chulmoon Yoo and Hirotaka Yoshino for discussions about vacuum bubbles. A.E. acknowledges support from the JSPS Postdoctoral Fellowships for Research in Japan (Graduate School of Sciences, Nagoya University). J.G. is supported by the Unit of Excellence MdM 2020-2023" award to the Institute of Cosmos Sciences (CEX2019-000918-M) and by grants AGAUR 2021-SGR00872, PID2019-105614GB-C22.

\newpage
\appendix

\section{Possible effect of mass accretion}
\label{sec:uper_bound}
\begin{figure}[H]
\centering
\includegraphics[width=3.5 in]{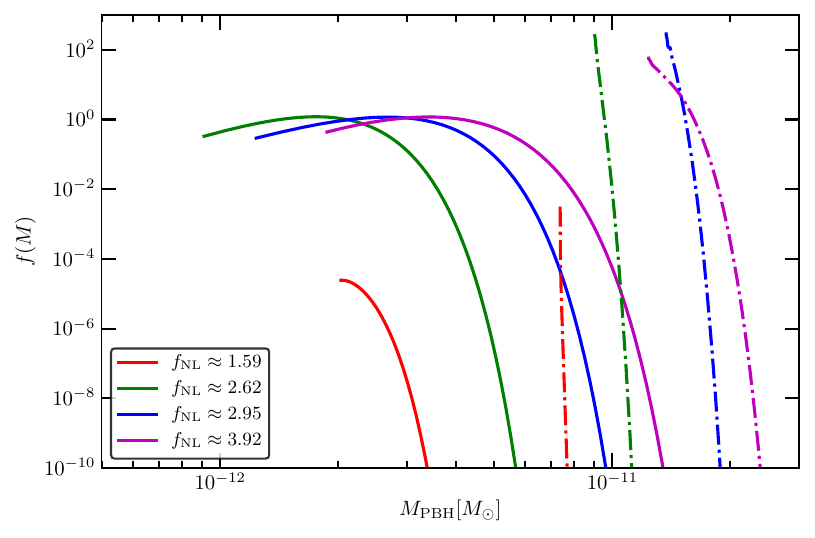}
\caption{Mass function for the bubble channel, considering the length-scale of the fluctuation $\zeta_{\rm tail}$ as $r_{\rm tail}$ like in Fig.\ref{fig:mass_function} (solid line) and with $r_{\rm tail,2}$ (point-dashed line). The mass range of $f(M)$ for the case $f_{\rm NL} \approx 2.95$ (blue line) and $f_{\rm NL} \approx 3.92$ (magenta line) have been shifted by a factor $\sim 2-4$ respectively for visualization purposes, like in Fig.\ref{fig:mass_function}.}
\label{fig:mass_function2}
\end{figure}

We have made another estimation using the averaged $\bar{\mathcal{C}}$ of the $\zeta_{\rm tail}$ to take into account the possible mas excess of the over-threshold averaged compaction function $\bar{\mathcal{C}}>2/5$ that can contribute of increasing the PBH mass due to accretion. To do that, we have considered a new length-scale $r_{\rm tail,2}$ as the scale for which the averaged $\bar{\mathcal{C}}_{\rm tail}$ integrated from $r_{\rm tail}$ (the location of the peak of $\mathcal{C}_{\rm tail}$) up to $r_{\rm tail,2}$ gives $\bar{\mathcal{C}}_{\rm tail}=2/5$. Our intuition is based on the fact that all the extra mass excess above the critical value $\bar{\mathcal{C}}_{\rm tail}=2/5$ will contribute to the PBH mass. In particular we find $r_{\rm tail,2}$ solving
\begin{equation}
\label{eq:criterio_average_2}
\frac{2}{5}=\frac{3}{ \left[ r^3_{\rm tail,2} e^{3 \zeta_{\rm tail}(r_{\rm tail,2})}-r^3_{\rm tail} e^{3 \zeta_{\rm tail}(r_{\rm tail})} \right]} \int_{r_{\rm tail}}^{r_{\rm tail,2}} \mathcal{C}_{\rm tail}(r)(1+r \zeta_{\rm tail}') e^{3 \zeta_{\rm tail}(r)} r^2 dr .
\end{equation}
The result is shown in Fig.\ref{fig:mass_function2}. The new mass function for the bubble channel is shifted to larger mass values. The contribution to the fraction of PBHs in the form of dark matter from the bubble channel is increased by a factor $\sim 5$, which indicates that it doesn't substantially change the result from our previous realistic estimation using $r_{\rm tail}$ as a length-scale. Numerical simulations will be needed to test this hypothesis.

\section{Numerical parameters data}

This appendix provides the numerical parameters that we have used to build the inflationary potential of Eq.\eqref{eq:pot_starobinsky} and proceed with the numerical simulations of bubble formation. In Table \ref{table:nuc}, we give the parameters for the cases fixing the critical threshold value for bubble formation $\nu_{b,c}$. In contrast, in Table \ref{table:fixedfraction}, we provide the parameters that give cases for what the fraction of PBHs in the form of dark matter is $f_{\rm PBH}^{\rm tot}= 1 \pm 0.03$. All parameters have been chosen to fulfill the CMB requirements at the pivot scale and to have the power spectrum peak $\mathcal{P}_{\zeta_G}(k_{\rm max})$, in a wave-mode scale $k_{\rm max}$ such that the corresponding horizon mass lies in the range $M_{k}(k_{\rm max})  \in [10^{-13},10^{-12}] M_{\odot}$.

\begin{table}[H]
\centering
\begin{tabular}{|c|c|c|c|c|c|c|c|}
\hline
$A ( \nu_{b,c} = 8 \pm 0.1)$ & $\frac{\sigma^2}{10^{-4}}$ & $\phi_{0}$ & $\frac{V_{0}}{10^{-3}}$ & $f_{\rm NL}$ & $\frac{\mathcal{P}_{\zeta_G}(k_{\rm pivot})}{10^{-9}}$ & $\frac{n_{s}}{10^{-1}}$ & $N_{\rm end}-N_{\rm pivot}$ \\ \hline
0.00234823544814031  & $4.2 $    & 3.925               & 3.53          & 0.69                                  & 2.09                                        & 9.59         & 54.19                                \\ \hline
0.00204353827240769  & 3              & 3.93                & 3.53          & 1.13                                  & 2.17                                        & 9.60         & 52.18                                \\ \hline
0.0018507625124141   & 2.25            & 3.925               & 3.54          & 1.59                                  & 2.24                                        & 9.60         & 51.28                                \\ \hline
0.00148802386561789  & 1.5             & 4                   & 3.53          & 2.08                                  & 2.30                                        & 9.61         & 51.39                                \\ \hline
0.00134343266747832  & 1.29            & 4.048               & 3.505         & 2.21                                  & 2.28                                        & 9.61         & 51.70                                \\ \hline
0.00126306699654353  & 1.05            & 4.048               & 3.5           & 2.62                                  & 2.26                                        & 9.61         & 51.30                                \\ \hline
0.00120933210369861  & 0.9             & 4.048               & 3.5           & 2.95                                  & 2.26                                        & 9.61         & 51.06                                \\ \hline
0.00115206525841421  & 0.75            & 4.048               & 3.505         & 3.36                                  & 2.28                                        & 9.61         & 50.83                                \\ \hline
0.00111555573420713  & 0.66            & 4.048               & 3.5           & 3.68                                  & 2.26                                        & 9.61         & 50.69                                \\ \hline
0.00109008478843295  & 0.6             & 4.048               & 3.5           & 3.92                                  & 2.26                                        & 9.61         & 50.60                                \\ \hline
0.00106811486865607  & 0.55            & 4.048               & 3.495         & 4.15                                  & 2.25                                        & 9.61         & 50.52                                \\ \hline
0.00102153747766813  & 0.45            & 4.048               & 3.495         & 4.72                                  & 2.25                                        & 9.61         & 50.37                                \\ \hline
0.000983754231379358 & 0.375           & 4.048               & 3.495         & 5.28                                  & 2.25                                        & 9.61         & 50.26                                \\ \hline
0.000942745243336801 & 0.3             & 4.048               & 3.495         & 6.04                                  & 2.25                                        & 9.61         & 50.15                                \\ \hline
\end{tabular}

\begin{tabular}{|l|l|}
\hline
\multicolumn{1}{|c|}{$A ( \nu_{b,c} = 10 \pm 0.05)$} & \multicolumn{1}{c|}{$A ( \nu_{b,c} = 12 \pm 0.05)$} \\ \hline
0.00234815594028443 & 0.00234808087098397 \\ \hline
0.00204342623486182 & 0.00204331588405784 \\ \hline
0.00185061586509428 & 0.00185046635285446 \\ \hline
0.00148783955909958 & 0.00148766634984688 \\ \hline
0.00134324912490089 & 0.00134307577203213 \\ \hline
0.0012628683408432 & 0.00126266364856562 \\ \hline
0.00120911430742841 & 0.00120887830114502 \\ \hline
0.00115180495570062 & 0.00115153762505408 \\ \hline
0.00111526841065318 & 0.00111498540297226 \\ \hline
0.00108980550543697 & 0.00108950683584457 \\ \hline
0.00106778894428082 & 0.00106749023768939 \\ \hline
0.00102120421557745 & 0.0010208741850228 \\ \hline
0.000983429190676518 & 0.00098309908693459 \\ \hline
0.000942412857224492 & 0.000942113906988106 \\ \hline
\end{tabular}
\caption{Parameters for the case fixing the value $\nu_{b,c} = 8 \pm 0.1$ (top table). The initial values chosen for solving the inflaton's homogeneous background dynamics are in the range $\phi_{\rm ini} \in [5.33-5.39]$. Parameter $A$ for the case fixing the value $\nu_{b,c} = 10 \pm 0.05$ and $\nu_{b,c} = 12 \pm 0.05$ (bottom table). The corresponding values $\sigma$, $\phi_{0}$ and $V_{0}$ are the same as in the top table. The other values $f_{\rm NL}$, $\frac{\mathcal{P}_{\zeta_G}(k_{\rm pivot})}{10^{-9}}$, $\frac{n_{s}}{10^{-1}}$ and $N_{\rm end}-N_{\rm pivot}$ for the cases $\nu_{b,c} \sim 10,12$ are very similar to the case $\nu_{b,c} \sim 8$ and are not shown.}
\label{table:nuc}
\end{table}

\begin{table}[H]
\centering
\begin{tabular}{|c|c|c|c|c|c|c|c|}
\hline
$A $ & $\frac{\sigma^2}{10^{-4}}$ & $\phi_{0}$ & $\frac{V_{0}}{10^{-3}}$ & $f_{\rm NL}$ & $\frac{\mathcal{P}_{\zeta_G}(k_{\rm pivot})}{10^{-9}}$ & $\frac{n_{s}}{10^{-1}}$ & $N_{\rm end}-N_{\rm pivot}$ \\ \hline
0.00185061 & 2.25 & 3.925 & 3.54 & 1.59 & 2.24 & 9.60 & 51.16 \\ \hline
0.0012630165 & 1.05 & 4.048 & 3.5 & 2.62 & 2.26 & 9.61 & 51.28 \\ \hline
0.001209285 & 0.9 & 4.048 & 3.5 & 2.95 & 2.26 & 9.61 & 51.04 \\ \hline
0.001090017 & 0.6 & 4.048 & 3.5 & 3.92 & 2.26 & 9.61 & 50.58 \\ \hline
\end{tabular}
\begin{tabular}{|c|c|c|c|c|c|}
\hline
$\nu_{b,c}$ & $\nu_{\star}$ & $\nu_{a,c}$ & $\mathcal{P}_{\zeta_G}(k_{\rm max})$ & $f_{\rm PBH}^{\rm tot (adiabatic)}$ & $f_{\rm PBH}^{\rm tot (bubble)}$ \\ \hline
10.09 & 10.31 & 8.57 & $2.28 \cdot 10^{-3}$  & 1.021 & $3.33 \cdot 10^{-6}$ \\ \hline
8.55 & 8.95 & 8.55 & $1.28 \cdot 10^{-3}$ & $7.36 \cdot 10^{-2}$ & 0.928 \\ \hline
8.51 & 9.08 & 8.85 & $1.00 \cdot 10^{-3}$ & $2.58 \cdot 10^{-3}$ & 1.025 \\ \hline
8.52 & 9.94 & 9.94 & $4.95 \cdot 10^{-4}$ & $3.57 \cdot 10^{-8}$ & 0.981 \\ \hline
\end{tabular}
\caption{Parameters for the case of fixing the total fraction of PBHs in the form of dark matter to be $f_{\rm PBH}^{\rm tot}=  1 \pm 0.03$. The initial values chosen for solving the inflaton's homogeneous background dynamics are in the range $\phi_{\rm ini} \in [5.33-5.39]$. }
\label{table:fixedfraction}
\end{table}

It is important to notice that in Table \ref{table:nuc} not all the digits for $A$ are significant. For instance, a change $\delta A \sim 10^{-8}$ in the amplitude of the barrier of the potential will reduce the fraction $f_{\rm PBH}^{\rm tot}$ by $10^{-1}$. But we show all the digits used in the simulations for clarity to the reader.

%\section{Table of data}

%\section{Convergence test of the simulations for bubble formation}
%In this appendix we briefly show some congernce test for the numercal solution of bubble formation

\newpage

\bibliographystyle{JHEP}
\bibliography{refs5_arreglar_duplicados.bib}

@article{1974MNRAS.168..399C,
	title        = {{Black holes in the early Universe}},
	author       = {{Carr}, B. and {Hawking}, S.},
	year         = 1974,
	month        = aug,
	journal      = {\mnras},
	volume       = 168,
	pages        = {399--416},
	doi          = {10.1093/mnras/168.2.399},
	adsurl       = {https://ui.adsabs.harvard.edu/abs/1974MNRAS.168..399C}
}

@article{musco2013,
	title        = {{Primordial black hole formation in the early universe: critical behaviour and self-similarity}},
	author       = {{Musco}, Ilia and {Miller}, John C.},
	year         = 2013,
	month        = jul,
	journal      = {Classical and Quantum Gravity},
	volume       = 30,
	number       = 14,
	pages        = 145009,
	doi          = {10.1088/0264-9381/30/14/145009},
	eid          = 145009,
	archiveprefix = {arXiv},
	eprint       = {1201.2379},
	primaryclass = {gr-qc},
	adsurl       = {https://ui.adsabs.harvard.edu/abs/2013CQGra..30n5009M}
}

@article{2002CQGra..19.3687H,
	title        = {{The dynamics of primordial black-hole formation}},
	author       = {{Hawke}, I. and {Stewart}, J.~M.},
	year         = 2002,
	month        = jul,
	journal      = {Classical and Quantum Gravity},
	volume       = 19,
	number       = 14,
	pages        = {3687--3707},
	doi          = {10.1088/0264-9381/19/14/310},
	adsurl       = {https://ui.adsabs.harvard.edu/abs/2002CQGra..19.3687H}
}

@article{PhysRevD.50.7173,
	title        = {Inflation and primordial black holes as dark matter},
	author       = {Ivanov, P. and Naselsky, P. and Novikov, I.},
	year         = 1994,
	month        = {Dec},
	journal      = {Phys. Rev. D},
	publisher    = {American Physical Society},
	volume       = 50,
	pages        = {7173--7178},
	doi          = {10.1103/PhysRevD.50.7173},
	url          = {https://link.aps.org/doi/10.1103/PhysRevD.50.7173},
	issue        = 12,
	numpages     = {0}
}

@article{Garriga,
	title        = {{PBH abundance from random Gaussian curvature perturbations and a local density threshold}},
	author       = {{Yoo}, Chul-Moon and {Harada}, Tomohiro and {Garriga}, Jaume and {Kohri}, Kazunori},
	year         = 2018,
	month        = may,
	journal      = {arXiv e-prints},
	pages        = {arXiv:1805.03946},
	eid          = {arXiv:1805.03946},
	archiveprefix = {arXiv},
	eprint       = {1805.03946},
	primaryclass = {astro-ph.CO},
	adsurl       = {https://ui.adsabs.harvard.edu/\#abs/2018arXiv180503946Y}
}

@article{germani-vicente,
	title        = {The role of non-gaussianities in primordial black hole formation},
	author       = {Vicente Atal and Cristiano Germani},
	year         = 2019,
	journal      = {Physics of the Dark Universe},
	volume       = 24,
	pages        = 100275,
	doi          = {https://doi.org/10.1016/j.dark.2019.100275},
	issn         = {2212-6864},
	url          = {http://www.sciencedirect.com/science/article/pii/S2212686418301997}
}

@article{1967SvA....10..602Z,
	title        = {{The Hypothesis of Cores Retarded during Expansion and the Hot Cosmological Model}},
	author       = {{Zel'dovich}, Ya. B. and {Novikov}, I.~D.},
	year         = 1967,
	month        = feb,
	journal      = {\sovast},
	volume       = 10,
	pages        = 602,
	adsurl       = {https://ui.adsabs.harvard.edu/abs/1967SvA....10..602Z}
}

@article{2019JCAP...09..073A,
	title        = {{Primordial black hole formation with non-Gaussian curvature perturbations}},
	author       = {Atal, Vicente and Garriga, Jaume and Marcos-Caballero, Airam},
	year         = 2019,
	journal      = {JCAP},
	volume       = {09},
	pages        = {073},
	doi          = {10.1088/1475-7516/2019/09/073},
	eprint       = {1905.13202},
	archiveprefix = {arXiv},
	primaryclass = {astro-ph.CO}
}

@article{1975ApJ...201....1C,
	title        = {{The primordial black hole mass spectrum.}},
	author       = {{Carr}, B.},
	year         = 1975,
	month        = oct,
	journal      = {\apj},
	volume       = 201,
	pages        = {1--19},
	doi          = {10.1086/153853},
	adsurl       = {https://ui.adsabs.harvard.edu/abs/1975ApJ...201....1C}
}

@article{2020JCAP...05..022A,
	title        = {{PBH in single field inflation: the effect of shape dispersion and non-Gaussianities}},
	author       = {{Atal}, Vicente and {Cid}, Judith and {Escriv{\`a}}, Albert and {Garriga}, Jaume},
	year         = 2020,
	month        = may,
	journal      = {\jcap},
	volume       = 2020,
	number       = 5,
	pages        = {022},
	doi          = {10.1088/1475-7516/2020/05/022},
	eid          = {022},
	archiveprefix = {arXiv},
	eprint       = {1908.11357},
	primaryclass = {astro-ph.CO},
	adsurl       = {https://ui.adsabs.harvard.edu/abs/2020JCAP...05..022A}
}

@article{Niemeyer1,
	title        = {Near-Critical Gravitational Collapse and the Initial Mass Function of Primordial Black Holes},
	author       = {Niemeyer, J. C. and Jedamzik, K.},
	year         = 1998,
	month        = jun,
	journal      = {Phys. Rev. Lett.},
	publisher    = {American Physical Society},
	volume       = 80,
	pages        = {5481--5484},
	doi          = {10.1103/PhysRevLett.80.5481},
	url          = {https://link.aps.org/doi/10.1103/PhysRevLett.80.5481},
	issue        = 25,
	numpages     = {0}
}

@article{refrencia-extra-jaume,
	title        = {Cosmological long-wavelength solutions and primordial black hole formation},
	author       = {Harada, Tomohiro and Yoo, Chul-Moon and Nakama, Tomohiro and Koga, Yasutaka},
	year         = 2015,
	month        = apr,
	journal      = {Phys. Rev. D},
	publisher    = {American Physical Society},
	volume       = 91,
	pages        = {084057},
	doi          = {10.1103/PhysRevD.91.084057},
	url          = {https://link.aps.org/doi/10.1103/PhysRevD.91.084057},
	issue        = 8,
	numpages     = 25
}

@article{2021JCAP...01..030E,
	title        = {{Analytical thresholds for black hole formation in general cosmological backgrounds}},
	author       = {{Escriv{\`a}}, Albert and {Germani}, Cristiano and {Sheth}, Ravi K.},
	year         = 2021,
	month        = jan,
	journal      = {\jcap},
	volume       = 2021,
	number       = 1,
	pages        = {030},
	doi          = {10.1088/1475-7516/2021/01/030},
	eid          = {030},
	archiveprefix = {arXiv},
	eprint       = {2007.05564},
	primaryclass = {gr-qc},
	adsurl       = {https://ui.adsabs.harvard.edu/abs/2021JCAP...01..030E}
}

@article{universal1,
	title        = {Universal threshold for primordial black hole formation},
	author       = {Escriv\`a, Albert and Germani, Cristiano and Sheth, Ravi K.},
	year         = 2020,
	month        = feb,
	journal      = {Phys. Rev. D},
	publisher    = {American Physical Society},
	volume       = 101,
	pages        = {044022},
	doi          = {10.1103/PhysRevD.101.044022},
	url          = {https://link.aps.org/doi/10.1103/PhysRevD.101.044022},
	issue        = 4,
	numpages     = 5
}

@article{1994PhRvL..72.1782E,
	title        = {{Critical phenomena and self-similarity in the gravitational collapse of radiation fluid}},
	author       = {{Evans}, Charles R. and {Coleman}, Jason S.},
	year         = 1994,
	month        = mar,
	journal      = {\prl},
	volume       = 72,
	number       = 12,
	pages        = {1782--1785},
	doi          = {10.1103/PhysRevLett.72.1782},
	archiveprefix = {arXiv},
	eprint       = {gr-qc/9402041},
	primaryclass = {gr-qc},
	adsurl       = {https://ui.adsabs.harvard.edu/abs/1994PhRvL..72.1782E}
}

@article{escriva_solo,
	title        = {Simulation of primordial black hole formation using pseudo-spectral methods},
	author       = {Albert Escrivà},
	year         = 2020,
	journal      = {Physics of the Dark Universe},
	volume       = 27,
	pages        = 100466,
	doi          = {https://doi.org/10.1016/j.dark.2020.100466},
	issn         = {2212-6864},
	url          = {http://www.sciencedirect.com/science/article/pii/S2212686419302845}
}

@article{1986ApJ...304...15B,
	title        = {{The Statistics of Peaks of Gaussian Random Fields}},
	author       = {{Bardeen}, J.~M. and {Bond}, J.~R. and {Kaiser}, N. and {Szalay}, A.~S.},
	year         = 1986,
	month        = may,
	journal      = {\apj},
	volume       = 304,
	pages        = 15,
	doi          = {10.1086/164143},
	adsurl       = {https://ui.adsabs.harvard.edu/abs/1986ApJ...304...15B}
}

@article{vacum_bubles,
	title        = {{Primordial black hole formation by vacuum bubbles}},
	author       = {Deng, Heling and Vilenkin, Alexander},
	year         = 2017,
	journal      = {JCAP},
	volume       = 12,
	pages        = {044},
	doi          = {10.1088/1475-7516/2017/12/044},
	eprint       = {1710.02865},
	archiveprefix = {arXiv},
	primaryclass = {gr-qc}
}

@article{2020ARNPS..70..355C,
	title        = {{Primordial Black Holes as Dark Matter: Recent Developments}},
	author       = {{Carr}, Bernard and {K{\"u}hnel}, Florian},
	year         = 2020,
	month        = oct,
	journal      = {Annual Review of Nuclear and Particle Science},
	volume       = 70,
	pages        = {355--394},
	doi          = {10.1146/annurev-nucl-050520-125911},
	archiveprefix = {arXiv},
	eprint       = {2006.02838},
	primaryclass = {astro-ph.CO},
	adsurl       = {https://ui.adsabs.harvard.edu/abs/2020ARNPS..70..355C}
}

@article{Germani:2017bcs,
	title        = {{On primordial black holes from an inflection point}},
	author       = {Germani, Cristiano and Prokopec, Tomislav},
	year         = 2017,
	journal      = {Phys. Dark Univ.},
	volume       = 18,
	pages        = {6--10},
	doi          = {10.1016/j.dark.2017.09.001},
	eprint       = {1706.04226},
	archiveprefix = {arXiv},
	primaryclass = {astro-ph.CO},
	reportnumber = {ICCUB-17-012}
}

@article{2018JCAP...05..012C,
	title        = {{Revisiting non-Gaussianity from non-attractor inflation models}},
	author       = {{Cai}, Yi-Fu and {Chen}, Xingang and {Namjoo}, Mohammad Hossein and {Sasaki}, Misao and {Wang}, Dong-Gang and {Wang}, Ziwei},
	year         = 2018,
	month        = may,
	journal      = {\jcap},
	volume       = 2018,
	number       = 5,
	pages        = {012},
	doi          = {10.1088/1475-7516/2018/05/012},
	eid          = {012},
	archiveprefix = {arXiv},
	eprint       = {1712.09998},
	primaryclass = {astro-ph.CO},
	adsurl       = {https://ui.adsabs.harvard.edu/abs/2018JCAP...05..012C}
}

@article{Kitajima:2021fpq,
	title        = {{Primordial black holes in peak theory with a non-Gaussian tail}},
	author       = {Kitajima, Naoya and Tada, Yuichiro and Yokoyama, Shuichiro and Yoo, Chul-Moon},
	year         = 2021,
	journal      = {JCAP},
	volume       = 10,
	pages        = {053},
	doi          = {10.1088/1475-7516/2021/10/053},
	eprint       = {2109.00791},
	archiveprefix = {arXiv},
	primaryclass = {astro-ph.CO},
	reportnumber = {TU-1130}
}

@article{Garriga:2015fdk,
	title        = {{Black holes and the multiverse}},
	author       = {Garriga, Jaume and Vilenkin, Alexander and Zhang, Jun},
	year         = 2016,
	journal      = {JCAP},
	volume       = {02},
	pages        = {064},
	doi          = {10.1088/1475-7516/2016/02/064},
	eprint       = {1512.01819},
	archiveprefix = {arXiv},
	primaryclass = {hep-th}
}

@article{Shibata:1999zs,
	title        = {{Black hole formation in the Friedmann universe: Formulation and computation in numerical relativity}},
	author       = {Shibata, Masaru and Sasaki, Misao},
	year         = 1999,
	journal      = {Phys. Rev. D},
	volume       = 60,
	pages        = {084002},
	doi          = {10.1103/PhysRevD.60.084002},
	eprint       = {gr-qc/9905064},
	archiveprefix = {arXiv},
	reportnumber = {OU-TAP-93}
}

@article{Ballesteros:2020qam,
	title        = {{Primordial black holes as dark matter and gravitational waves from single-field polynomial inflation}},
	author       = {Ballesteros, Guillermo and Rey, Juli\'an and Taoso, Marco and Urbano, Alfredo},
	year         = 2020,
	journal      = {JCAP},
	volume       = {07},
	pages        = {025},
	doi          = {10.1088/1475-7516/2020/07/025},
	eprint       = {2001.08220},
	archiveprefix = {arXiv},
	primaryclass = {astro-ph.CO}
}

@article{Gundlach:2007gc,
	title        = {{Critical phenomena in gravitational collapse}},
	author       = {Gundlach, Carsten and Martin-Garcia, Jose M.},
	year         = 2007,
	journal      = {Living Rev. Rel.},
	volume       = 10,
	pages        = 5,
	doi          = {10.12942/lrr-2007-5},
	eprint       = {0711.4620},
	archiveprefix = {arXiv},
	primaryclass = {gr-qc}
}

@article{Boudaud:2018hqb,
	title        = {{Voyager 1 $e^\pm$ Further Constrain Primordial Black Holes as Dark Matter}},
	author       = {Boudaud, Mathieu and Cirelli, Marco},
	year         = 2019,
	journal      = {Phys. Rev. Lett.},
	volume       = 122,
	number       = 4,
	pages        = {041104},
	doi          = {10.1103/PhysRevLett.122.041104},
	eprint       = {1807.03075},
	archiveprefix = {arXiv},
	primaryclass = {astro-ph.HE}
}

@article{Niikura:2017zjd,
	title        = {{Microlensing constraints on primordial black holes with Subaru/HSC Andromeda observations}},
	author       = {Niikura, Hiroko and others},
	year         = 2019,
	journal      = {Nature Astron.},
	volume       = 3,
	number       = 6,
	pages        = {524--534},
	doi          = {10.1038/s41550-019-0723-1},
	eprint       = {1701.02151},
	archiveprefix = {arXiv},
	primaryclass = {astro-ph.CO}
}

@article{2017JCAP...04..050D,
	title        = {{Primordial black hole and wormhole formation by domain walls}},
	author       = {Deng, Heling and Garriga, Jaume and Vilenkin, Alexander},
	year         = 2017,
	journal      = {JCAP},
	volume       = {04},
	pages        = {050},
	doi          = {10.1088/1475-7516/2017/04/050},
	eprint       = {1612.03753},
	archiveprefix = {arXiv},
	primaryclass = {gr-qc}
}

@article{Nakama:2017xvq,
	title        = {{Limits on primordial black holes from $\mu$ distortions in cosmic microwave background}},
	author       = {Nakama, Tomohiro and Carr, Bernard and Silk, Joseph},
	year         = 2018,
	journal      = {Phys. Rev. D},
	volume       = 97,
	number       = 4,
	pages        = {043525},
	doi          = {10.1103/PhysRevD.97.043525},
	eprint       = {1710.06945},
	archiveprefix = {arXiv},
	primaryclass = {astro-ph.CO}
}

@article{Tanaka:2007gh,
	title        = {{Gradient expansion approach to nonlinear superhorizon perturbations. II. A Single scalar field}},
	author       = {Tanaka, Yoshiharu and Sasaki, Misao},
	year         = 2007,
	journal      = {Prog. Theor. Phys.},
	volume       = 118,
	pages        = {455--473},
	doi          = {10.1143/PTP.118.455},
	eprint       = {0706.0678},
	archiveprefix = {arXiv},
	primaryclass = {gr-qc},
	reportnumber = {YITP-07-31}
}

@article{Lyth:2004gb,
	title        = {{A General proof of the conservation of the curvature perturbation}},
	author       = {Lyth, David H. and Malik, Karim A. and Sasaki, Misao},
	year         = 2005,
	journal      = {JCAP},
	volume       = {05},
	pages        = {004},
	doi          = {10.1088/1475-7516/2005/05/004},
	eprint       = {astro-ph/0411220},
	archiveprefix = {arXiv},
	reportnumber = {YITP-04-67}
}

@article{2019PhRvD.100b3537T,
	title        = {{Primordial black hole tower: Dark matter, earth-mass, and LIGO black holes}},
	author       = {{Tada}, Yuichiro and {Yokoyama}, Shuichiro},
	year         = 2019,
	month        = jul,
	journal      = {\prd},
	volume       = 100,
	number       = 2,
	pages        = {023537},
	doi          = {10.1103/PhysRevD.100.023537},
	eid          = {023537},
	archiveprefix = {arXiv},
	eprint       = {1904.10298},
	primaryclass = {astro-ph.CO},
	adsurl       = {https://ui.adsabs.harvard.edu/abs/2019PhRvD.100b3537T}
}

@article{2018CQGra..35f3001S,
	title        = {{Primordial black holes{\textemdash}perspectives in gravitational wave astronomy}},
	author       = {{Sasaki}, Misao and {Suyama}, Teruaki and {Tanaka}, Takahiro and {Yokoyama}, Shuichiro},
	year         = 2018,
	month        = mar,
	journal      = {Classical and Quantum Gravity},
	volume       = 35,
	number       = 6,
	pages        = {063001},
	doi          = {10.1088/1361-6382/aaa7b4},
	eid          = {063001},
	archiveprefix = {arXiv},
	eprint       = {1801.05235},
	primaryclass = {astro-ph.CO},
	adsurl       = {https://ui.adsabs.harvard.edu/abs/2018CQGra..35f3001S}
}

@article{2021JPhG...48d3001G,
	title        = {{Primordial black holes as a dark matter candidate}},
	author       = {{Green}, Anne M. and {Kavanagh}, Bradley J.},
	year         = 2021,
	month        = apr,
	journal      = {Journal of Physics G Nuclear Physics},
	volume       = 48,
	number       = 4,
	pages        = {043001},
	doi          = {10.1088/1361-6471/abc534},
	eid          = {043001},
	archiveprefix = {arXiv},
	eprint       = {2007.10722},
	primaryclass = {astro-ph.CO},
	adsurl       = {https://ui.adsabs.harvard.edu/abs/2021JPhG...48d3001G}
}

@article{2020JCAP...09..023D,
	title        = {{Primordial black hole formation by vacuum bubbles. Part II}},
	author       = {{Deng}, Heling},
	year         = 2020,
	month        = sep,
	journal      = {\jcap},
	volume       = 2020,
	number       = 9,
	pages        = {023},
	doi          = {10.1088/1475-7516/2020/09/023},
	eid          = {023},
	archiveprefix = {arXiv},
	eprint       = {2006.11907},
	primaryclass = {astro-ph.CO},
	adsurl       = {https://ui.adsabs.harvard.edu/abs/2020JCAP...09..023D}
}

@article{Garriga:1992nm,
	title        = {{Black holes from nucleating strings}},
	author       = {Garriga, Jaume and Vilenkin, Alexander},
	year         = 1993,
	journal      = {Phys. Rev. D},
	volume       = 47,
	pages        = {3265--3274},
	doi          = {10.1103/PhysRevD.47.3265},
	eprint       = {hep-ph/9208212},
	archiveprefix = {arXiv},
	reportnumber = {TUTP-92-6, NSF-ITP-92-1161}
}

@article{Carr:2020erq,
	title        = {{Constraints on Stupendously Large Black Holes}},
	author       = {Carr, Bernard and Kuhnel, Florian and Visinelli, Luca},
	year         = 2021,
	journal      = {Mon. Not. Roy. Astron. Soc.},
	volume       = 501,
	number       = 2,
	pages        = {2029--2043},
	doi          = {10.1093/mnras/staa3651},
	eprint       = {2008.08077},
	archiveprefix = {arXiv},
	primaryclass = {astro-ph.CO}
}

@article{Escriva:2021pmf,
	title        = {{Effects of the shape of curvature peaks on the size of primordial black holes}},
	author       = {Escriv\`a, Albert and Romano, Antonio Enea},
	year         = 2021,
	journal      = {JCAP},
	volume       = {05},
	pages        = {066},
	doi          = {10.1088/1475-7516/2021/05/066},
	eprint       = {2103.03867},
	archiveprefix = {arXiv},
	primaryclass = {gr-qc},
	reportnumber = {ICCUB-21-003}
}

@article{Kusenko:2020pcg,
	title        = {{Exploring Primordial Black Holes from the Multiverse with Optical Telescopes}},
	author       = {Kusenko, Alexander and Sasaki, Misao and Sugiyama, Sunao and Takada, Masahiro and Takhistov, Volodymyr and Vitagliano, Edoardo},
	year         = 2020,
	journal      = {Phys. Rev. Lett.},
	volume       = 125,
	pages        = 181304,
	doi          = {10.1103/PhysRevLett.125.181304},
	eprint       = {2001.09160},
	archiveprefix = {arXiv},
	primaryclass = {astro-ph.CO},
	reportnumber = {IPMU20-0006, YITP-20-11}
}

@article{Kopp:2010sh,
	title        = {{Separate Universes Do Not Constrain Primordial Black Hole Formation}},
	author       = {Kopp, Michael and Hofmann, Stefan and Weller, Jochen},
	year         = 2011,
	journal      = {Phys. Rev. D},
	volume       = 83,
	pages        = 124025,
	doi          = {10.1103/PhysRevD.83.124025},
	eprint       = {1012.4369},
	archiveprefix = {arXiv},
	primaryclass = {astro-ph.CO}
}

@article{Atal:2020yic,
	title        = {{NANOGrav signal as mergers of Stupendously Large Primordial Black Holes}},
	author       = {Atal, Vicente and Sanglas, Albert and Triantafyllou, Nikolaos},
	year         = 2021,
	journal      = {JCAP},
	volume       = {06},
	pages        = {022},
	doi          = {10.1088/1475-7516/2021/06/022},
	eprint       = {2012.14721},
	archiveprefix = {arXiv},
	primaryclass = {astro-ph.CO}
}

@article{Atal:2021jyo,
	title        = {{Probing non-Gaussianities with the high frequency tail of induced gravitational waves}},
	author       = {Atal, Vicente and Dom\`enech, Guillem},
	year         = 2021,
	journal      = {JCAP},
	volume       = {06},
	pages        = {001},
	doi          = {10.1088/1475-7516/2021/06/001},
	eprint       = {2103.01056},
	archiveprefix = {arXiv},
	primaryclass = {astro-ph.CO}
}

@article{Starobinsky:1980te,
	title        = {{A New Type of Isotropic Cosmological Models Without Singularity}},
	author       = {Starobinsky, Alexei A.},
	year         = 1980,
	journal      = {Phys. Lett. B},
	volume       = 91,
	pages        = {99--102},
	doi          = {10.1016/0370-2693(80)90670-X},
	editor       = {Khalatnikov, I. M. and Mineev, V. P.}
}

@article{Sasaki:1983kd,
	title        = {{Gauge Invariant Scalar Perturbations in the New Inflationary Universe}},
	author       = {Sasaki, Misao},
	year         = 1983,
	journal      = {Prog. Theor. Phys.},
	volume       = 70,
	pages        = 394,
	doi          = {10.1143/PTP.70.394},
	reportnumber = {RIFP-506}
}

@article{1988ZhETF..94....1M,
	title        = {{The quantum theory of gauge-invariant cosmological perturbations}},
	author       = {{Mukhanov}, V.~F.},
	year         = 1988,
	month        = jul,
	journal      = {Zhurnal Eksperimentalnoi i Teoreticheskoi Fiziki},
	volume       = 94,
	pages        = {1--11},
	adsurl       = {https://ui.adsabs.harvard.edu/abs/1988ZhETF..94....1M}
}

@article{Mishra:2019pzq,
	title        = {{Primordial Black Holes from a tiny bump/dip in the Inflaton potential}},
	author       = {Mishra, Swagat S. and Sahni, Varun},
	year         = 2020,
	journal      = {JCAP},
	volume       = {04},
	pages        = {007},
	doi          = {10.1088/1475-7516/2020/04/007},
	eprint       = {1911.00057},
	archiveprefix = {arXiv},
	primaryclass = {gr-qc}
}

@article{Escriva:2022duf,
	title        = {{Primordial Black Holes}},
	author       = {Escriv\`a, Albert and Kuhnel, Florian and Tada, Yuichiro},
	year         = 2022,
	month        = 11,
	eprint       = {2211.05767},
	archiveprefix = {arXiv},
	primaryclass = {astro-ph.CO}
}

@inproceedings{2017JPhCS.840a2032G,
	title        = {{Massive Primordial Black Holes as Dark Matter and their detection with Gravitational Waves}},
	author       = {{Garc{\'\i}a-Bellido}, Juan},
	year         = 2017,
	month        = may,
	booktitle    = {Journal of Physics Conference Series},
	series       = {Journal of Physics Conference Series},
	volume       = 840,
	pages        = {012032},
	doi          = {10.1088/1742-6596/840/1/012032},
	eid          = {012032},
	archiveprefix = {arXiv},
	eprint       = {1702.08275},
	primaryclass = {astro-ph.CO},
	adsurl       = {https://ui.adsabs.harvard.edu/abs/2017JPhCS.840a2032G}
}

@article{doi:10.1098/rspa.1978.0060,
	title        = {Quantum field theory in de Sitter space: renormalization by point-splitting},
	author       = {Bunch, T. S.  and Davies, P. C. W.  and Penrose, Roger},
	year         = 1978,
	journal      = {Proceedings of the Royal Society of London. A. Mathematical and Physical Sciences},
	volume       = 360,
	number       = 1700,
	pages        = {117--134},
	doi          = {10.1098/rspa.1978.0060},
	url          = {https://royalsocietypublishing.org/doi/abs/10.1098/rspa.1978.0060},
	eprint       = {https://royalsocietypublishing.org/doi/pdf/10.1098/rspa.1978.0060}
}

@article{2016PhRvD..94h3504C,
	title        = {{Primordial black holes as dark matter}},
	author       = {{Carr}, Bernard and {K{\"u}hnel}, Florian and {Sandstad}, Marit},
	year         = 2016,
	month        = oct,
	journal      = {\prd},
	volume       = 94,
	number       = 8,
	pages        = {083504},
	doi          = {10.1103/PhysRevD.94.083504},
	adsurl       = {https://ui.adsabs.harvard.edu/abs/2016PhRvD..94h3504C},
	archiveprefix = {arXiv},
	eid          = {083504},
	eprint       = {1607.06077},
	primaryclass = {astro-ph.CO}
}

@article{Deng:2021edw,
	title        = {{\ensuremath{\mu}-distortion around stupendously large primordial black holes}},
	author       = {Deng, Heling},
	year         = 2021,
	journal      = {JCAP},
	volume       = 11,
	number       = 11,
	pages        = {054},
	doi          = {10.1088/1475-7516/2021/11/054},
	eprint       = {2106.09817},
	archiveprefix = {arXiv},
	primaryclass = {astro-ph.CO}
}

@article{Kohri:2014lza,
	title        = {{Testing scenarios of primordial black holes being the seeds of supermassive black holes by ultracompact minihalos and CMB $\mu$-distortions}},
	author       = {Kohri, Kazunori and Nakama, Tomohiro and Suyama, Teruaki},
	year         = 2014,
	journal      = {Phys. Rev. D},
	volume       = 90,
	number       = 8,
	pages        = {083514},
	doi          = {10.1103/PhysRevD.90.083514},
	eprint       = {1405.5999},
	archiveprefix = {arXiv},
	primaryclass = {astro-ph.CO},
	reportnumber = {KEK-TH-1736, KEK-COSMO-146, RESCEU-9-14}
}

@article{Carr:2021bzv,
	title        = {{Primordial black holes as dark matter candidates}},
	author       = {Carr, Bernard and K{\"u}hnel, Florian},
	year         = 2022,
	journal      = {SciPost Phys. Lect. Notes},
	volume       = 48,
	pages        = 1,
	doi          = {10.21468/SciPostPhysLectNotes.48},
	eprint       = {2110.02821},
	archiveprefix = {arXiv},
	primaryclass = {astro-ph.CO}
}

@article{Murgia:2019duy,
	title        = {{Lyman-\ensuremath{\alpha} Forest Constraints on Primordial Black Holes as Dark Matter}},
	author       = {Murgia, Riccardo and Scelfo, Giulio and Viel, Matteo and Raccanelli, Alvise},
	year         = 2019,
	journal      = {Phys. Rev. Lett.},
	volume       = 123,
	number       = 7,
	pages        = {071102},
	doi          = {10.1103/PhysRevLett.123.071102},
	eprint       = {1903.10509},
	archiveprefix = {arXiv},
	primaryclass = {astro-ph.CO},
	reportnumber = {CERN-TH-2019-029}
}

@article{Planck:2018vyg,
	title        = {{Planck 2018 results. VI. Cosmological parameters}},
	author       = {Aghanim, N. and others},
	year         = 2020,
	journal      = {Astron. Astrophys.},
	volume       = 641,
	pages        = {A6},
	doi          = {10.1051/0004-6361/201833910},
	note         = {[Erratum: Astron.Astrophys. 652, C4 (2021)]},
	collaboration = {Planck},
	eprint       = {1807.06209},
	archiveprefix = {arXiv},
	primaryclass = {astro-ph.CO}
}

@article{Carr:2020gox,
	title        = {{Constraints on primordial black holes}},
	author       = {Carr, Bernard and Kohri, Kazunori and Sendouda, Yuuiti and Yokoyama, Jun'ichi},
	year         = 2021,
	journal      = {Rept. Prog. Phys.},
	volume       = 84,
	number       = 11,
	pages        = 116902,
	doi          = {10.1088/1361-6633/ac1e31},
	eprint       = {2002.12778},
	archiveprefix = {arXiv},
	primaryclass = {astro-ph.CO},
	reportnumber = {RESCEU-03/20; KEK-Cosmo-249; KEK-TH-2199; IPMU20-0024}
}

@article{Hawking:1971ei,
	title        = {{Gravitationally collapsed objects of very low mass}},
	author       = {Hawking, Stephen},
	year         = 1971,
	journal      = {Mon. Not. Roy. Astron. Soc.},
	volume       = 152,
	pages        = 75
}

@article{Chapline:1975ojl,
	title        = {{Cosmological effects of primordial black holes}},
	author       = {Chapline, George F.},
	year         = 1975,
	journal      = {Nature},
	volume       = 253,
	number       = 5489,
	pages        = {251--252},
	doi          = {10.1038/253251a0}
}

@article{2021arXiv211103606T,
	title        = {{GWTC-3: Compact Binary Coalescences Observed by LIGO and Virgo During the Second Part of the Third Observing Run}},
	author       = {Abbott, R. and others and {The LIGO Scientific Collaboration} and {the Virgo Collaboration} and {the KAGRA Collaboration}},
	year         = 2021,
	month        = nov,
	journal      = {arXiv e-prints},
	pages        = {arXiv:2111.03606},
	doi          = {10.48550/arXiv.2111.03606},
	eid          = {arXiv:2111.03606},
	archiveprefix = {arXiv},
	eprint       = {2111.03606},
	primaryclass = {gr-qc},
	adsurl       = {https://ui.adsabs.harvard.edu/abs/2021arXiv211103606T}
}

@article{2016PhRvX...6d1015A,
	title        = {{Binary Black Hole Mergers in the First Advanced LIGO Observing Run}},
	author       = {Abbott, B. and others and {LIGO Scientific Collaboration} and {Virgo Collaboration}},
	year         = 2016,
	month        = oct,
	journal      = {Physical Review X},
	volume       = 6,
	number       = 4,
	pages        = {041015},
	doi          = {10.1103/PhysRevX.6.041015},
	eid          = {041015},
	archiveprefix = {arXiv},
	eprint       = {1606.04856},
	primaryclass = {gr-qc},
	adsurl       = {https://ui.adsabs.harvard.edu/abs/2016PhRvX...6d1015A}
}

@article{Ballesteros:2017fsr,
	title        = {{Primordial black hole dark matter from single field inflation}},
	author       = {Ballesteros, Guillermo and Taoso, Marco},
	year         = 2018,
	journal      = {Phys. Rev. D},
	volume       = 97,
	number       = 2,
	pages        = {023501},
	doi          = {10.1103/PhysRevD.97.023501},
	eprint       = {1709.05565},
	archiveprefix = {arXiv},
	primaryclass = {hep-ph}
}

@article{2022JCAP...05..012E,
	title        = {{Simulation of primordial black holes with large negative non-Gaussianity}},
	author       = {{Escriv{\`a}}, Albert and {Tada}, Yuichiro and {Yokoyama}, Shuichiro and {Yoo}, Chul-Moon},
	year         = 2022,
	month        = may,
	journal      = {\jcap},
	volume       = 2022,
	number       = 5,
	pages        = {012},
	doi          = {10.1088/1475-7516/2022/05/012},
	eid          = {012},
	archiveprefix = {arXiv},
	eprint       = {2202.01028},
	primaryclass = {astro-ph.CO},
	adsurl       = {https://ui.adsabs.harvard.edu/abs/2022JCAP...05..012E}
}

@article{Pi:2022ysn,
	title        = {{Logarithmic Duality of the Curvature Perturbation}},
	author       = {Pi, Shi and Sasaki, Misao},
	year         = 2022,
	month        = 11,
	eprint       = {2211.13932},
	archiveprefix = {arXiv},
	primaryclass = {astro-ph.CO},
	reportnumber = {IPMU22-0060, YITP-22-144}
}

@article{2022Univ....8...66E,
	title        = {{PBH Formation from Spherically Symmetric Hydrodynamical Perturbations: A Review}},
	author       = {{Escriv{\`a}}, Albert},
	year         = 2022,
	month        = jan,
	journal      = {Universe},
	volume       = 8,
	number       = 2,
	pages        = 66,
	doi          = {10.3390/universe8020066},
	archiveprefix = {arXiv},
	eprint       = {2111.12693},
	primaryclass = {gr-qc},
	adsurl       = {https://ui.adsabs.harvard.edu/abs/2022Univ....8...66E}
}

@article{Garriga:2012bc,
	title        = {{Watchers of the multiverse}},
	author       = {Garriga, Jaume and Vilenkin, Alexander},
	year         = 2013,
	journal      = {JCAP},
	volume       = {05},
	pages        = {037},
	doi          = {10.1088/1475-7516/2013/05/037},
	eprint       = {1210.7540},
	archiveprefix = {arXiv},
	primaryclass = {hep-th}
}

@article{Planck:2018jri,
	title        = {{Planck 2018 results. X. Constraints on inflation}},
	author       = {Akrami, Y. and others},
	year         = 2020,
	journal      = {Astron. Astrophys.},
	volume       = 641,
	pages        = {A10},
	doi          = {10.1051/0004-6361/201833887},
	collaboration = {Planck},
	eprint       = {1807.06211},
	archiveprefix = {arXiv},
	primaryclass = {astro-ph.CO}
}

@article{Haro:2020qos,
	title        = {{Note on the reheating temperature in Starobinsky-type potentials}},
	author       = {Haro, Jaume and Arest\'e Sal\'o, Llibert},
	year         = 2020,
	journal      = {Gen. Rel. Grav.},
	volume       = 52,
	number       = 12,
	pages        = 116,
	doi          = {10.1007/s10714-020-02770-3},
	eprint       = {2005.14653},
	archiveprefix = {arXiv},
	primaryclass = {gr-qc}
}

@article{Liddle:2003as,
	title        = {{How long before the end of inflation were observable perturbations produced?}},
	author       = {Liddle, Andrew R and Leach, Samuel M},
	year         = 2003,
	journal      = {Phys. Rev. D},
	volume       = 68,
	pages        = 103503,
	doi          = {10.1103/PhysRevD.68.103503},
	eprint       = {astro-ph/0305263},
	archiveprefix = {arXiv}
}

@article{Dodelson:2003vq,
	title        = {{A Horizon ratio bound for inflationary fluctuations}},
	author       = {Dodelson, Scott and Hui, Lam},
	year         = 2003,
	journal      = {Phys. Rev. Lett.},
	volume       = 91,
	pages        = 131301,
	doi          = {10.1103/PhysRevLett.91.131301},
	eprint       = {astro-ph/0305113},
	archiveprefix = {arXiv},
	reportnumber = {FERMILAB-PUB-03-188-A}
}

@article{harris2020array,
	title        = {Array programming with {NumPy}},
	author       = {Charles R. Harris and K. Jarrod Millman and St{\'{e}}fan J. van der Walt and Ralf Gommers and Pauli Virtanen and David Cournapeau and Eric Wieser and Julian Taylor and Sebastian Berg and Nathaniel J. Smith and Robert Kern and Matti Picus and Stephan Hoyer and Marten H. van Kerkwijk and Matthew Brett and Allan Haldane and Jaime Fern{\'{a}}ndez del R{\'{i}}o and Mark Wiebe and Pearu Peterson and Pierre G{\'{e}}rard-Marchant and Kevin Sheppard and Tyler Reddy and Warren Weckesser and Hameer Abbasi and Christoph Gohlke and Travis E. Oliphant},
	year         = 2020,
	month        = sep,
	journal      = {Nature},
	publisher    = {Springer Science and Business Media {LLC}},
	volume       = 585,
	number       = 7825,
	pages        = {357--362},
	doi          = {10.1038/s41586-020-2649-2},
	url          = {https://doi.org/10.1038/s41586-020-2649-2}
}

@article{2020SciPy-NMeth,
	title        = {{{SciPy} 1.0: Fundamental Algorithms for Scientific Computing in Python}},
	author       = {Virtanen, Pauli and Gommers, Ralf and Oliphant, Travis E. and Haberland, Matt and Reddy, Tyler and Cournapeau, David and Burovski, Evgeni and Peterson, Pearu and Weckesser, Warren and Bright, Jonathan and {van der Walt}, St{\'e}fan J. and Brett, Matthew and Wilson, Joshua and Millman, K. Jarrod and Mayorov, Nikolay and Nelson, Andrew R. J. and Jones, Eric and Kern, Robert and Larson, Eric and Carey, C J and Polat, {\.I}lhan and Feng, Yu and Moore, Eric W. and {VanderPlas}, Jake and Laxalde, Denis and Perktold, Josef and Cimrman, Robert and Henriksen, Ian and Quintero, E. A. and Harris, Charles R. and Archibald, Anne M. and Ribeiro, Ant{\^o}nio H. and Pedregosa, Fabian and {van Mulbregt}, Paul and {SciPy 1.0 Contributors}},
	year         = 2020,
	journal      = {Nature Methods},
	volume       = 17,
	pages        = {261--272},
	doi          = {10.1038/s41592-019-0686-2},
	adsurl       = {https://rdcu.be/b08Wh}
}

@article{ZhengRuiFeng:2021zoz,
	title        = {{On primordial black holes and secondary gravitational waves generated from inflation with solo/multi-bumpy potential *}},
	author       = {Ruifeng Zheng and Shi Jiaming and Taotao Qiu},
	year         = 2022,
	journal      = {Chin. Phys. C},
	volume       = 46,
	number       = 4,
	pages        = {045103},
	doi          = {10.1088/1674-1137/ac42bd},
	eprint       = {2106.04303},
	archiveprefix = {arXiv},
	primaryclass = {astro-ph.CO}
}

@article{Wang:2021kbh,
	title        = {{Primordial black holes from the perturbations in the inflaton potential in peak theory}},
	author       = {Wang, Qing and Liu, Yi-Chen and Su, Bing-Yu and Li, Nan},
	year         = 2021,
	journal      = {Phys. Rev. D},
	volume       = 104,
	number       = 8,
	pages        = {083546},
	doi          = {10.1103/PhysRevD.104.083546},
	eprint       = {2111.10028},
	archiveprefix = {arXiv},
	primaryclass = {astro-ph.CO}
}

@article{Rezazadeh:2021clf,
	title        = {{Non-Gaussianity and secondary gravitational waves from primordial black holes production in $\alpha $-attractor inflation}},
	author       = {Rezazadeh, Kazem and Teimoori, Zeinab and Karimi, Saeid and Karami, Kayoomars},
	year         = 2022,
	journal      = {Eur. Phys. J. C},
	volume       = 82,
	number       = 8,
	pages        = 758,
	doi          = {10.1140/epjc/s10052-022-10735-w},
	eprint       = {2110.01482},
	archiveprefix = {arXiv},
	primaryclass = {gr-qc}
}

@article{Iacconi:2021ltm,
	title        = {{Revisiting small-scale fluctuations in \ensuremath{\alpha}-attractor models of inflation}},
	author       = {Iacconi, Laura and Assadullahi, Hooshyar and Fasiello, Matteo and Wands, David},
	year         = 2022,
	journal      = {JCAP},
	volume       = {06},
	number       = {06},
	pages        = {007},
	doi          = {10.1088/1475-7516/2022/06/007},
	eprint       = {2112.05092},
	archiveprefix = {arXiv},
	primaryclass = {astro-ph.CO}
}

@book{Mukhanov:2005sc,
	title        = {{Physical Foundations of Cosmology}},
	author       = {Mukhanov, V.},
	year         = 2005,
	publisher    = {Cambridge University Press},
	address      = {Oxford},
	doi          = {10.1017/CBO9780511790553},
	isbn         = {978-0-521-56398-7}
}

@article{Kristiano:2022maq,
	title        = {{Ruling Out Primordial Black Hole Formation From Single-Field Inflation}},
	author       = {Kristiano, Jason and Yokoyama, Jun'ichi},
	year         = 2022,
	month        = 11,
	eprint       = {2211.03395},
	archiveprefix = {arXiv},
	primaryclass = {hep-th},
	reportnumber = {RESCEU-20/22}
}

@article{Acharya:2020jbv,
	title        = {{CMB and BBN constraints on evaporating primordial black holes revisited}},
	author       = {Acharya, Sandeep Kumar and Khatri, Rishi},
	year         = 2020,
	journal      = {JCAP},
	volume       = {06},
	pages        = {018},
	doi          = {10.1088/1475-7516/2020/06/018},
	eprint       = {2002.00898},
	archiveprefix = {arXiv},
	primaryclass = {astro-ph.CO}
}

@article{Chluba:2020oip,
	title        = {{Thermalization of large energy release in the early Universe}},
	author       = {Chluba, Jens and Ravenni, Andrea and Acharya, Sandeep Kumar},
	year         = 2020,
	journal      = {Mon. Not. Roy. Astron. Soc.},
	volume       = 498,
	number       = 1,
	pages        = {959--980},
	doi          = {10.1093/mnras/staa2131},
	eprint       = {2005.11325},
	archiveprefix = {arXiv},
	primaryclass = {astro-ph.CO}
}

@article{Sugiyama:2012tj,
	title        = {{$\delta$N formalism}},
	author       = {Sugiyama, Naonori S. and Komatsu, Eiichiro and Futamase, Toshifumi},
	year         = 2013,
	journal      = {Phys. Rev. D},
	volume       = 87,
	number       = 2,
	pages        = {023530},
	doi          = {10.1103/PhysRevD.87.023530},
	eprint       = {1208.1073},
	archiveprefix = {arXiv},
	primaryclass = {gr-qc}
}

@article{Wainwright:2013lea,
	title        = {{Simulating the universe(s): from cosmic bubble collisions to cosmological observables with numerical relativity}},
	author       = {Wainwright, Carroll L. and Johnson, Matthew C. and Peiris, Hiranya V. and Aguirre, Anthony and Lehner, Luis and Liebling, Steven L.},
	year         = 2014,
	journal      = {JCAP},
	volume       = {03},
	pages        = {030},
	doi          = {10.1088/1475-7516/2014/03/030},
	eprint       = {1312.1357},
	archiveprefix = {arXiv},
	primaryclass = {hep-th}
}

@article{1984JCoPh..53..484B,
	title        = {{Adaptive Mesh Refinement for Hyperbolic Partial Differential Equations}},
	author       = {{Berger}, Marsha J. and {Oliger}, Joseph},
	year         = 1984,
	month        = mar,
	journal      = {Journal of Computational Physics},
	volume       = 53,
	number       = 3,
	pages        = {484--512},
	doi          = {10.1016/0021-9991(84)90073-1},
	adsurl       = {https://ui.adsabs.harvard.edu/abs/1984JCoPh..53..484B}
}

@article{Unal:2020mts,
	title        = {{Multimessenger probes of inflationary fluctuations and primordial black holes}},
	author       = {\"Unal, Caner and Kovetz, Ely D. and Patil, Subodh P.},
	year         = 2021,
	journal      = {Phys. Rev. D},
	volume       = 103,
	number       = 6,
	pages        = {063519},
	doi          = {10.1103/PhysRevD.103.063519},
	eprint       = {2008.11184},
	archiveprefix = {arXiv},
	primaryclass = {astro-ph.CO}
}

@article{2009ApJ...706L..91V,
	title        = {{Detectability of the Effect of Inflationary Non-Gaussianity on Halo Bias}},
	author       = {{Verde}, Licia and {Matarrese}, Sabino},
	year         = 2009,
	month        = nov,
	journal      = {\apjl},
	volume       = 706,
	number       = 1,
	pages        = {L91-L95},
	doi          = {10.1088/0004-637X/706/1/L91},
	archiveprefix = {arXiv},
	eprint       = {0909.3224},
	primaryclass = {astro-ph.CO},
	adsurl       = {https://ui.adsabs.harvard.edu/abs/2009ApJ...706L..91V}
}

@article{Maldacena:2002vr,
	title        = {{Non-Gaussian features of primordial fluctuations in single field inflationary models}},
	author       = {Maldacena, Juan Martin},
	year         = 2003,
	journal      = {JHEP},
	volume       = {05},
	pages        = {013},
	doi          = {10.1088/1126-6708/2003/05/013},
	eprint       = {astro-ph/0210603},
	archiveprefix = {arXiv}
}

@article{Mukhanov:1990me,
	title        = {{Theory of cosmological perturbations. Part 1. Classical perturbations. Part 2. Quantum theory of perturbations. Part 3. Extensions}},
	author       = {Mukhanov, Viatcheslav F. and Feldman, H. A. and Brandenberger, Robert H.},
	year         = 1992,
	journal      = {Phys. Rept.},
	volume       = 215,
	pages        = {203--333},
	doi          = {10.1016/0370-1573(92)90044-Z},
	reportnumber = {BROWN-HET-796, BROWN-HET-800, BROWN-HET-780}
}

@article{Tasinato:2020vdk,
	title        = {{An analytic approach to non-slow-roll inflation}},
	author       = {Tasinato, Gianmassimo},
	year         = 2021,
	journal      = {Phys. Rev. D},
	volume       = 103,
	number       = 2,
	pages        = {023535},
	doi          = {10.1103/PhysRevD.103.023535},
	eprint       = {2012.02518},
	archiveprefix = {arXiv},
	primaryclass = {hep-th}
}

@article{Carr:2009jm,
    author = "Carr, B. J. and Kohri, Kazunori and Sendouda, Yuuiti and Yokoyama, Jun'ichi",
    title = "{New cosmological constraints on primordial black holes}",
    eprint = "0912.5297",
    archivePrefix = "arXiv",
    primaryClass = "astro-ph.CO",
    reportNumber = "RESCEU-31-09, TU-852, YITP-09-112",
    doi = "10.1103/PhysRevD.81.104019",
    journal = "Phys. Rev. D",
    volume = "81",
    pages = "104019",
    year = "2010"
}

@article{Carr:2016hva,
    author = "Carr, B. J. and Kohri, Kazunori and Sendouda, Yuuiti and Yokoyama, Jun'ichi",
    title = "{Constraints on primordial black holes from the Galactic gamma-ray background}",
    eprint = "1604.05349",
    archivePrefix = "arXiv",
    primaryClass = "astro-ph.CO",
    reportNumber = "RESCEU-16-16, KEK-TH-1895, KEK-COSMO-193",
    doi = "10.1103/PhysRevD.94.044029",
    journal = "Phys. Rev. D",
    volume = "94",
    number = "4",
    pages = "044029",
    year = "2016"
}

@article{Domenech:2016zxn,
    author = "Domenech, Guillem and Gong, Jinn-Ouk and Sasaki, Misao",
    title = "{Consistency relation and inflaton field redefinition in the \ensuremath{\delta}N formalism}",
    eprint = "1606.03343",
    archivePrefix = "arXiv",
    primaryClass = "astro-ph.CO",
    reportNumber = "APCTP-PRE2016-014, YITP-16-70",
    doi = "10.1016/j.physletb.2017.04.014",
    journal = "Phys. Lett. B",
    volume = "769",
    pages = "413--417",
    year = "2017"
}

@article{Fumagalli:2023hpa,
    author = "Fumagalli, Jacopo",
    title = "{Absence of one-loop effects on large scales from small scales in non-slow-roll dynamics}",
    eprint = "2305.19263",
    archivePrefix = "arXiv",
    primaryClass = "astro-ph.CO",
    month = "5",
    year = "2023"
}

@article{Kawaguchi:2023mgk,
    author = "Kawaguchi, Ryodai and Fujita, Tomohiro and Sasaki, Misao",
    title = "{Highly asymmetric probability distribution from a finite-width upward step during inflation}",
    eprint = "2305.18140",
    archivePrefix = "arXiv",
    primaryClass = "astro-ph.CO",
    reportNumber = "WUCG-23-07, YITP-23-69",
    month = "5",
    year = "2023"
}

@article{Cai:2022erk,
    author = "Cai, Yi-Fu and Ma, Xiao-Han and Sasaki, Misao and Wang, Dong-Gang and Zhou, Zihan",
    title = "{Highly non-Gaussian tails and primordial black holes from single-field inflation}",
    eprint = "2207.11910",
    archivePrefix = "arXiv",
    primaryClass = "astro-ph.CO",
    reportNumber = "YITP-22-48",
    doi = "10.1088/1475-7516/2022/12/034",
    journal = "JCAP",
    volume = "12",
    pages = "034",
    year = "2022"
}

@book{van2009python,
 author = {Van Rossum, Guido and Drake, Fred L.},
 title = {Python 3 Reference Manual},
 year = {2009},
 isbn = {1441412697},
 publisher = {CreateSpace},
 address = {Scotts Valley, CA}
}

@article{He:2023yvl,
    author = "He, Jibin and Deng, Heling and Piao, Yun-Song and Zhang, Jun",
    title = "{Implications of GWTC-3 on primordial black holes from vacuum bubbles}",
    eprint = "2303.16810",
    archivePrefix = "arXiv",
    primaryClass = "astro-ph.CO",
    month = "3",
    year = "2023"
}

@article{Deng:2018cxb,
    author = "Deng, Heling and Vilenkin, Alexander and Yamada, Masaki",
    title = "{CMB spectral distortions from black holes formed by vacuum bubbles}",
    eprint = "1804.10059",
    archivePrefix = "arXiv",
    primaryClass = "gr-qc",
    doi = "10.1088/1475-7516/2018/07/059",
    journal = "JCAP",
    volume = "07",
    pages = "059",
    year = "2018"
}

@ARTICLE{1979A&A....80..104N,
       author = {{Novikov}, I.~D. and {Polnarev}, A.~G. and {Starobinskii}, A.~A. and {Zeldovich}, Ia. B.},
        title = "{Primordial black holes}",
      journal = {\aap},
         year = 1979,
        month = nov,
       volume = {80},
       number = {1},
        pages = {104-109},
       adsurl = {https://ui.adsabs.harvard.edu/abs/1979A&A....80..104N}
}

@article{Motohashi:2023syh,
    author = "Motohashi, Hayato and Tada, Yuichiro",
    title = "{Squeezed bispectrum and one-loop corrections in transient constant-roll inflation}",
    eprint = "2303.16035",
    archivePrefix = "arXiv",
    primaryClass = "astro-ph.CO",
    doi = "10.1088/1475-7516/2023/08/069",
    journal = "JCAP",
    volume = "08",
    pages = "069",
    year = "2023"
}

@article{Tada:2023rgp,
    author = "Tada, Yuichiro and Terada, Takahiro and Tokuda, Junsei",
    title = "{Cancellation of quantum corrections on the soft curvature perturbations}",
    eprint = "2308.04732",
    archivePrefix = "arXiv",
    primaryClass = "hep-th",
    reportNumber = "CTPU-PTC-23-31",
    month = "8",
    year = "2023"
}

\end{document}